\documentclass[english,aps,superscriptaddress]{revtex4}
\usepackage[T1]{fontenc}
\usepackage[latin9]{inputenc}
\usepackage{amstext}
\usepackage{graphicx}
\usepackage{esint}

\makeatletter

\DeclareRobustCommand{\greektext}{%
  \fontencoding{LGR}\selectfont\def\encodingdefault{LGR}}
\DeclareRobustCommand{\textgreek}[1]{\leavevmode{\greektext #1}}
\DeclareFontEncoding{LGR}{}{}
\DeclareTextSymbol{\~}{LGR}{126}

\@ifundefined{textcolor}{}
{%
 \definecolor{BLACK}{gray}{0}
 \definecolor{WHITE}{gray}{1}
 \definecolor{RED}{rgb}{1,0,0}
 \definecolor{GREEN}{rgb}{0,1,0}
 \definecolor{BLUE}{rgb}{0,0,1}
 \definecolor{CYAN}{cmyk}{1,0,0,0}
 \definecolor{MAGENTA}{cmyk}{0,1,0,0}
 \definecolor{YELLOW}{cmyk}{0,0,1,0}
}
\usepackage{color}

\makeatother

\usepackage{babel}
\begin{document}

\title{Physics perspectives of heavy-ion collisions at very high energy}

\author{Ning-bo Chang}
\affiliation{Key Laboratory of Quark and Lepton Physics (MOE) and Institute of
Particle Physics, Central China Normal University, Wuhan 430079, China}

\author{Shanshan Cao}
\affiliation{Nuclear Science Division MS70R0319, Lawrence Berkeley National Laboratory, Berkeley, CA 94720}

\author{Bao-yi Chen}
\affiliation{Physics Department, Tsinghua University, Beijing 100084, China}

\author{Shi-yong Chen}
\affiliation{Key Laboratory of Quark and Lepton Physics (MOE) and Institute of
Particle Physics, Central China Normal University, Wuhan 430079, China}

\author{Zhen-yu Chen}
\affiliation{Physics Department, Tsinghua University, Beijing 100084, China}

\author{Heng-Tong Ding}
\affiliation{Key Laboratory of Quark and Lepton Physics (MOE) and Institute of
Particle Physics, Central China Normal University, Wuhan 430079, China}

\author{Min He}
\affiliation{Department of Applied Physics, Nanjing University of Science and Technology, Nanjing 210094, China}

\author{Zhi-quan Liu}
\affiliation{Key Laboratory of Quark and Lepton Physics (MOE) and Institute of
Particle Physics, Central China Normal University, Wuhan 430079, China}

\author{Long-gang Pang}
\affiliation{Key Laboratory of Quark and Lepton Physics (MOE) and Institute of
Particle Physics, Central China Normal University, Wuhan 430079, China}

\author{Guang-you Qin}
\affiliation{Key Laboratory of Quark and Lepton Physics (MOE) and Institute of
Particle Physics, Central China Normal University, Wuhan 430079, China}

\author{Ralf Rapp}
\affiliation{Cyclotron Institute and Department of Physics and Astronomy, Texas A\&M University, College Station, TX 77843, USA}

\author{Bj\"orn Schenke}
\address{Physics Department, Brookhaven National Laboratory, Upton, NY 11973, USA}

\author{Chun Shen}
\address{Department of Physics, McGill University, Montreal, Quebec, H3A 2T8, Canada}

\author{Huichao Song}
\affiliation{Department of Physics and State Key Laboratory of Nuclear Physics and Technology, Peking
University, Beijing 100871, China}

%\affiliation{Collaborative Innovation Center of Quantum Matter, Beijing 100871, China}
%\affiliation{Center for High Energy Physics, Peking University, Beijing100871, China}

\author{Hao-jie Xu}
\affiliation{Department of Modern Physics, University of Science and Technology
of China, Hefei, Anhui 230026, China}

\author{Qun Wang}
\affiliation{Department of Modern Physics, University of Science and Technology
of China, Hefei, Anhui 230026, China}

\author{Xin-Nian Wang}
%\email[Corresponding author: ]{xnwang@mail.ccnu.edu.cn}
\affiliation{Key Laboratory of Quark and Lepton Physics (MOE) and Institute of
Particle Physics, Central China Normal University, Wuhan 430079, China}
\affiliation{Nuclear Science Division MS70R0319, Lawrence Berkeley National Laboratory, Berkeley, CA 94720}

\author{Ben-wei Zhang}
\affiliation{Key Laboratory of Quark and Lepton Physics (MOE) and Institute of
Particle Physics, Central China Normal University, Wuhan 430079, China}

\author{Han-zhong Zhang}
\affiliation{Key Laboratory of Quark and Lepton Physics (MOE) and Institute of
Particle Physics, Central China Normal University, Wuhan 430079, China}

\author{Xiangrong Zhu}
\affiliation{Department of Physics and State Key Laboratory of Nuclear Physics and Technology, Peking University, Beijing 100871, China}

\author{Peng-fei Zhuang}
\affiliation{Physics Department, Tsinghua University, Beijing 100084, China}

\begin{abstract}

Heavy-ion collisions at very high colliding energies are expected to produce a 
quark-gluon plasma (QGP) at the highest temperature obtainable in a laboratory setting. 
Experimental studies of these reactions can provide an unprecedented range of information 
on properties of the QGP at high temperatures. We report theoretical investigations of the 
physics perspectives of heavy-ion collisions at a future high-energy collider. These include 
initial parton production, collective expansion of the dense medium, jet quenching, 
heavy-quark transport, dissociation and regeneration of quarkonia, photon and dilepton production. 
We illustrate the potential of future experimental studies of the initial particle production 
and formation of QGP at the highest temperature to provide constraints on properties of  
strongly interaction matter. 

\end{abstract}

\maketitle

\section{Introduction}

The fundamental theory of strong interactions among quarks and gluons is Quantum Chromodynamics
(QCD). Because of the non-Abelian nature of the strong interaction caharacterized by the $SU(3)$ 
gauge symmetry in QCD, quarks and gluons are confined within the realm of hadrons which are the 
only stable vacuum excitations. The approximate chiral symmetry among light quarks is spontaneously 
broken in the vacuum giving rise to non-zero quark condensates and the light pions as Goldstone 
bosons. The approximate conformal symmetry is also broken by quantum effects leading to a 
non-vanishing gluon condensate and a running strong coupling constant.
Under conditions of extremely high temperature and/or density, one expects the boundary between 
hadrons to disappear and quark and gluon degrees of freedom are liberated to form a new state of 
matter called quark gluon plasma (QGP). According to lattice-discretized numerical studies of 
QCD (LQCD)~\cite{Ding:2015ona}, a rapid cross-over from hadronic matter to QGP 
occurs around a pseudo-critical temperature $T_{\rm c}\approx 155$ MeV at zero baryon 
chemical potential, characterized by restoration of the chiral symmetry. 
Below $T_{\rm c}$, quarks and gluons are confined in color-neutral hadrons  
in the form of a hadronic resonance matter. These hadrons melt during the deconfinement phase 
transition. At temperatures above $T_{\rm c}$ quarks and gluons can roam freely throughout 
a volume much larger than the nucleon size. The deconfinement phase transition is caused by 
breaking of the $Z_3$ center symmetry (which becomes exact in pure gauge QCD, {\it i.e.}, without
quark fields) at high temperature, which is characterized by a rapid change of the corresponding 
order parameter, the expectation value of the Polyakov loop.
%The chiral symmetry is also approximately restored at these high temperatures above $T_c$.

Such a new state of matter of very high temperatures and densities prevailed in the early 
Universe as the quark epoch from $10^{-12}$ to $10^{-6}$ seconds after the Big Bang.
It might still exist today in compact stellar objects such as neutron stars. In order to 
create this new state of matter in the laboratory, one accelerates two heavy nuclei close 
to the speed of light and collides them head-on. In these high-energy heavy-ion collisions, 
a large fraction of the colliding energy is converted into an initial matter of extremely 
high temperatures and densities, well beyond the phase transition region to form a QGP.
Currently, two major facilities for high-energy heavy-ion collision experiments are being 
operated, the Relativistic Heavy-Ion Collider (RHIC) at Brookhaven National Laboratory (BNL)
and the Large Hadron Collider (LHC) at the European Organization for Nuclear Research (CERN).
From its start in 2000 until 2010, RHIC was the highest-energy heavy-ion collider in the world.
In November 2010 the LHC took the lead as the heavy-ion collider running at the highest energy.

Remarkable discoveries have been made at RHIC since commencing its operation in 
2000~\cite{Adcox:2004mh,Adams:2005dq,Jacobs:2004qv}, with multiple evidence pointing at 
the formation of a strongly-coupled QGP (sQGP) in central Au+Au collisions at its maximum
energy. One surprising discovery is that the hot and dense QCD matter created in relativistic
heavy-ion collisions develops a strong collective flow characteristic of a strongly-coupled 
liquid, rather than of a weakly-coupled gas of quarks and gluons. In fact, the shear viscosity 
to entropy density ratio extracted from comparisons between experimental data and viscous 
hydrodynamic calculations is so low~\cite{Heinz:2013th} that it has been termed 
\textquotedbl{}the perfect liquid\textquotedbl{}. The second discovery at RHIC is the 
observation of substantial jet quenching~\cite{Wang:2004dn}, indicating that the matter is 
virtually opaque to energetic quarks and gluons. Differences in the yields and flow of 
baryons versus mesons indicate that hadron formation at intermediate transverse momenta 
proceeds via coalescence of constituent quarks, providing evidence for partonic collectivity 
in the observed hadron spectra~\cite{Adams:2005dq}. In fact, even heavy quarks were found to 
exhibit substantial collectivity and suppression indicating their approach to thermalization with a 
small diffusion coefficient in the strongly interacting medium~\cite{Adare:2006nq}. 
The STAR experiment has also identified 
anti-hypertriton and anti-alpha production in Au+Au collisions, the first ever observation 
of an anti-hypernucleus and anti-alpha~\cite{Abelev:2010rv}.

With more than one order of magnitude higher colliding energy, many of the proposed signals 
for the QGP became much stronger and easier to observe at the LHC~\cite{Abreu:2007kv}. The 
dense matter created in heavy-ion collisions at LHC energies is much hotter and has a longer 
lifetime of its dynamical evolution. The QGP matter has also a smaller net baryon density as 
compared to that at RHIC. With increased colliding energy, the rates of hard processes are 
much higher than at RHIC making them much better and easily accessible probes of the QGP matter.  
Recent experimental data from heavy-ion collisions at LHC unambiguously confirmed all 
experimental evidences of the QGP as first observed at RHIC \cite{Muller:2012zq}. The collective 
phenomena as manifested in anisotropic flows and a ridge structure with a large pseudo-rapidity 
gap in hadron production yields in the most central Pb+Pb collisions point to a QGP at high 
temperatures with small specific shear viscosity. Jet quenching phenomena are clearly observed 
with jet energies up to hundreds of GeV both in the single-inclusive hadron spectra and 
reconstructed jets. The mass dependence of the quark energy loss is observed for the first 
time according to high $p_T$ suppression of charm mesons and non-prompt $J/\psi$ originating from
bottom mesons. The centrality dependence of $J/\psi$ production clearly shows the increasing 
fraction of $J/\psi$'s from recombination charm and anti-charm quarks in the QGP medium. 
Recent data also bear strong evidence for collectivity in high-multiplicity events of p+Pb 
collisions at the LHC.

In the near future, the focus of heavy-ion collisions at RHIC and LHC will be on a
quantitative characterization of the strongly coupled QGP using rare probes such as 
large transverse momentum jets, heavy flavor particles, real and virtual photons and 
quarkonia states. Studies of collective phenomena using detailed multiple particle 
correlations can provide precision constraints on the bulk transport coefficients of the QGP.  
Since existing RHIC and LHC data have already provided tantalizing hints on the weakening 
of the interaction strength both among bulk partons \cite{Song:2011qa,Gale:2012rq} and 
between hard probes and the bulk medium \cite{Burke:2013yra}, it will be extremely 
interesting to see whether such trends continue at future even higher collider energies 
and eventually reach the weakly interacting scenario as predicted by pQCD. 

Given the state of the accelerator technology and interests in particle physics going beyond 
the discovery of the Higgs boson, new proposals for hadron and heavy-ion colliders at tens of 
TeV center of mass energy per nucleon pair have been envisioned~\cite{vlhc,cepc}. One can address 
many important questions in future heavy-ion collision experiments in the energy range from tens 
to hundreds of TeV. These include:

(a) What is the equation of state (EoS) for the strongly interacting matter
at high temperatures?  Do effects of charm quarks start to become significant in the EoS?

(b) What is the thermalization mechanism, and how does the thermalization time 
depend on the colliding energy?

(c) What are the transport properties of strongly interacting matter
at the highest temperatures probed by high-energy jets and collective phenomena? 
Are they approaching the weak coupling values as predicted by perturbative QCD?

(d) What is the nature of the initial state and its fluctuations in nuclear collisions?

(e) Can we find other exotic hadrons or nuclei such as light multi-\textgreek{L}
hyper-nuclei, bound states of (\textgreek{LL}) or the H di-baryon?

(f) What are the fundamental symmetries of QCD at high temperatures? 
How does the restoration of the spontaneously broken chiral symmetry manifest
itself in the electromagnetic radiation from the medium?
Is the axial $U_A(1)$ symmetry effectively restored and what are the possible
consequences in the hadron yields?

The answers to these important questions in strong interactions rely on both theoretical 
advances and experimental programs of high-energy electron-nuclei (proton) and heavy-ion 
collisions at future high-energy collider facilities. In this report, we will give a brief 
review of the physics potentials of heavy-ion collision at energy scales of tens or hundreds 
of TeV. The scope of this report is limited to a few selected topics listed above. A more 
comprehensive report will need a much more concerted and dedicated effort.

\section{QCD and strong interaction matter }

The quantum chromodynamics (QCD), as a non-Abelian quantum gauge field theory, has been very successful in describing the strong interaction among quarks and gluons that are the fundamental constituents of visible matter in nature. The asymptotic freedom of QCD at short distances renders the possibility of calculating  hard processes via perturbative methods.  On the other hand, its non-perturbative features at long distances are only systematically computable using numerical simulations in a path-integral representation of QCD.  Many of our current theoretical understanding of properties of dense matter at high temperature and baryon density are based on lattice QCD. Though experiments at RHIC and LHC have confirmed the existence of a new form of matter, strongly coupled quark-gluon plasma (sQGP),  in relativistic heavy-ion collisions, its properties is not yet fully understood. This requires future efforts from both experimental and theoretical studies.  In addition, lattice QCD calculations can provide crucial inputs to phenomenological studies of QGP properties.

Lattice QCD is a discretized version of QCD in the Euclidean space and time which reproduces QCD in the continuum limit when the lattice spacing goes to zero. Most lattice QCD calculations which are relevant to heavy-ion collisions have been
performed using non-chiral fermions which recover the flavor or chiral symmetry of QCD only in the continuum limit, e.g. staggered and
Wilson fermions. Chiral fermions are generally much more expensive to work with. However, with continued increase of 
the available computing power owing to Moore's law,  these actions are also currently used and start to produce interesting 
results in QCD thermodynamics, e.g. the confirmation of the value of the crossover temperature $T_c$ ~\cite{Bhattacharya:2014ara}
and investigations of the restoration of $U(1)_A$ symmetry ~\cite{Cossu:2013uua,Bazavov:2012qja, Buchoff:2013nra}.

\subsection{QCD transition and QCD equation of state}

%{\em QCD Transition and the equation of state}

\begin{figure}
\includegraphics[scale=1.0]{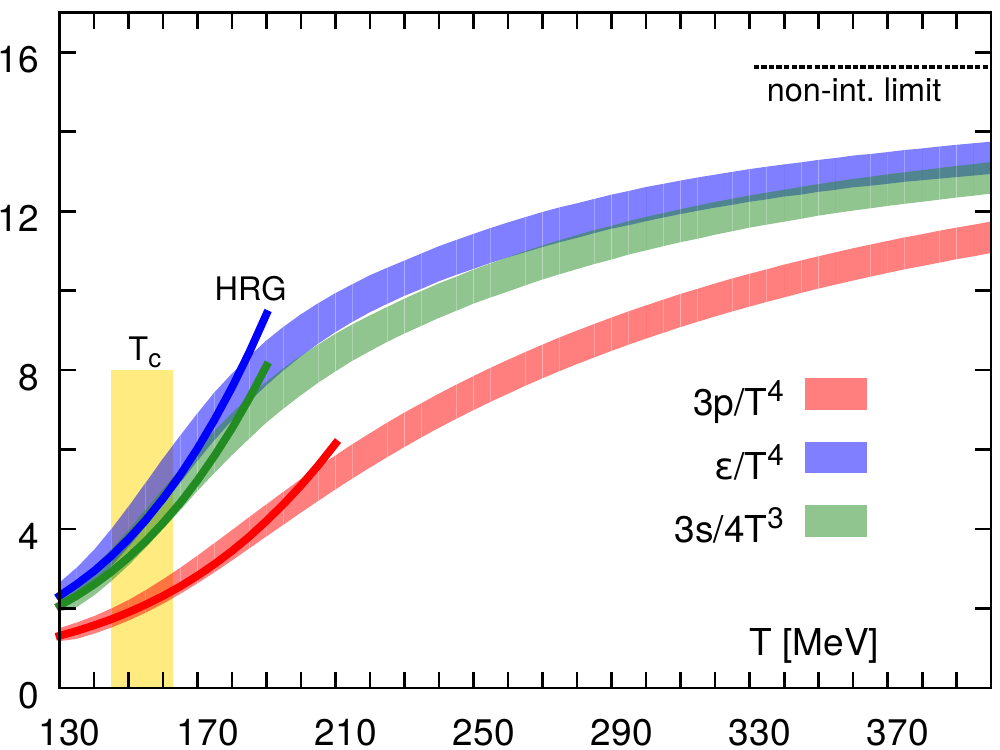}
\caption{The pressure, energy and entropy density (scaled by $T^4$) as functions of the temperature from lattice QCD calculation by the HotQCD Collaboration \cite{Bazavov:2014pvz} (shaded bands) as compared to hadron resonance gas (HRG) model results (solid lines).}
\label{fig:hotqcd_eos}
\end{figure}

The equation of state (EoS) of QCD matter contains information about the change of degrees of freedom
in different regimes of temperature and baryon density. It is one of the important ingredients to model
the evolution of the fireball produced in heavy-ion collisions through classical hydrodynamic equations.
Computation of the QCD EoS has been one of the major goals in the field of lattice QCD since 1980~\cite{Engels:1980ty}. At zero baryon number density it has been shown very recently with lattice calculations for $N_f=2+1$ that the QCD equation of state obtained from the HotQCD and Wuppertal-Budapest collaborations by using two different discretization schemes agree very well~\cite{Bazavov:2014pvz,Borsanyi:2013bia}.  Shown in Fig.~\ref{fig:hotqcd_eos} are energy density, entropy density and pressure as functions of temperature from the HotQCD Collaboration \cite{Bazavov:2014pvz} (shaded bands). There is apparently a rapid transition from low to high temperature. It has been established from the analysis of chiral condensates that this transition in QCD with its physical mass spectrum is a rapid crossover at zero baryon density. The pseudo critical temperature of the QCD transition is confirmed to be $T_c \simeq 155~$MeV~\cite{Bazavov:2011nk,Borsanyi:2010bp,Bhattacharya:2014ara}. Below and around this crossover, the EoS can be described well by a hadron resonance gas model (solid lines). In the high temperature region, lattice QCD calculations of  EoS and other observables, e. g.  fluctuations of conserved charges can be compared to perturbative calculations~\cite{Bazavov:2013uja,Bazavov:2014pvz,Ding:2015fca,Bellwied:2015lba}. Such comparisons can provide the window of applicability for  perturbation calculations and test whether the system is in the weakly coupled regime at high temperatures. In the case of $N_f=2+1+1$ QCD,  the inclusion of charm quarks may have some effects on the QCD equation of state which might be noticeable at higher temperatures reached in heavy-ion collisions at 30 TeV scale~\cite{Philipsen:2012nu,Borsanyi:2014rza}.

There is also some evidence that 2 or 2+1-flavor QCD in the ``chiral'' limit, {\it i.e.} vanishing light quark masses with
the strange quark mass being at its physical value, is second-order and belongs to the universality class  of the three-dimensional $O(N)$ spin models~\cite{Ejiri:2009ac,Ding:2013lfa}. If confirmed, this would be in accordance with the picture of Pisarski and Wilczek~\cite{Pisarski:1983ms}.  However, existing studies of $O(N)$ scaling have been performed on rather coarse lattices with
staggered fermion actions that are no longer state-of-the-art. They lead to large taste violations.
Therefore the order of the QCD phase transition in the chiral limit is still under
debate and arguments in favor of a first-order transition have been put forward~\cite{Aoki:2012yj}.

Properties of light mesons (e.g. $\rho$) and heavy quarkonia (e.g. $J/\psi$ and $\Upsilon$) as measured via the dilepton channel can serve as useful probes for the chiral symmetry restoration and deconfinement transition in the QCD medium, respectively.
Theoretical study of these hadron properties at finite temperature requires the computation of two-point correlation functions on the lattice and extraction of hadron spectral functions. These hadron spectral functions are directly related to thermal dilepton rates,
the dissociation of quarkonia states as well as transport properties of the medium, e.g. electrical conductivity and heavy quark diffusion coefficients.

The most current lattice QCD study of hadron spectral functions suggests that all charmonia 
dissociate at $T\gtrsim1.5~T_c$ in gluonic plasma~\cite{Ding:2012sp}. Very recent lattice QCD studies including 
dynamic quarks on screening mass and spectral functions suggest the same picture~\cite{Bazavov:2014cta,Borsanyi:2014vka}.
 Due to the large value of heavy bottom quark mass, a direct study of bottomonia is very hard on the lattice since the 
 lattice spacing $a$ has to be much smaller than the inverse of the heavy quark mass. Effective theories,
e.g. non-relativistic QCD (NRQCD), have been put on the lattice to study properties of bottomonia. 
It has been found that all S wave states exist at temperatures up to at least 2 $T_c$ and
P wave states melt just above $T_c$ \cite{Aarts:2010ek,Aarts:2013kaa,Aarts:2012ka,Aarts:2014cda,Aarts:2011sm}.
However, a different observation is found in Ref.~\cite{Kim:2014iga} that P waves states might stay bounded at 
higher temperatures above $T_c$ by using a new inversion method.
It has also been realized that the potential of static quarks in the medium is complex~\cite{Laine:2006ns,Beraudo:2007ky,Brambilla:2008cx} whose computation on the lattice has been carried out \cite{Burnier:2014ssa,Burnier:2013nla,Rothkopf:2011db,Burnier:2012az}.

The fate of heavy-light mesons or baryons also reflects the change of relevant degrees of freedom in the strong interaction matter.
For instance, the abundance of strange hadrons is considered as one of the signals for the formation of QGP.
Investigations of fluctuations and correlations of electrical charge and baryon number with strangeness and charm 
found that both open strange and open charm hadrons start to dissociate in the temperature region
of the chiral crossover ~\cite{Bazavov:2013dta,Bazavov:2014yba,Bellwied:2013cta}.

As proposed recently in Ref.~\cite{Bazavov:2012vg} , hadron chemical freeze-out temperatures and baryon chemical potentials
can be determined by matching lattice QCD computations with those measured in heavy-ion collisions.
An upper band of freeze-out temperature is found to be 148$\pm4~$MeV~\cite{Borsanyi:2014ewa}. An indirect evidence
of experimentally yet unobserved open strange and open charm hadrons has been found~\cite{Bazavov:2014yba,Bazavov:2014xya}.
These unobserved hadrons bring down the freeze-out temperature in the strange hadron 
sector by $\sim 5-8$~MeV~\cite{Bazavov:2014xya}.

\subsection{Transport coefficients}

Transport properties of the hot QCD medium are also the focus of future experimental studies through collective phenomena of both light and heavy flavor hadrons and electromagnetic emissions.  Currently there are only a limited number of results on transport coefficients from lattice-QCD calculations with dynamical quarks.
Most calculations have been performed in the quenched limit at vanishing net-baryon number density~\cite{Ding:2011hr,Ding:2012ar,Ding:2014xha}. It proves difficult to extract transport coefficients directly from imaginary-time two-point correlation functions. Currently, the maximum entropy method (MEM) is a commonly used technique to achieve this goal~\cite{Asakawa:2000tr}.
The determination of the electrical conductivity and the heavy-quark diffusion coefficient
in full QCD is rather straightforward and is mainly limited by computational resources. However, the determination of fluid-dynamical transport coefficients, e.g. viscosities, is hampered by large noise-to-signal ratios. For QCD in the quenched approximation, noise reduction techniques are known and are applied while for full QCD computations  such algorithms still need to be developed.

Electrical conductivity has been computed in the continuum limit in quenched QCD at three temperatures above $T_c$~\cite{Ding:2010ga,Ding:2014dua}. Recently computation has also been performed on the lattice with dynamic quarks~\cite{Brandt:2012jc,Aarts:2014nba,Amato:2013naa}. The charm-quark diffusion coefficient has been obtained at one value of the lattice cutoff
and three temperatures in the deconfined phase~\cite{Ding:2011hr}. Currently, there are no lattice results on bottom-quark diffusion coefficients which are very important in heavy-quark physics at LHC energies and beyond.
The heavy-quark diffusion coefficients have also been studied on the lattice by measuring proposed observables in heavy-quark effective theory~\cite{CaronHuot:2009uh}. Results on heavy-quark diffusion coefficients obtained in this approach are close to the charm-quark diffusion coefficients~\cite{Francis:2011gc,Banerjee:2011ra,Kaczmarek:2014jga}. However,
most of these results are also obtained at a finite lattice cutoff, so a reliable extraction of diffusion coefficients needs to be performed.

Shear and bulk viscosities have been calculated a few years ago on rather coarse and small lattices, without a continuum extrapolation\cite{Meyer:2007ic,Meyer:2007dy}. In order to obtain better results, the number of gauge field configurations needs to be increased by an order of magnitude. However, algorithms like multi-level updates to improve the signal-to-noise ratio~\cite{Meyer:2003hy} of two-point correlators of the energy-momentum tensor currently used in the
quenched approximation are not applicable in full QCD. Recently, there have been efforts \cite{Moore:2010bu,Moore:2012tc,Denicol:2012cn}  to determine some of the 2nd-order transport coefficients from a first-principles calculation on lattice.

\section{Bulk properties of matter in heavy-ion collisions}

In the study of QGP properties in high-energy heavy-ion collisions, the space-time evolution of the bulk matter
underpins all experimental and phenomenological studies since it will affect all the expected final observables from which
one extracts medium properties of the QGP.  Whether it is an effective theory such as relativistic hydrodynamics or a Monte
Carlo model for parton and hadron transport, one always needs the basic information of initial parton production. The
initial parton production determines the initial energy density or temperature at the thermalization time
and its fluctuation in both transverse area and longitudinal direction. Given these initial conditions, one can
then use the hydrodynamical model or parton-hadron transport model for the space-time evolution of the bulk medium. Through comparisons between hydrodynamic or transport results and experimental data on the final hadron spectra and their azimuthal anisotropy or multiple hadron correlations, one can extract values of the bulk transport coefficients such as shear and bulk viscosity. For the study of other hard and electromagnetic signals, one also has to reply on the space-time evolution of the bulk medium to understand the experimental measurements and extract medium properties such as initial temperature, flow velocity and jet transport
coefficients.

\subsection{Multiplicity}

\label{subsec-mult}

\begin{figure}
\includegraphics[scale=0.6]{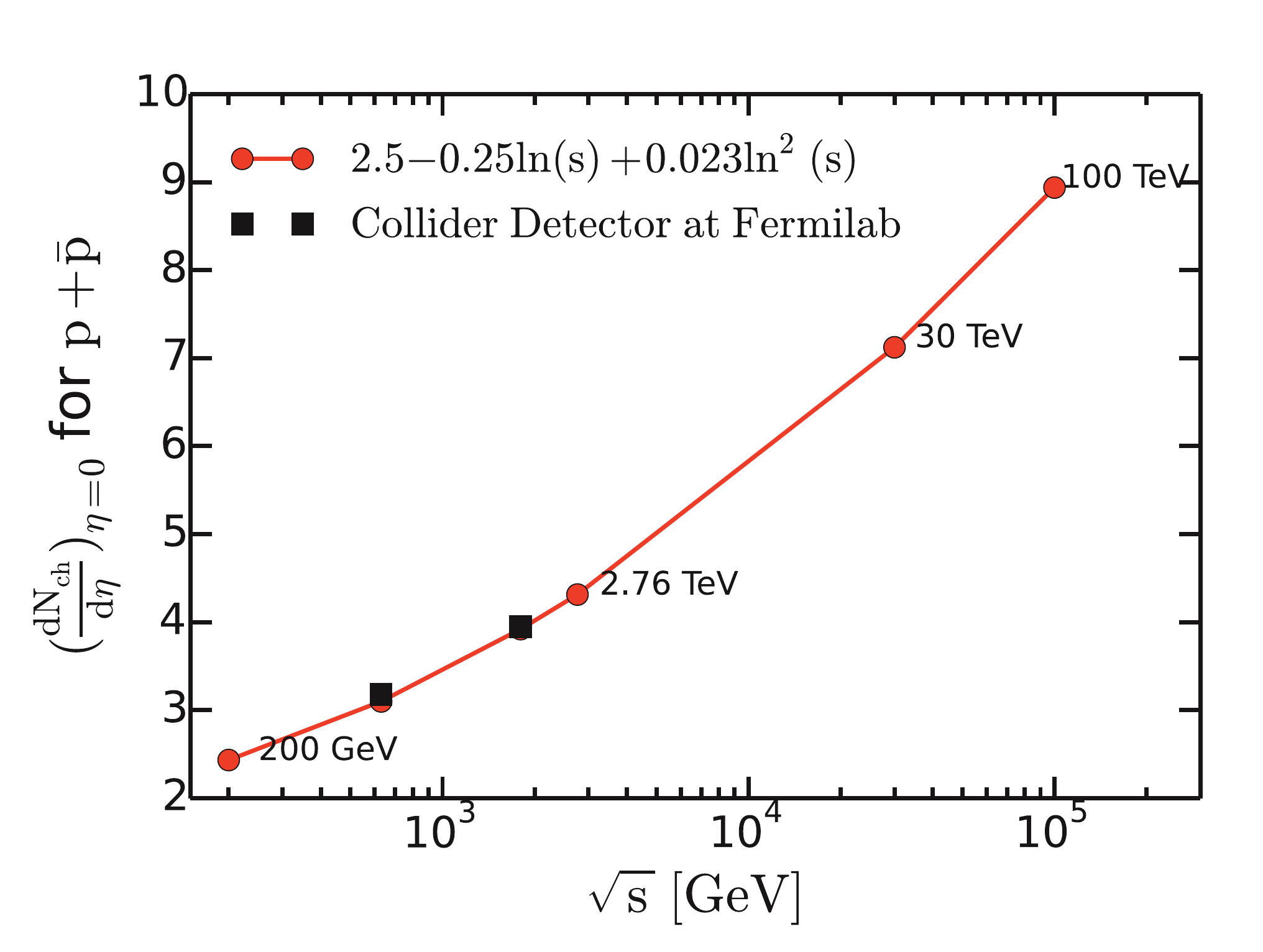}
\caption{The charged multiplicity for $p+p(\bar{p})$ collisions as a function of the colliding energy, as extrapolated from
experimental data at Fermilab Tevatron, BN RHIC and CERN LHC.}
 \label{fig:dndeta_vs_pp}
\end{figure}

The mechanism of initial parton production has been one of the  fundamental problems in heavy-ion collisions and strong interaction in general. It is determined by the properties of strong interaction at high energy where non-linear aspects of QCD are at play and it is also the focus of research at the future electron-ion colliders (EIC). Shown in Fig. \ref{fig:dndeta_vs_pp} is the charged hadron multiplicity in $p+p(\bar p)$ collisions as a function of the colliding energy as extrapolated from experimental data at Fermilab Tevatron \cite{Abe:1989td}, BNL RHIC \cite{Nouicer:2004ke} and CERN LHC \cite{Aamodt:2009aa} to very high energies. This extrapolation is also consistent with HIJING calculations \cite{Deng:2010mv} in which the rise of the multiplicity in the central rapidity region at high colliding energy is mainly caused by the increase of gluonic mini-jet production with the large initial gluon distribution inside the beam proton at small momentum fraction.

There are currently two types of  pQCD based models for the description of initial parton production in heavy-ion collisions.  HIJING Monte Carlo model \cite{Wang:1991hta,Gyulassy:1994ew,Deng:2010mv} employs the Glauber model for multiple interaction in high-energy nucleon-nucleus and nucleus-nucleus collisions. It includes both the incoherent hard and semi-hard parton scattering that are described by pQCD and the coherent soft interaction via excitation of remanent strings between valence quarks and diquarks. Initial parton production from incoherent hard or semi-hard parton scatterings is proportional to the number of binary nucleon-nucleon collisions $N_{\mathrm{coll}}$ while the soft parton production from string excitation is proportional to the number of participant nucleons $N_{\mathrm{part}}$ in a given centrality. One should also take into account the impact-parameter-dependent nuclear modification of parton distributions in the semi-hard parton interaction. This will introduce additional impact-parameter dependence of the parton production in the hard or semi-hard parton scattering. The final centrality dependence  of the initial parton multiplicity from both soft and semi-hard processes will therefore be a linear combination of $N_{\mathrm{part}}$ and $N_{\rm coll}$.

\begin{figure}
\includegraphics[scale=1.0]{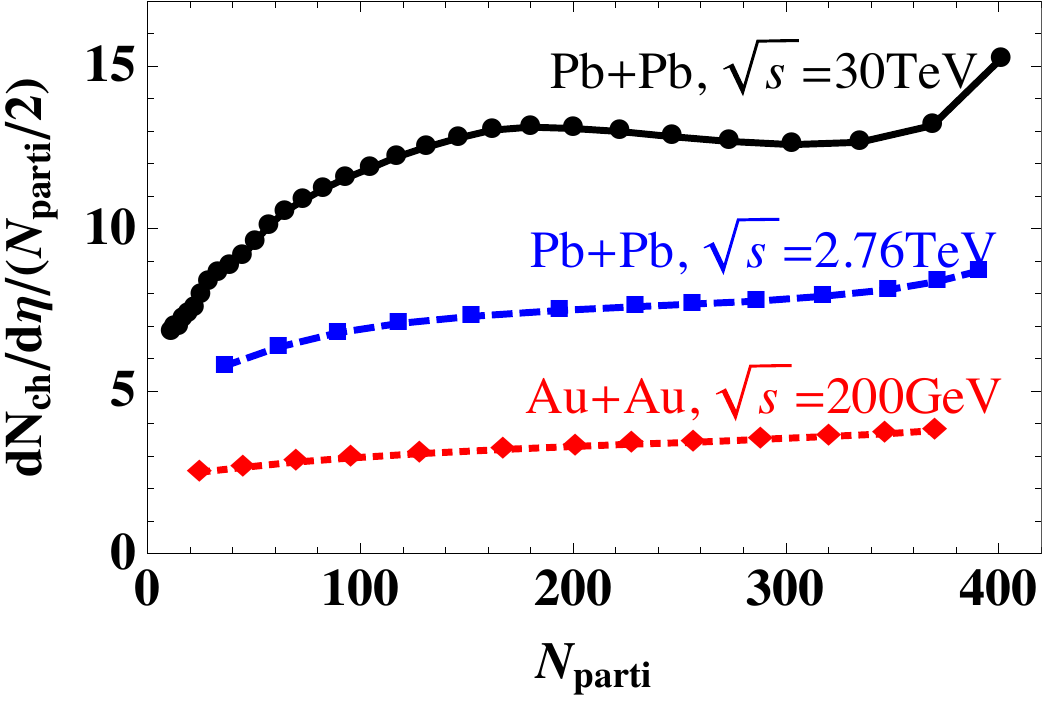}
\caption{Rapidity density of charged hadrons in the central rapidity per participant pair as functions of the number of participant nucleons in Au+Au collisions at RHIC, Pb+Pb collisions at LHC and at $\sqrt{s}=30$ TeV, from HIJING Monte Carlo simulations.}
 \label{fig:dndetahijing}
\end{figure}

The average number of participant nucleons or wounded nucleons $N_{\mathrm{part}}$ in  heavy-ion collisions as a function of the impact-parameter can be calculated within the Glauber model in terms the overlapping functions of two nuclei \cite{Kolb:2000sd}. It  can reach the limit of the total number of nucleons within the overlap region of two colliding nuclei. It therefore has a very weak energy dependence in very high energies. The number of binary collisions depends almost linearly on the total inelastic cross section and therefore has a strong energy dependence. Correspondingly, the final hadron multiplicity per participant pair should increase faster as a function of energy as compared to $p+p$ collisions. Similarly, the final hadron multiplicity in the central rapidity region per participant pair at fixed colliding energy should increase with $N_{\rm part}$ towards more central collisions as shown by the HIJING simulations in Fig.~\ref{fig:dndetahijing}. The exact behavior of the final hadron multiplicity per participant pair as a function of the centrality or $N_{\rm part}$ is controlled mainly by the impact-parameter dependence of the parton shadowing in heavy nuclei which can also be addressed by experiments at future high energy electron-ion colliders.

%We take heavy ion collisions at $\sqrt{s_{NN}}=20$ TeV as an example.
%For the proton-proton collisions at the same energy, the total/elastic/inelastic
%cross sections $\sigma_{\mathrm{tot}}/\sigma_{\mathrm{el}}/\sigma_{\mathrm{inel}}$
%are $ $116.3/32.2/84 mb respectively. From inelastic cross sections,
%we can determine the average interaction distance $d\sim\sqrt{\sigma_{\mathrm{inel}}/\pi}$.
%We can use this parameter in the Monte Carlo Glauber model to determine
%the number of participants $N_{\mathrm{part}}$ and the number of
%binary collisions $N_{\mathrm{coll}}$ in heavy ion collisions at
%this energy. In Fig. \ref{fig:NpartNcoll} we show $N_{\mathrm{part}}$
%and $N_{\mathrm{coll}}$ distributions at $\sqrt{s_{NN}}=$0.2, 2.76
%and 20 TeV. From the distribution of $N_{\mathrm{part}}$ in the left
%panel of Fig. \ref{fig:NpartNcoll}, we see that its collisional energy
%dependence is weak. But the distribution of $N_{\mathrm{coll}}$ strongly
%depends on the collision energy: the average value of $N_{\mathrm{coll}}$
%increases with the collision energy, see the right panel of Fig. \ref{fig:NpartNcoll}.

%\begin{figure}
%\begin{centering}
%\includegraphics[scale=0.4]{Npart}\includegraphics[scale=0.4]{Ncoll}
%\par\end{centering}

%\caption{Distributions of the number of participants and number of binary collisions
%at three colliding energies 0.2, 2.76 and 20 TeV. \label{fig:NpartNcoll}}
%\end{figure}

The second type of models for initial particle production is based on the approach of interacting
semi-classical gluonic fields or the Color Glass Condensate model \cite{McLerran:1993ni}. There
are many variants of the model including KLN \cite{Kharzeev:2000ph,Kharzeev:2001gp,Drescher:2006ca},
rcBK \cite{Balitsky:1995ub,Kovchegov:1999yj,ALbacete:2010ad,Albacete:2012xq} and
IP-Glasma \cite{Schenke:2012wb,Schenke:2012fw,Schenke:2013dpa}. One can calculate
initial gluon multiplicity in heavy-ion collisions and assume parton-hadron duality to obtain
the final hadron multiplicity.
The IP-Glasma model combines the impact parameter-dependent  saturation model
for high-energy nucleon and nuclear wave function with classical Yang-Mills
dynamics of Glasma fields in heavy ion collisions. It can be used to estimate the initial energy
density event by event. In the rcBK model, the $k_{T}$-factorization is assumed which involves an integral over
unintegrated gluon distributions whose evolution can be obtained by
solving the nonlinear Balitisky-Kovchegov (BK) equation with the running
coupling kernel (rcBK) \cite{Balitsky:1995ub,Kovchegov:1999yj,ALbacete:2010ad,Albacete:2012xq}.

\begin{figure}
\begin{centering}
\includegraphics[scale=0.6]{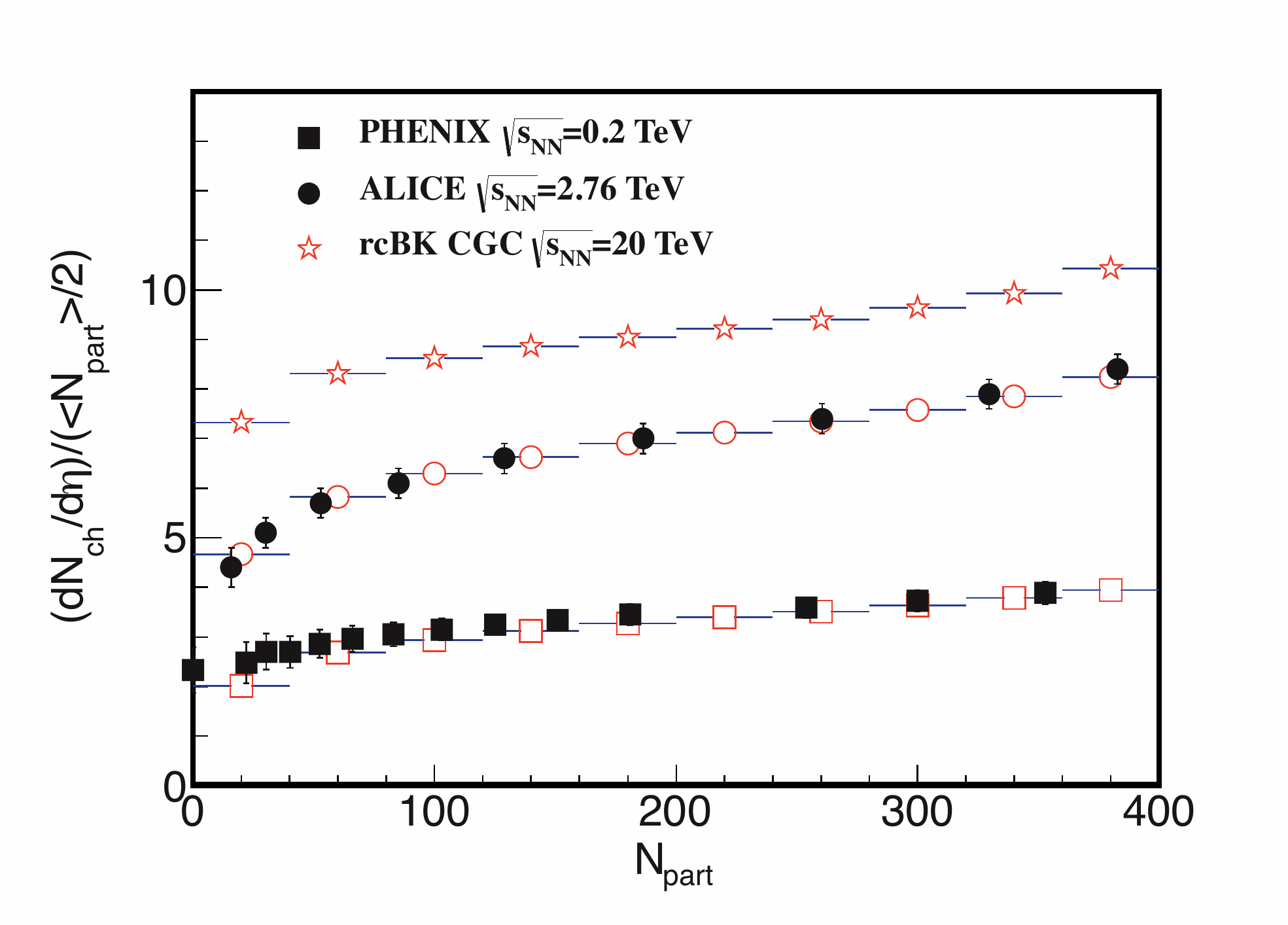}
\par\end{centering}
\caption{Centrality dependences of charged particle multiplicity at $\sqrt{s_{NN}}=$
0.2, 2.76 and 20 TeV, respectively from the rcBK model \cite{Balitsky:1995ub,Kovchegov:1999yj,ALbacete:2010ad,Albacete:2012xq}. \label{fig:multiplicity}}
\end{figure}

Shown in Fig. \ref{fig:multiplicity} is the centrality dependence of charged particle multiplicity
at three collision energies using the rcBK model (open symbols) which can reproduce experimental results at RHIC
and LHC energies (solid symbols). The hadron multiplicity in the most central Pb+Pb collisions at $\sqrt{s}=20$ TeV from the rcBK
model estimate is comparable to the HIJING estimate (shown in Fig.~\ref{fig:dndetahijing} is for $\sqrt{s}=30$ TeV).
 Notice that a cross section parameter of hard valence charges is assumed as energy-independent
($\sigma_{0}=4.2$ fm$^{2}$).  If an energy-dependent cross section parameter is used, one will get a flatter curve for the centrality
dependence.  The mechanism and consequences of gluon saturation is also one of the main topics at future high-energy
electron-ion colliders.

Together with the transverse distribution of participant nucleons and binary collisions, the above models
for initial particle production can provide the initial energy density distributions which can fluctuate from
event to event. These fluctuating initial energy density distributions in turn will provide the initial conditions for hydrodynamic
or transport evolution of the bulk matter in heavy-ion collisions. The initial energy density and temperature at the center of heavy-ion collisions 
at an initial thermalization time $\tau_0=0.6$ fm/$c$ are listed in Table~\ref{tab:inicond}.

%\begin{table}
%\begin{tabular}{r|l|||l|l} 		
%$\sqrt{s_{NN}}\ [TeV]$ & $\varepsilon_0$ [GeV/fm$^3$] & $T_0\ [MeV]$ & $\tau_0\ [fm]$ & $dN_{\rm ch}/d\eta$ \\ \hline 	
%0.20	(Au+Au) &   30 & $360$ & $0.6$ &  $720$  \\
%2.76 (Pb+Pb) &  77&  $470$ & $0.6$ &  $1600$  \\
%30 (Pb+Pb) &  136 & $560$ & $0.6$ &  $2700$   	
%\end{tabular}
%\caption{The initial energy density $\varepsilon_0$, temperature $T_0$ at the center of heavy-ion collisions, thermalization time $\tau_0$ and the final charged hadron rapidity density at different colliding energies.}
%\label{tab:inicond}
%\end{table}

\begin{table}
\begin{tabular}{r|l|l|l|l} 		
$\sqrt{s_{NN}}\ [TeV]$ &  $\varepsilon_0$ [GeV/fm$^3$]   & $T_0\ [MeV]$ & $\tau_0\ [fm]$ & $dN_{\rm ch}/d\eta$ \\ \hline 	
0.20	(Au+Au) &  30 &  $357$ & $0.6$ &  $745$  \\
2.76 (Pb+Pb) & 77 &  $449$ & $0.6$ &  $1700$  \\
30 (Pb+Pb) &  136 & $517$ & $0.6$ &  $2700$   	
\end{tabular}
\caption{The initial energy density $\varepsilon_0$, temperature $T_0$ at the center of heavy-ion collisions, thermalization time $\tau_0$ and the final charged hadron rapidity density at different colliding energies.}
\label{tab:inicond}
\end{table}

%\begin{tabular}{l|l|l|l|l} 		
%$\sqrt{s_{NN}}\ [GeV]$ & $\varepsilon_0$ $[GeV]$ & $T_0\ [MeV]$ & $\tau_0\ [fm]$ & $dN/dY$ \\ \hline 	
%30	 & $30$  &   $360$ & $0.6$ &  $700$  \\
%2760 & $108$  &   $450$ & $0.6$ &  $1800$  \\
%30000 & $190$  &   $560$ & $0.6$ &  $2800$   	
%\end{tabular}
%\label{table1}
%\end{table}

%\begin{figure}
%\caption{The impact parameter dependence of charged multiplicity for Au+Au
%collisions at $\sqrt{s_{NN}}=200$ GeV (blue circles), Pb+Pb collisions
%at $\sqrt{s_{NN}}=2.76$ TeV (green squares) and Pb+Pb collisions
%at $\sqrt{s_{NN}}=30$ TeV (red diamonds). \label{fig:dNdEta_vs_b}}
%\includegraphics[scale=0.5]{dNdEta_vs_b}
%\end{figure}

\subsection{Collective expansion and anisotropic flow}
\label{subsec-flow}

One of the evidences for the formation of sQGP in heavy-ion collisions at RHIC and LHC is the observation of
strong anisotropic flow due to collective expansion driven by the initial high energy density and pressure in
the overlapping region of the collisions \cite{Adcox:2004mh,Adams:2005dq}.  During the last decade of both experimental
and theoretical exploration of this phenomenon, a rather detailed picture of the collective expansion of the anisotropic
fireball in heavy-ion collisions emerges. During the early stage of high-energy heavy-ion collisions, the local transverse energy density is
governed by the initial wave functions of the colliding nuclei, the interaction strength of beam partons and the quantum
process of parton production. These different aspects of initial parton production determine
the event-by-event transverse as well as longitudinal energy density distributions during the early stage of the
heavy-ion collisions. Due to the thermalization processes whose mechanism is still under intense theoretical
investigation \cite{Gelis:2014tda}, these initial states of fluctuating energy density distributions achieve local
equilibrium and the subsequent collective expansion can be approximately described by relativistic viscous hydrodynamic
 equations with an effective EoS as obtained from lattice QCD results \cite{Huovinen:2009yb}. After hydrodynamic
 expansion over a finite period of time, the spatial anisotropies  of the initial energy density distributions are converted
 into anisotropies of the final hadron spectra in momentum space \cite{Ollitrault:1992bk}. One can characterize the momentum anisotropies
 in terms of the Fourier coefficients of the final hadron azimuthal distribution or two-particle azimuthal correlation in each event.
 One normally refers to these Fourier coefficients as anisotropic flows $v_n$ with the corresponding order $n$ of the Fourier
 expansion. Comparisons of the experimental measurements of  the anisotropic flows at RHIC and results from viscous
 hydrodynamic model simulations point to rather small values of the shear viscosity to entropy density
 ratio $\eta/s$ \cite{Romatschke:2007mq,Song:2010mg} that is very close to the quantum mechanics bound \cite{Policastro:2001yc}.

\begin{figure}
\begin{centering}
\includegraphics[scale=0.7]{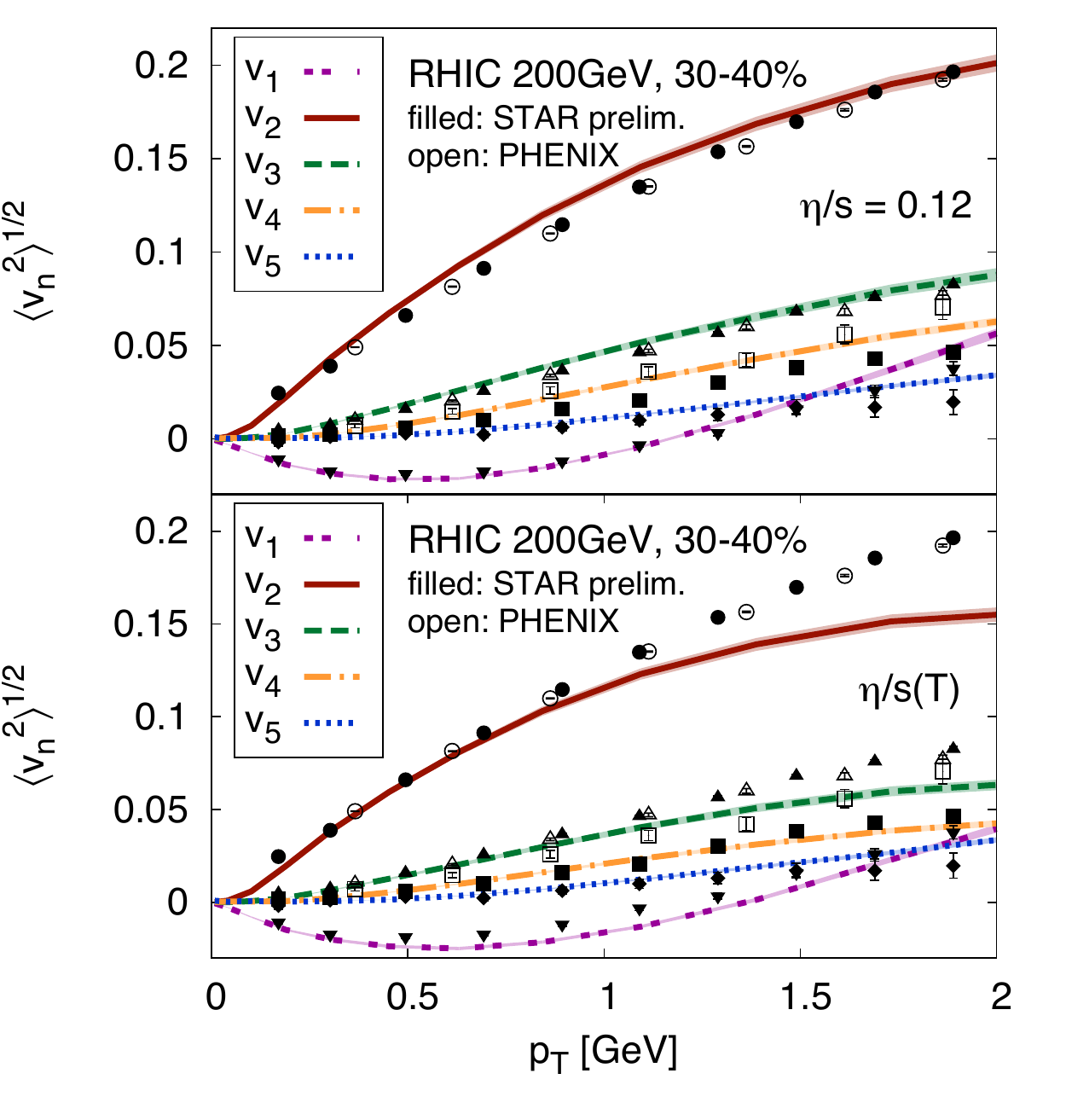}
\includegraphics[scale=0.7]{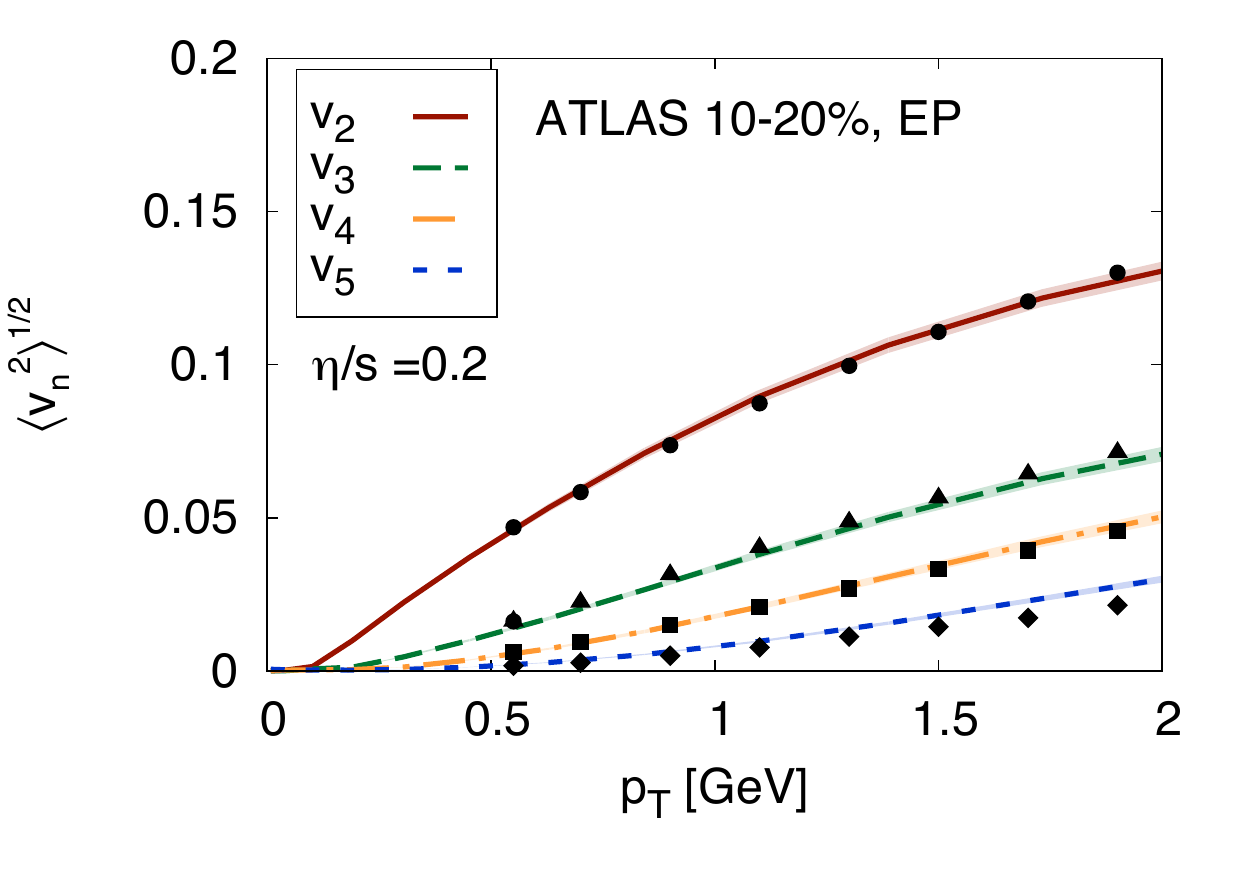}
\par\end{centering}

\caption{Anisotropic flows as functions of $p_T$ from viscous hydrodynamic model simulations with event-by-event
fluctuating initial condition from the IP-Glasma model \cite{Gale:2012rq} as compared to experimental data from PHENIX \cite{Adare:2011tg} and
 STAR \cite{Pandit:2012mq} at RHIC and ATLAS \cite{ATLAS:2012at} at LHC.
\label{fig:bj-hydro}}
\end{figure}

Shown in Fig.~\ref{fig:bj-hydro} are the calculated anisotropic flows from the state of art 2+1D viscous hydrodynamic simulations \cite{Gale:2012rq} that employs the IP-Glasma model for initial gluon production with both event-by-event geometric fluctuations in nucleon positions and the sub-nucleon color-charge fluctuations. Hydrodynamic results describe extremely well the experimental data on the anisotropic flows up to the fifth order in heavy-ion collisions at both RHIC and LHC. There is also an indication that the shear viscosity to entropy ratio decreases slightly from RHIC to LHC. This points to the direction of theoretical estimate that the QGP at higher temperatures might transit from a strongly coupled to weakly coupled one as described by pQCD calculations. Heavy-ion collisions at the very high energy region can reach even higher initial temperatures (see Table~\ref{tab:inicond}) and therefore approach closer to such a weak coupling limit.

\begin{figure}
\includegraphics[scale=0.5]{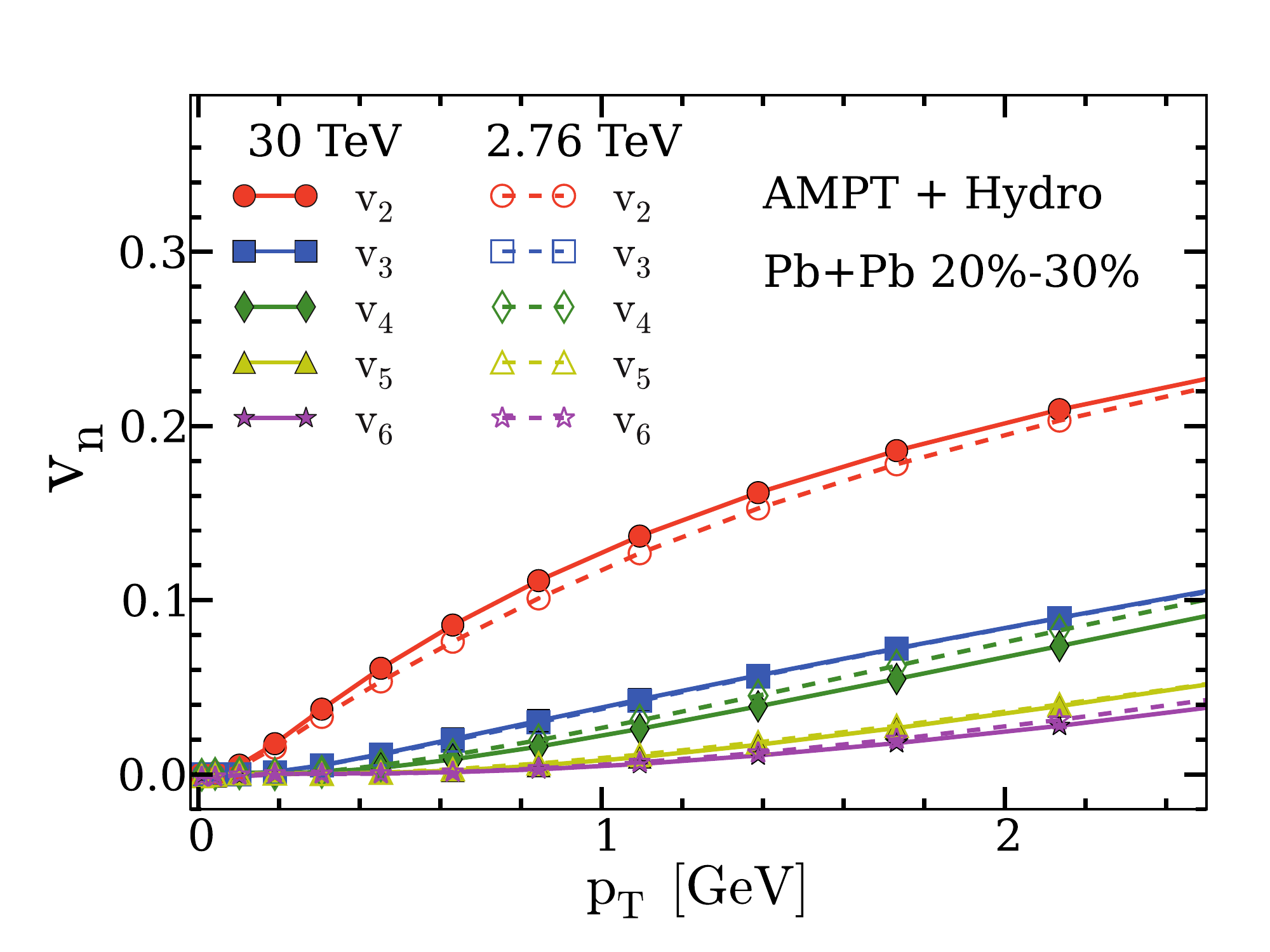}
\caption{Anisotropic flows $v_{n}$ for 20-30\% central Pb+Pb collisions at $\sqrt{s_{NN}}=30$ TeV from 3+1D ideal
hydrodynamic simulations with full fluctuating initial conditions from HIJING and AMPT model.}
\label{fig:vn-hijing-ampt}
\end{figure}

Assuming the same values of shear viscosity to entropy density ratio as in heavy-ion collisions at the LHC, the differential anisotropic flows at $\sqrt{s}=20-30$ TeV should remain roughly the same. However, due to increased radial flow and the flattening of the transverse momentum spectra, the integrated anisotropic flows should continue to increase with the colliding energy.  Shown in Fig.~\ref{fig:vn-hijing-ampt} are differential anisotropic flows calculated from 3+1D ideal hydrodynamic simulations \cite{Pang:2012he} with fluctuating initial conditions
from HIJING \cite{Wang:1991hta,Gyulassy:1994ew,Deng:2010mv} and AMPT \cite{Zhang:1999bd} models for 20-30\% central Pb+Pb
collisions at 30 TeV (solid lines) as compared to those at 2.76 TeV (dashed lines). For over all normalization of the final hadron multiplicity we have rescaled the initial energy density from the AMPT
model by a factor so that the multiplicity in 20-30\% central Pb+Pb collisions at 30 TeV is close to most central Pb+Pb
collisions at $\sqrt{s_{NN}}=2.76$ TeV. In these initial conditions fluctuations in the longitudinal direction are also considered that should
affect the final state anisotropic flow in the central rapidity region \cite{Pang:2012he}. Harmonic flow coefficients $v_{n}$ at both energies show
a normal ordering that decreases with the order of harmonics at the same $p_T$.  The second harmonic caused mainly by the
initial geometry for non-central collisions at 30 TeV is slightly higher than at 2.76 TeV because of increased initial energy density
and longer duration of expansion. The higher harmonics which are caused by initial fluctuations remains almost the same at 30 TeV
as at 2.76 TeV. This indicates that the relative transverse fluctuations do not change at higher colliding energies.

AMPT+hydro model introduces longitudinal fluctuations which will result in decorrelation of event planes for particles
with large pseudo rapidity gaps. Recent studies \cite{Pang:2014pxa} show that the decorrelation of anisotropic flows along longitudinal direction is stronger for lower energy collisions due to bigger fluctuations.

\begin{figure}[h]
\includegraphics[width=0.6\textwidth]{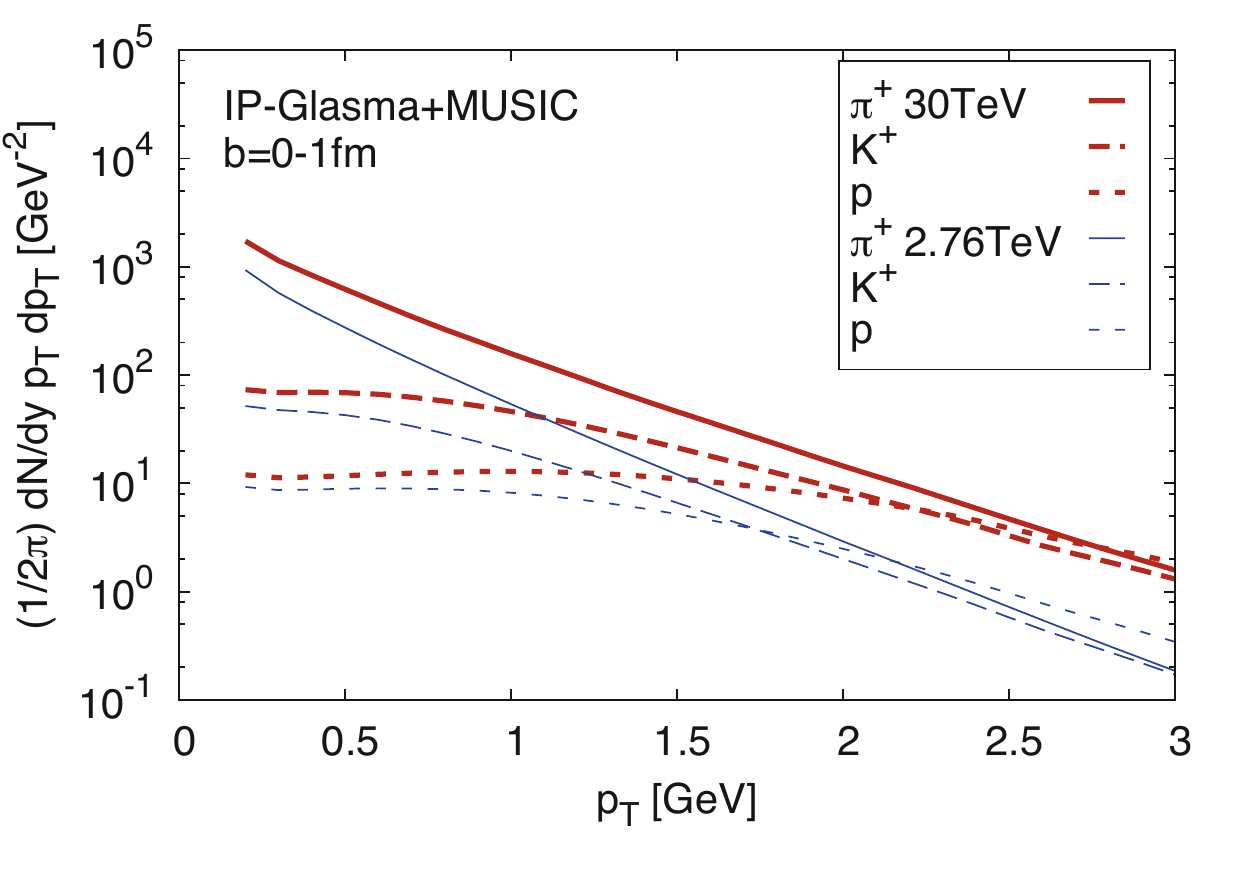}\\
\caption{Transverse momentum spectra of positive pions, kaons, and protons for impact parameter $b=0-1\,{\rm fm}$ in $2.76\,{\rm TeV}$ (thin lines) and $30\,{\rm TeV}$ (thick lines) Pb+Pb collisions computed using IP-Glasma + \textsc{Music} with $\eta/s=0.2$.}
\label{fig_ipglasmaPT}
\end{figure}

\begin{figure}[h]
\includegraphics[width=0.6\textwidth]{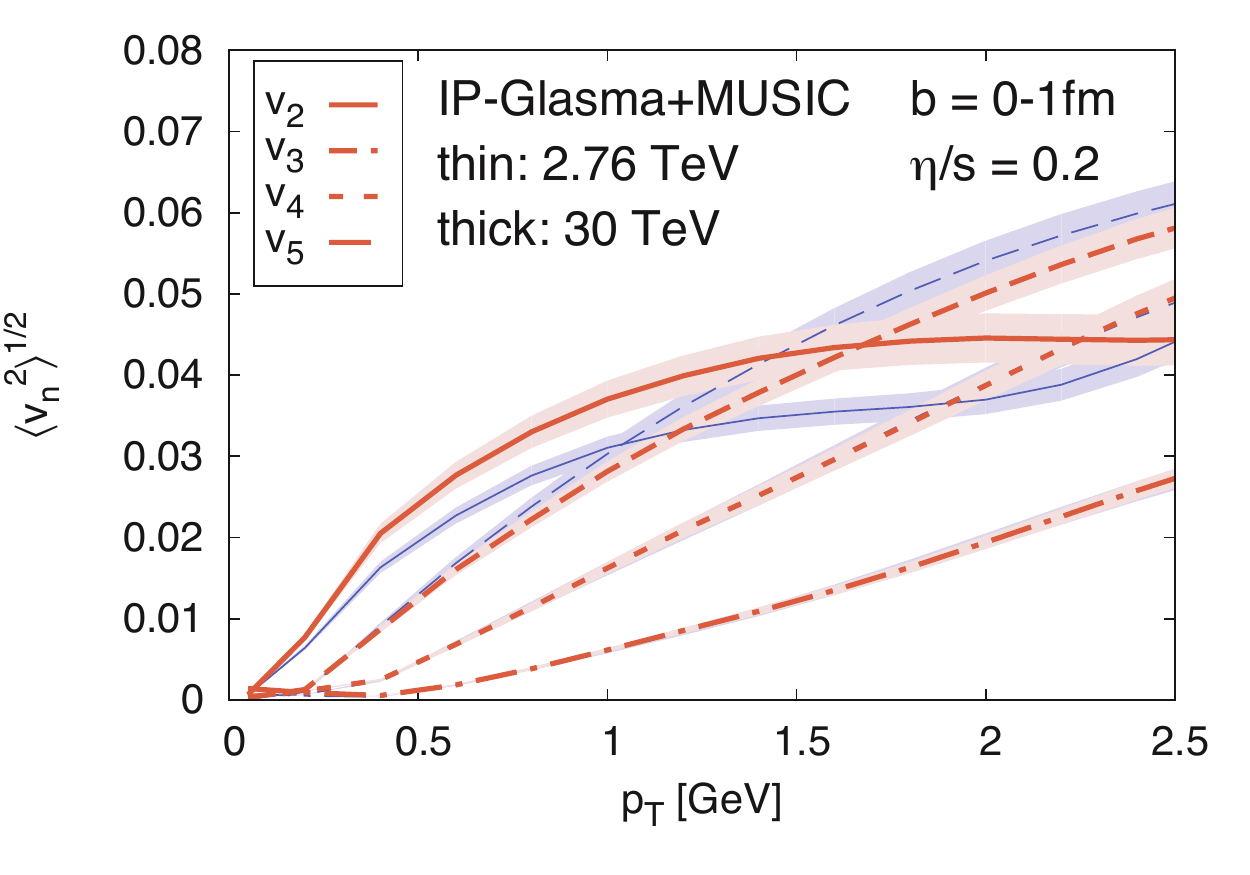}\\
\caption{Root-mean-square harmonic flow coefficients $v_n(p_T)$ of charged hadrons for impact parameter $b=0-1\,{\rm fm}$ in $2.76\,{\rm TeV}$ (thin lines) and $30\,{\rm TeV}$ (thick lines) Pb+Pb collisions computed using IP-Glasma + \textsc{Music} with $\eta/s=0.2$.}
\label{fig_ipglasmaVN}
\end{figure}

To study the effect of shear viscosity in heavy-ion collisions at very high colliding energies, we show
predictions for transverse momentum spectra and anisotropic flows $v_n(p_T)$ of pions, kaons, and protons and charged hadrons in central ($b=0-1\,{\rm fm}$) Pb+Pb collisions at $30\,{\rm TeV}$ compared to those at $2.76\,{\rm TeV}$, from IP-Glasma \cite{Schenke:2012wb,Schenke:2012fw} + \textsc{Music} simulations \cite{Gale:2012rq,Gale:2013da}. The IP-Glasma model relates the nuclear dipole cross-sections constrained by the deeply inelastic scattering (DIS) data to the initial classical dynamics of highly occupied gluon fields produced in a nuclear collision. Given an initial distribution of color charges in the high-energy nuclear wave-functions, the strong multiple scatterings of gluon fields are computed by event-by-event solutions of Yang-Mills equations. This includes both fluctuations of nucleon positions and subnucleonic color charge distributions. 
The scale of the resulting fluctuating structure of the gluon fields is given on the average by the nuclear saturation scale $Q_s$. Typically, this length scale $1/Q_s$ is smaller than the nucleon size \cite{Kowalski:2007rw}. A detailed description of the IP-Glasma model can be found in Refs.~\cite{Schenke:2012wb,Schenke:2012fw,Schenke:2013dpa}. 

The IP-Glasma model provides the initial conditions for fluid dynamic calculations at a given time $\tau_0$. The initial energy density $\varepsilon$ and flow velocities $u^\mu$ are extracted from the gluon fields' energy-momentum tensor $T^{\mu\nu}$ at every transverse position via the relation $u_\mu T^{\mu\nu} = \varepsilon u^\nu$. 
%In the results presented in Figs. \ref{fig_ipglasmaPT} and \ref{fig_ipglasmaVN}, 
The viscous part of the energy-momentum tensor is set to zero at the initial time of the fluid dynamic simulation $\tau=0.2\,{\rm fm}$. 

The fluid dynamic simulation used is the 3+1 dimensional viscous relativistic simulation \textsc{Music} \cite{Schenke:2010nt,Schenke:2010rr,Schenke:2011bn} employing a lattice equation of state with partial chemical equilibrium as described in Ref.~\cite{Gale:2012rq} and a shear viscosity to entropy density ratio $\eta/s=0.2$, which led to a good description of the experimentally measured flow harmonics in Pb+Pb collisions at $2.76\,{\rm TeV}$ \cite{Gale:2012rq}. When employing IP-Glasma initial conditions the spatial dimensions are reduced to 2, assuming boost-invariance of the initial condition. The calculation of particle spectra and the analysis of the azimuthal anisotropy follows the same steps as discussed above for the ideal hydrodynamic simulations.

As shown in Fig.~\ref{fig_ipglasmaPT}, we find that the transverse momentum spectra at $30\,{\rm TeV}$ are significantly harder than at $2.76\,{\rm TeV}$ and that the $p_T$-integrated multiplicity at mid-rapidity after viscous fluid dynamic evolution is approximately a factor of 2.8 greater. This increase of multiplicity from 2.76 to 30 TeV is somewhat higher than other model predictions as shown in Sec.~\ref{subsec-mult}. The initial gluon multiplicity obtained in Coulomb gauge only increases by approximately a factor of 2.3. This change depends on the energy dependence of $Q_s$ in the IP-Saturation model as well as the implementation of the running coupling. The multiplicity scales with $\alpha_s^{-1}$ and we have chosen the scale of the running coupling to be the average minimum $Q_s$ value. The possibility to choose another scale introduces a logarithmic uncertainty on the overall multiplicity. The additional relative increase in multiplicity from Pb+Pb collisions at $2.76\,{\rm TeV}$ to $30\,{\rm TeV}$ can be attributed to increased entropy production due to larger gradients and the approximately  $4\,{\rm fm}/c$ longer evolution time.

As in the ideal hydrodynamic calculations, $v_n$ coefficients in Pb+Pb collisions at $\sqrt{s}=30$ TeV are not changed significantly from those at 2.76 TeV as a function of $p_T$ when using a fixed shear viscosity to entropy density ratio as shown in Fig. \ref{fig_ipglasmaVN}.  The second harmonics $v_2$ is slighty higher in the higher energy collisions, while all higher harmonics agree within the statistical errors shown. It would thus be a clear indication of a change of transport coefficients with collision energy, should significantly different values of the flow harmonics be measured in the $30\,{\rm TeV}$ collisions. Precision measurements of these anisotropic flows at future very high energy heavy-ion collisions can therefore shed light on the temperature dependence of the shear viscosity to entropy ratio and whether one is approaching a weakly coupling limit as given by pQCD at higher colliding energies.

\subsection{Flavor dependence of hadron spectra and elliptic flow}

\begin{figure}
 \includegraphics[width=0.4\linewidth,height=8cm]{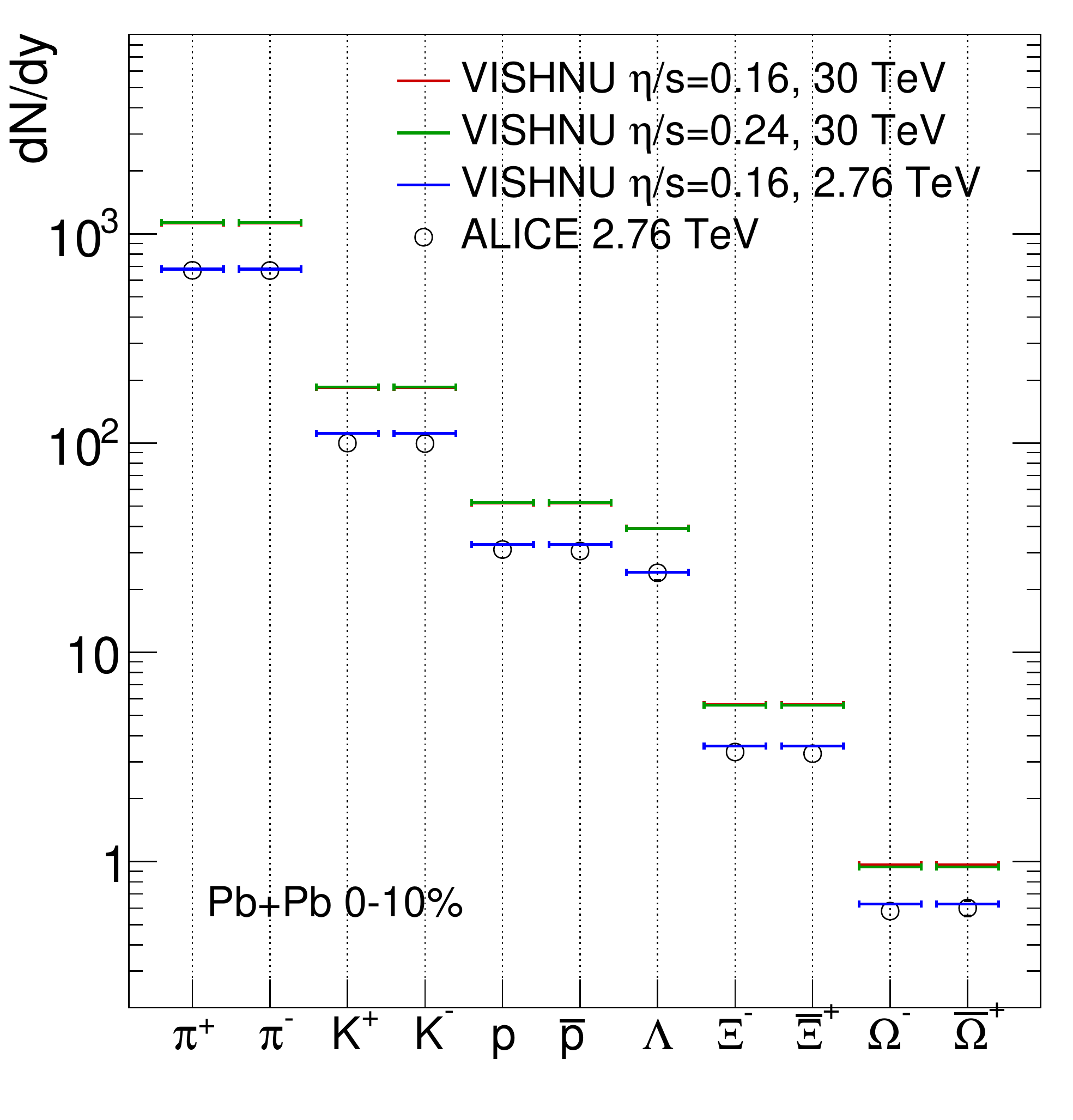}
 \caption{(Color online) Multtiplicity density, $dN_{\rm ch}/dy$, for various hadron species in the most central Pb+Pb
 collisions at $\sqrt{s_{\mathrm{NN}}}$=2.76 TeV and at $\sqrt{s_{\mathrm{NN}}}$=30 TeV. The {\tt VISHNU} results and the ALICE measuremnts at the LHC are respectively taken from~\cite{Zhu:2015dfa} and~\cite{Abelev:2013vea, Abelev:2013xaa, ABELEV:2013zaa}.
 \label{fig:dNdy}}
\end{figure}

After the QCD phase transition, the succeeding hadronic evolution involves frequent elastic, semi-elastic, inelastic collisions and resonance decays. When most of the inelastic collisions and resonance decays cease, yields of various hadrons no longer change. The system is considered to reach chemical freeze-out. Therefore, yields of soft identified hadrons can also provide information on the chemical freeze-out of the evolving dense matter.  In the statistical
model~\cite{BraunMunzinger:2001ip,Andronic:2005yp,Letessier:2005qe}, the chemical freeze-out temperature $T_{\rm ch}\sim $ 165 MeV  and chemical potential $\mu_{b}\sim $ 24 MeV are extracted from particle yields in Au+Au collisions at the top RHIC energy~\cite{BraunMunzinger:2001ip}. With $T_{\rm ch}\sim $ 165 MeV and $\mu_{b}\sim 0$, the statistical model can also describe yields of many identified hadrons in Pb+Pb collisions at $\sqrt{s_{\mathrm{NN}}} =$ 2.76 TeV, but over-predicts the protons/antiprotons data. Recent hybrid model simulations  indicated that the chemical freeze-out temperature might depend on hadron species~\cite{Zhu:2015dfa,Song:2013qma}. With baryon-antibaryon ($B$-$\bar{B}$) annihilations that delay the chemical freeze-out of baryon and antibaryons, {\tt VISHNU}~\cite{Song:2010aq,Song:2010mg} largely improves the description of protons/anti-protons data, which also fits particle yields of other identified hadron species well in Pb+Pb collisions at the LHC~\cite{Zhu:2015dfa}.

At much higher collision energies, the created QGP fireball could reach even higher temperatures, leading to more frequent $B$-$\bar{B}$ annihilations in the succeeding hadronic evolution. Therefore, future measurements of soft hadron yields in Pb+Pb collisions at $\sqrt{s_{\mathrm{NN}}}$=30 TeV will provide more information for the chemical freeze-out process of the hot QCD system. Meanwhile, measurements of spectra and elliptic flow of identified soft hadrons will help us to understand the interplay of radial and elliptic flow at much higher collision energies and provide additional constraints for the extracted QGP viscosity at higher temperatures.
\begin{figure}[t]
  \includegraphics[width=0.65\linewidth]{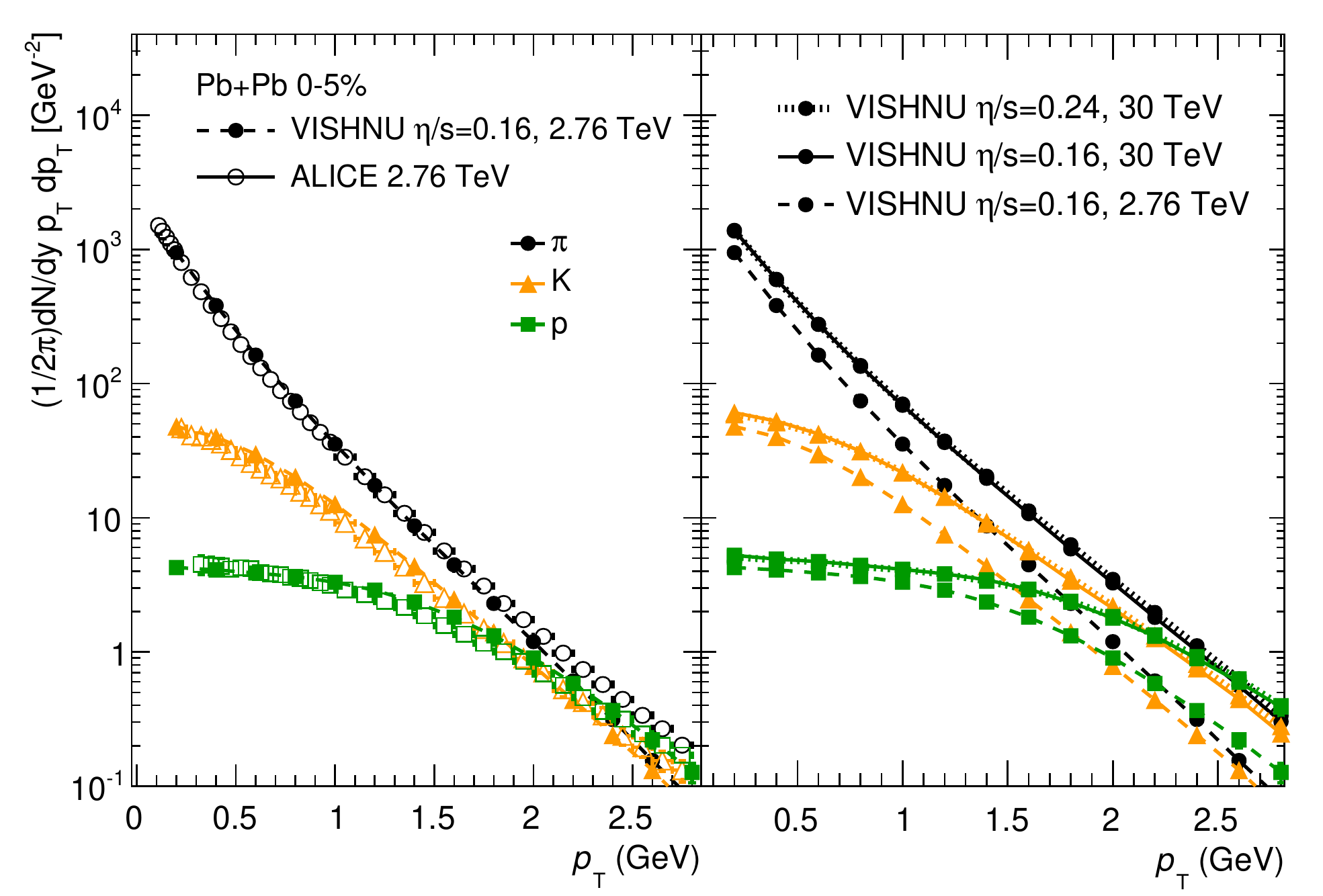}
  \caption{(Color online) Transverse momentum spectra of $\pi$, $K$, and $p$ in the most central Pb+Pb collisions
  at $\sqrt{s_{\mathrm{NN}}}$=2.76 TeV and at $\sqrt{s_{\mathrm{NN}}}$=30 TeV. The {\tt VISHNU} results and the ALICE measurements at the LHC are respectively taken from~\cite{Song:2013qma} and~\cite{Abelev:2013vea}.
 \label{fig:PtpiKp}}
\end{figure}

In this subsection, we predict multiplicity, spectra and elliptic flow of identified hadrons in Pb+Pb
collisions at $\sqrt{s_{\mathrm{NN}}}$=30 TeV, using the {\tt VISHNU} hybrid model.  {\tt VISHNU}~\cite{Song:2010aq,Song:2010mg}
combines (2+1)-D relativistic viscous
hydrodynamics ({\tt VISH2+1})~\cite{Song:2007fn,Song:2007ux} for the QGP fluid expansion with a microscopic hadronic transport
model ({\tt UrQMD})~\cite{Bass:1998ca,Bleicher:1999xi} for the hadron resonance gas evolution. The transition from hydrodynamics to
 hadron cascade occurs on a switching hyper-surface with a constant temperature. Generally, the switching temperature $T_{\rm sw}$ is set
to 165 MeV, which is close to the QCD phase transition 
temperature~\cite{Aoki:2006br,Aoki:2009sc,Borsanyi:2010bp,Borsanyi:2012rr,Bazavov:2011nk}.
For hydrodynamic evolution above $T_{\rm sw}$, we use an equation of state ({\tt EoS}) constructed from recent lattice
QCD data~\cite{Huovinen:2009yb,Shen:2010uy}.  Hydrodynamic simulations start at $\tau_0=0.9~{\rm fm}/c$ with the MC-KLN initial conditions.
For computational efficiency, we implement single-shot simulations~\cite{Song:2013qma,Zhu:2015dfa} with smooth initial entropy density profiles generated from the MC-KLN model through averaging over a large number of events within specific centrality bins.
Considering the approximate liner relationship between initial entropy and final multiplicity of all charged hadrons,
we cut centrality bins through the distribution of initial entropies obtained from MC-KLN. In these calculations, the
QGP specific shear viscosity $(\eta/s)_{QGP}$ is set to 0.16 and 0.24. The normalization factor of the initial entropy density is tuned to fit the estimated multiplicity density of all charged hadrons in the most central collisions ($\sim$ 2700). To simplify the theoretical calculations, we set the bulk viscosity to zero and neglect net
baryon density and the heat conductivity.

\begin{figure}[t]
 \includegraphics[width=0.8\linewidth]{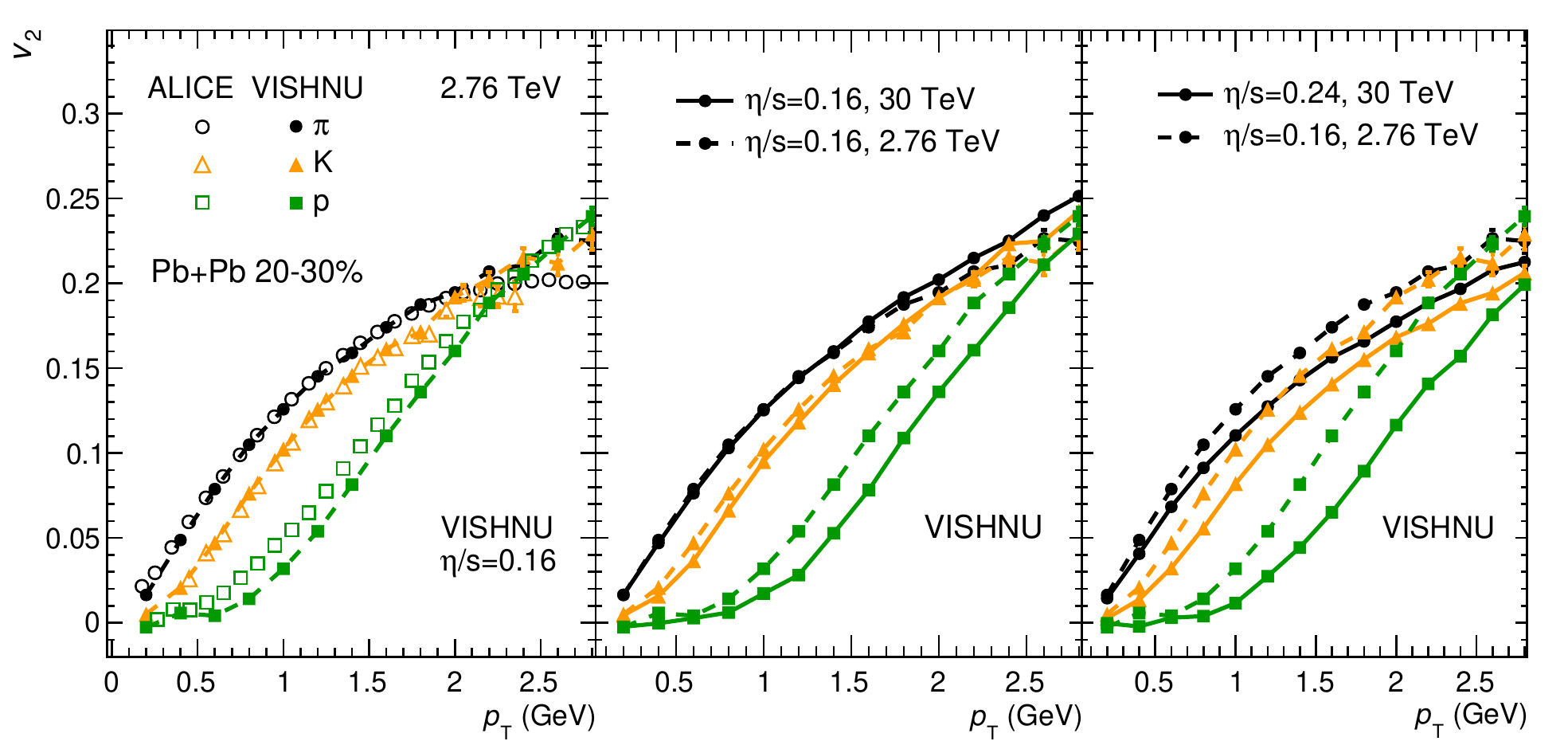}
  \caption{(Color online) Differential elliptic flow of $\pi$, $K$, and $p$ in semi-central Pb+Pb collisions at $\sqrt{s_{\mathrm{NN}}}$=2.76 TeV and at $\sqrt{s_{\mathrm{NN}}}$=30 TeV. The {\tt VISHNU} results and the ALICE measurements at the LHC are respectively taken from~\cite{Song:2013qma} and~\cite{Abelev:2014pua}.
 \label{fig:V2piKp}}
\end{figure}

Fig.~\ref{fig:dNdy} shows the {\tt VISHNU} prediction for the multiplicity densities $dN/dy$ of $\pi$, $K$, $p$, $\Lambda$,
$\Xi$, and $\Omega$ in the most central Pb+Pb collisions at $\sqrt{s_{\mathrm{NN}}}$=30 TeV. Compared with results at the
LHC~\cite{Zhu:2015dfa} energy, particle yields of various hadron species increase significantly in heavy-ion collisions at the future colliding energy. These results also indicate that the $B$-$\bar{B}$ annihilations during the hadronic evolution are significantly enhanced at the high collision energy of 30 TeV.

Fig.~\ref{fig:PtpiKp} presents {\tt VISHNU} calculations for the transverse momentum spectra
of $\pi$, $K$, and $p$ in the most central Pb+Pb collisions at $\sqrt{s_{\mathrm{NN}}}$=2.76 TeV and at $\sqrt{s_{\mathrm{NN}}}$=30 TeV.
Compared with results at the LHC energy~\cite{Song:2013qma}, the predicted spectra of $\pi$, $K$, and $p$ at $\sqrt{s}=30$ TeV are flatter with higher integrated yields. This result indicates that the amount of radial flow generated from 30 A TeV collisions is larger than the one generated from collisions at 2.76 TeV.

Besides the multiplicity and spectra, we also predict the differential elliptic flow of pions, kaons and protons in semi-central Pb-Pb collisions at $\sqrt{s_{\mathrm{NN}}}$=30 TeV.  Fig.~\ref{fig:V2piKp} shows that the differential elliptic flows $v_2(p_T)$ have clear mass ordering at different collision energies.  As the collision energy increases from 2.76 TeV to 30 TeV, the splitting  of $v_2$  between pions and protons also increases due to the larger radial flow developed at higher collision energies. Meanwhile, Fig.~\ref{fig:V2piKp} (middle and right panels) also shows that the elliptic flow of identified hadrons are sensitive to the QGP viscosity. Larger QGP shear viscosity leads to larger suppression of $v_2$.  The measurement of elliptic flow of identified hadrons in the future heavy-ion collider will thus provide  detailed information for the evolution of the QGP fireball, and help us to constrain the temperature dependence of the QGP shear viscosity.

\section{Jet quenching in heavy-ion collisions}

In high-energy heavy-ion collisions, hard scattering of beam partons can produce energetic partons with very large transverse momentum.
These energetic partons will fragment into large transverse momentum hadrons and appear in the detector as clusters of collimated hadrons
which can be reconstructed as jets in experimental measurements. These initial energetic partons are produced in the very early
stage of heavy-ion collisions and will certainly interact with soft partons from the bulk QGP that is formed over large volume of space.
The interaction between jet partons and the QGP medium will lead to elastic and radiative energy loss and therefore suppression of
the final state jets or large transverse momentum hadrons. These phenomena of jet quenching was originally proposed as one of the signatures of the QGP matter in high-energy heavy-ion collisions \cite{Wang:1991xy} which were first observed in heavy-ion collisions at RHIC \cite{Adcox:2001jp}. After more than a decade of both theoretical and experimental studies at RHIC and LHC \cite{Muller:2012zq},  jet quenching has become a powerful tool to study properties of the dense medium in heavy-ion collisions such as the jet transport parameter, defined as the broadening of averaged transverse momentum squared per unit length which is also related to the local gluon number density,
 \begin{equation}
 \hat q= \frac{4\pi^2\alpha_s C_R}{N_c^2-1}\int\frac{dy^-}{\pi}\langle F^{\sigma +}(0) F_{\sigma}^{\,+}(y)\rangle=\frac{4\pi^2\alpha_s C_R}{N_c^2-1}\rho_A xG_N(x)|_{x\rightarrow 0}.
 \end{equation}

\begin{figure}
\includegraphics[scale=0.5]{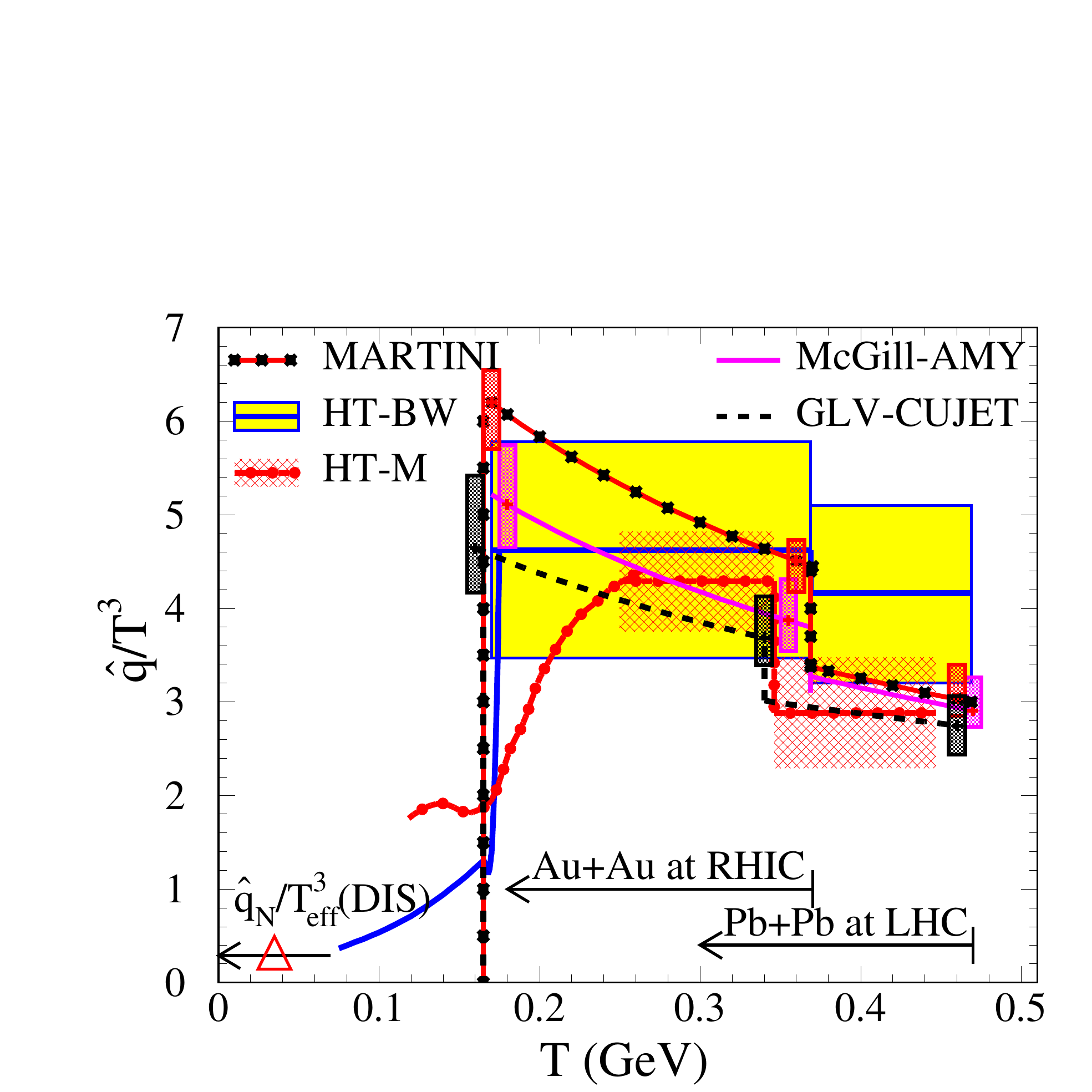}
\caption{Scaled jet transport parameter $\hat q/T^3$ for an initial quark jet with energy $E=10$ GeV at the center of the most central A+A collisions at an initial time $\tau_0=0.6$ fm/$c$ constrained from recent analysis by the JET Collaboration \cite{Burke:2013yra} with $\chi^2$ fits to the suppression factors of single inclusive hadron spectra at RHIC and LHC.  Errors from the fits are indicated by filled boxes at three separate temperatures at RHIC and LHC, respectively. The arrows indicate the range of temperatures at the center of the most central A+A collisions. The triangle indicates the
value of $\hat q_N/T^3_{\rm eff}$ in cold nuclei from DIS experiments. }
\label{fig:qhat1}
\end{figure}

\begin{figure}
\includegraphics[scale=0.5]{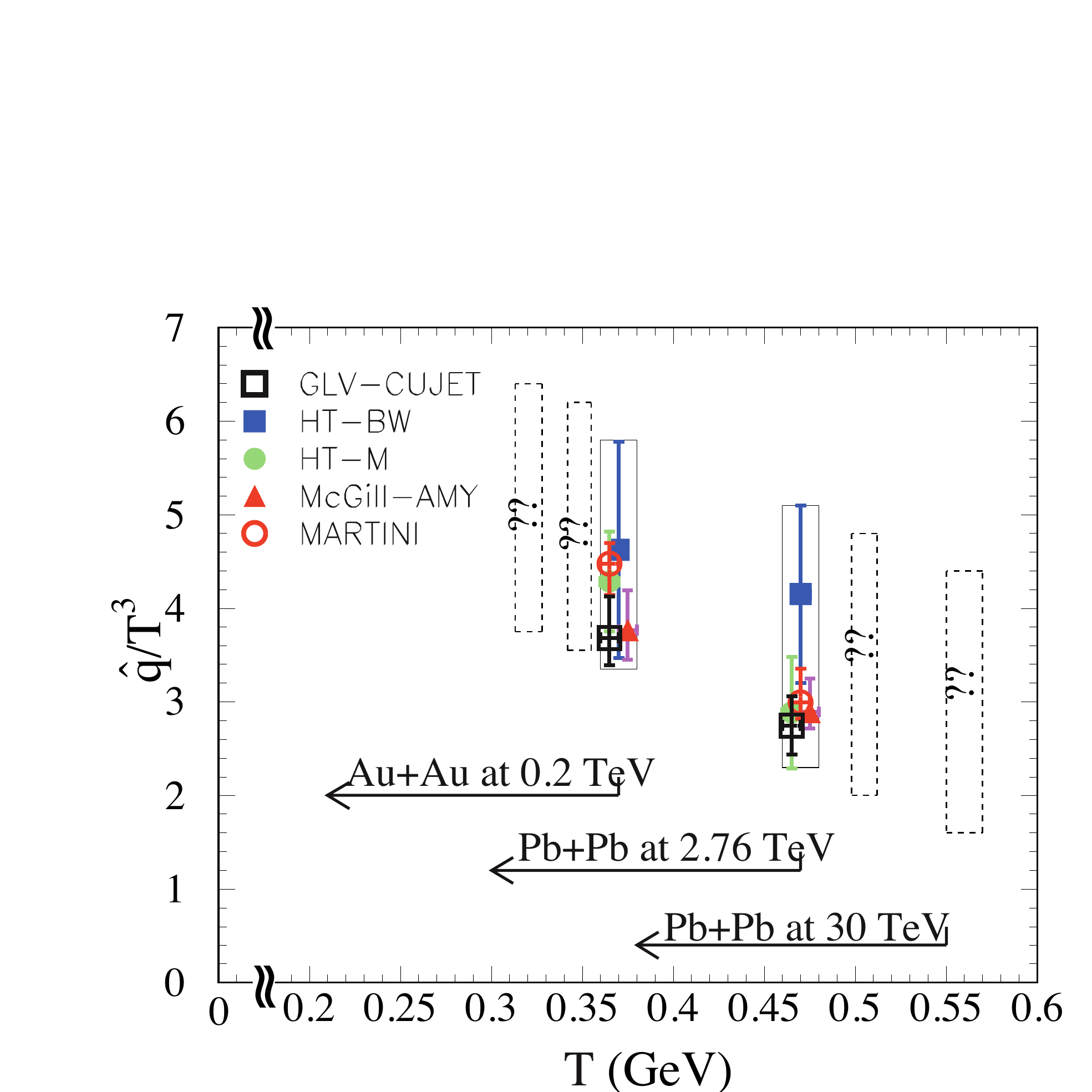}
\caption{Scaled jet transport parameter $\hat q/T^3$ for an initial quark jet with energy $E=10$ GeV at the center of the most central A+A collisions at an initial time $\tau_0=0.6$ fm/$c$ constrained from recent analysis by the JET Collaboration \cite{Burke:2013yra}. The dashed boxes indicate expected values in A+A collisions at $\sqrt{s}=0.063, 0.130$, 5.5 and 30 TeV/n, assuming the initial entropy is proportional to the final measured charged hadron rapidity density.
The arrows indicate the range of temperatures at the center of the most central A+A collisions at different colliding energies.}
\label{fig:qhat2}
\end{figure}

\subsection{Temperature dependence of jet transport parameter}

Since the hot bulk QGP medium is transient with a very short life-time and rapid expansion, the dynamical evolution of the bulk medium has to be taken into account for accurate descriptions of jet quenching phenomena. The hydrodynamical models as discussed in the previous section become necessary for jet quenching studies. One therefore needs a framework for the study that combines bulk medium evolution and jet quenching for extraction of jet transport parameter.  A recent effort has been carried out by the JET topical collaboration to create a comprehensive Monte Carlo package which combines the most advanced model for bulk medium evolution, up-to-date models for parton propagation in medium and final hadronization of jet shower partons and jet-induced medium excitation. A comprehensive study has been carried out that surveyed five different approaches to parton energy loss combined with bulk medium evolution from 2+1D and 3+1D hydrodynamic models that have been constrained by the bulk hadron spectra \cite{Burke:2013yra}.
 Through $\chi^2$-fitting of the single
inclusive hadron spectra at both RHIC and LHC with five different approaches to parton energy loss: GLV \cite{Gyulassy:2000er} and its recent CUJET implementation \cite{Buzzatti:2011vt}, the high-twist (HT) approaches (HT-BW and HT-M) \cite{Chen:2011vt,Majumder:2011uk} and the MARTINI \cite{Schenke:2009gb} and McGill-AMY \cite{Qin:2007rn} model, one obtained the most up-to-date constraints on the values of the jet transport parametersas shown in
Fig.~\ref{fig:qhat1} \cite{Burke:2013yra}.  Analyses of RHIC and LHC data with the YaJEM model \cite{Renk:2013rla} give similar constraints as shown
in Fig.~\ref{fig:qhat1}.  The jet transport parameter extracted from these analyses are $\hat q \approx 1.2\pm 0.3$ and $1.9\pm 0.7$ GeV$^2$/fm in the center of the most central Au+Au collisions at $\sqrt{s}=200$ GeV and Pb+Pb collisions at $\sqrt{s}=2.67$ TeV, respectively, at an initial time $\tau_0=0.6$ fm/$c$ for a quark jet with an initial energy of 10 GeV/$c$.  When scaled by $T^3$, the natural scale in a QGP at high temperature, $\hat q/T^3$ represents the interaction strength between jets and the medium. Current values at RHIC and LHC indicate a possible gradual weakening toward higher colliding energies where the initial temperatures are also higher.  At $\sqrt{s}=30$ TeV, one expects to reach even higher initial temperatures in the center of Pb+Pb collisions and further weakening of the jet-medium interaction.  Shown in Fig.~\ref{fig:qhat2} as open boxes with question marks are the predicted values of $\hat q$ at this energy, higher LHC energy and lower energies of the beam energy scan program at RHIC. Together with the current values at the LHC and RHIC energy, one can provide a glimpse to the temperature dependence of $\hat q/T^3$.

\subsection{Suppression of single hadron spectra}

For an estimate of the suppression of single inclusive hadron spectra in heavy-ion collisions at very high future collider energy, we
use both the higher-twist (HT) \cite{Chen:2011vt} and McGill-AMY \cite{Qin:2007rn} model.

Within the HT approach, the effect of parton energy on the final hadron spectra is implemented through effective medium-modified
fragmentation functions (FF) \cite{Wang:2003mm,Zhang:2007ja,Zhang:2009rn},
\begin{eqnarray}
D_{h/c}(z_{c},\Delta E_{c},\mu^{2}) & = & (1-e^{-\langle N_g^c \rangle})\left[\frac{z_{c}^{\prime}}{z_{c}}D_{h/c}^{0}(z_{c}^{\prime},\mu^{2})\right.\nonumber \\
 &  & \left.+\langle N_g^c \rangle\frac{z_{g}^{\prime}}{z_{c}}D_{h/g}^{0}(z_{g}^{\prime},\mu^{2})\right]
+e^{-\langle N_g^c \rangle}D_{h/c}^{0}(z_{c},\mu^{2}),\label{eq:modfrag}
\end{eqnarray}
where $z_{c}^{\prime}=p_{T}/(p_{Tc}-\Delta E_{c})$, $z_{g}^{\prime}=\langle L/\lambda\rangle p_{T}/\Delta E_{c}$
are the rescaled momentum fractions, $\Delta E_{c}$ is the radiative
parton energy loss and $\langle N_g^c \rangle$ is the average number of induced gluon emissions. The FFs in vacuum $D_{h/c}^{0}(z_{c},\mu^{2})$ is given
by the AKK08 parameterizations \cite{Albino:2008fy}.  The total parton energy loss within the HT approach in a finite and expanding medium can
be expressed as \cite{Wang:2009qb},
\begin{eqnarray}
\frac{\Delta E_a}{E} & = & C_{A}\frac{\alpha_{s}}{2\pi}\int dy^{-}\int_{0}^{\mu^{2}}\frac{dl_{T}^{2}}{l_{T}^{4}}\int dz[1+(1-z)^{2}]\nonumber \\
 &  & \times\hat{q}_{a}(y)4\sin^{2}\left[ \frac{y^-l_T^2}{4Ez(1-z)}\right],
\end{eqnarray}
in terms of the jet transport parameter $\hat q_a$ for a jet parton $a$. The jet transport
parameter for a gluon is $9/4$ times of a quark and therefore the
radiative energy loss of a gluon jet is also $9/4$ times larger than
that of a quark jet. According to the definition of jet transport
parameter, we can assume that it is proportional to the local parton
density in a QGP medium. In a dynamical evolving medium,
it can be expressed in general as \cite{Chen:2010te,Chen:2011vt,CasalderreySolana:2007sw},
\begin{equation}
\hat{q}(\tau,r)=\hat{q}_{0}\frac{\rho_{QGP}(\tau,r)}{\rho_{QGP}(\tau_{0},0)}\cdot\frac{p^{\mu}u_{\mu}}{p_{0}}\,,\label{q-hat-qgph}
\end{equation}
In our calculation, we use a full (3+1)D ideal hydrodynamics
\cite{Pang:2012he,Pang:2013pma} to describe the space-time evolution
of the local temperature and flow velocity in the bulk medium along
the jet propagation path in heavy-ion collisions. Here $\rho_{QGP}(\tau,r)$
is the parton density in the comoving frame of the fluid cell in hydrodynamics , and $\rho_{QGP}(\tau_{0},0)$
is the initial parton density at the time $\tau_{0}=0.6$ fm/$c$
in the center of the hot system, $p^{\mu}$ is the four momentum of
the jet and $u^{\mu}$ is the four flow velocity in the collision
frame, $\hat{q}_{0}$ denotes the jet transport parameter at the center
of the bulk medium in the QGP phase at the initial time $\tau_{0}$.

The averaged number of gluon emissions $\langle N_g^a \rangle$ from the propagating parton ($a=q,g$)
within the high-twist approach of parton energy loss \cite{Chang:2014fba} is given by,
\begin{eqnarray}
\langle N_{g}^{a}(\mu^{2}) \rangle & = & C_{A}\frac{\alpha_{s}}{2\pi}\int dy^{-}\int_{0}^{\mu^{2}}\frac{dl_{T}^{2}}{l_{T}^{4}}\int \frac{dz}{z}[1+(1-z)^{2}]\nonumber \\
 &  & \times\hat{q}_{a}(y)4\sin^{2}\left[ \frac{y^-l_T^2}{4Ez(1-z)}\right].
\label{eq:dngdz}
\end{eqnarray}

\begin{figure}
\includegraphics[width=85mm]{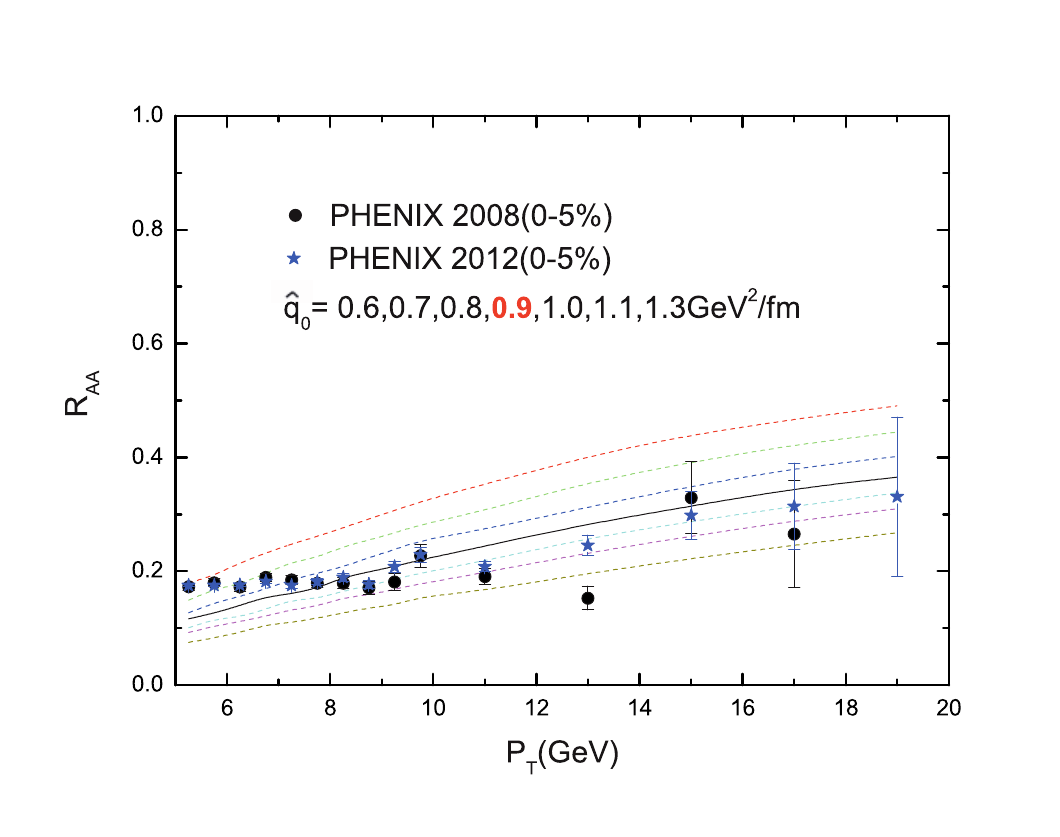}
\caption{Nuclear modification factor at mid-rapidity for $\pi^{0}$ spectra
in $0-5\%$ central Au+Au collisions at $\sqrt{s_{NN}}=200$ GeV with
a range of values of initial quark jet transport parameter $\hat{q}_{0}$
at $\tau_{0}=0.6$ fm/$c$ in the center of the most central collisions
(from top to bottom), as compared to PHENIX data \cite{Adare:2008qa,Adare:2012wg}
at RHIC.
\label{fig:rhic-raa}}
\end{figure}

\begin{figure}
\centering{}\includegraphics[width=85mm]{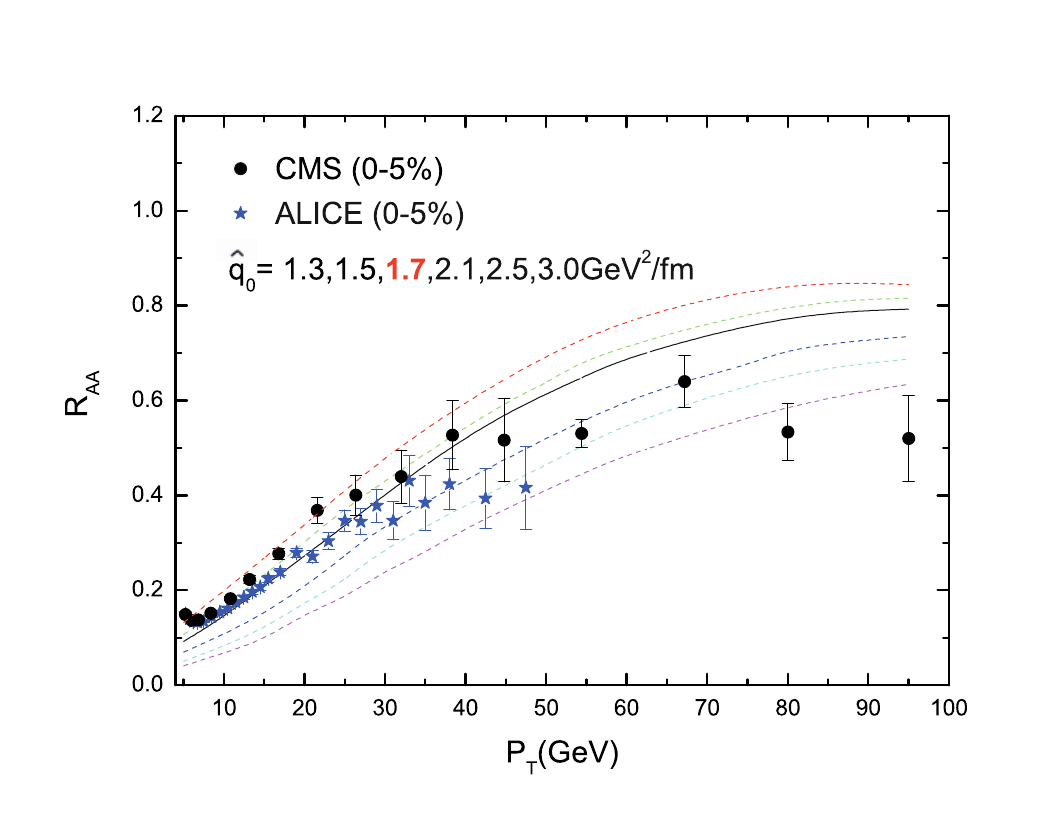}
\begin{centering}
\caption{Nuclear modification factor at mid-rapidity for changed hadron spectra
in $0-5\%$ central Pb+Pb collisions at $\sqrt{s_{NN}}=2.76$ TeV/n
with a range of values of initial quark jet transport parameter $\hat{q}_{0}$
at $\tau_{0}=0.6$ fm/$c$ in the center of the most central collisions(from
top to bottom), as compared to ALICE \cite{Abelev:2012hxa} and CMS
data \cite{CMS:2012aa} at LHC.
\label{fig:lhc-raa}}
\par
\end{centering}
\end{figure}

Using the above medium modified FFs with the collinear next-to-leading order (NLO) pQCD parton model \cite{Kidonakis:2000gi,Harris:2001sx}
and the CTEQ5 parameterization of parton distributions, one can calculate the final hadron spectra in both heavy-ion and p+p collisions.
Shown in Fig. \ref{fig:rhic-raa} and Fig. \ref{fig:lhc-raa} are the nuclear modification factors,
\begin{eqnarray}
R_{AA} & = & \frac{d\sigma_{AA}/dp_{T}^{2}dy}{\int d^{2}b\, T_{AA}({\bf b})d\sigma_{NN}/dp_{T}^{2}dy},\label{eqn:modifactoer}
\end{eqnarray}
for the charged hadron spectra as compared to the RHIC/LHC data on central collisions with different
values for the jet transport parameter.  The values of $\hat q$ from best $\chi^{2}$ fits are $\hat{q}_{0}=0.7-1.0$
GeV$^{2}$/fm at RHIC energy and $\hat{q}_{0}=1.3-2.0$ GeV$^{2}$/fm at LHC. This is consistent with HIJING 2.0 prediction \cite{Deng:2010mv}
and the JET analyses \cite{Burke:2013yra}.

\begin{figure}
\includegraphics[width=85mm]{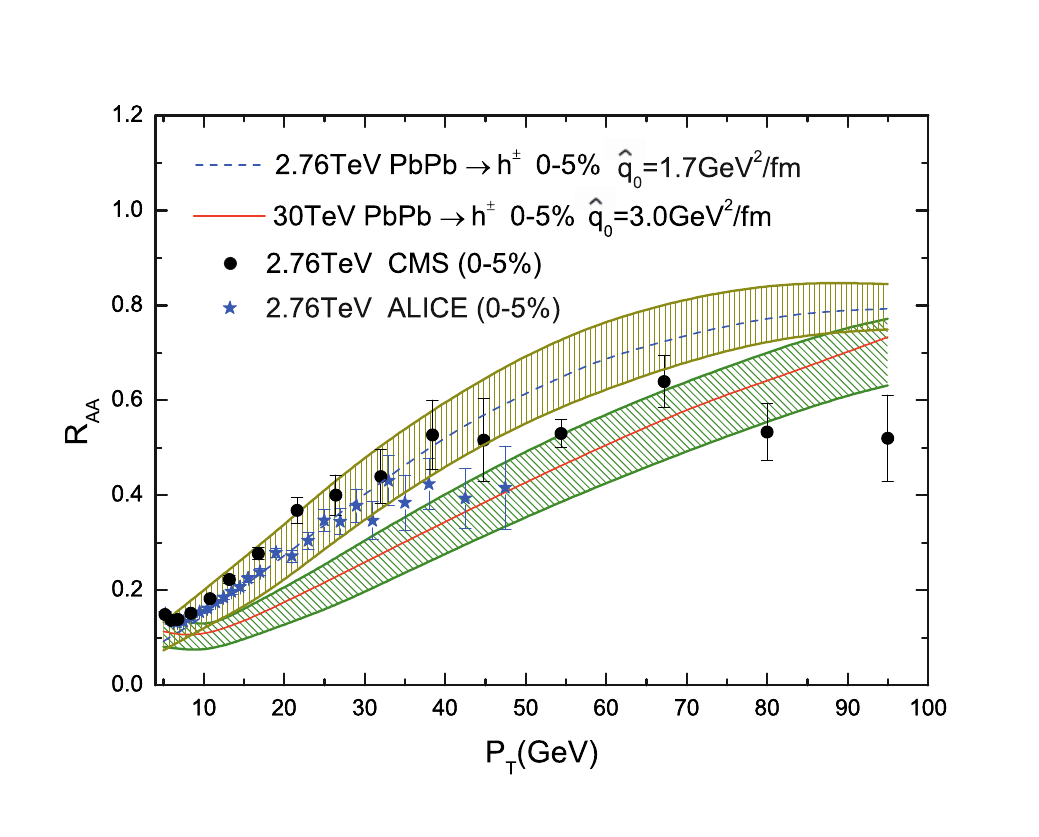}
\caption{Nuclear modification factor at mid-rapidity for changed hadron spectra
in $0-5\%$ central Pb+Pb collisions at $\sqrt{s_{NN}}=30.0$ TeV/n
with a range of values of initial quark jet transport parameter $\hat{q}_{0}=2.6-4.0$
GeV$^{2}$/fm at $\tau_{0}=0.6$ fm/$c$ in the center of the most
central collisions(from top to bottom), as compared to Pb+Pb collisions
at $\sqrt{s_{NN}}=2.76$ TeV/n with the value of $\hat{q}_{0}$ is from
1.3 GeV$^{2}$/fm to 2.0 GeV$^{2}$/fm and ALICE \cite{Abelev:2012hxa}
and CMS data \cite{CMS:2012aa} at LHC as shown in Fig. \ref{fig:lhc-raa}.
\label{fig:30t-raa} }
\end{figure}

Since the jet transport parameter $\hat{q}_{0}$ is proportional to the initial parton number density which in turn is
proportional to the final charged hadron multiplicity, we can assume $\hat{q}_{0}=2.6-4.0$ GeV$^{2}$/fm for a quark jet
in central $Pb+Pb$ collisions at $\sqrt{s_{NN}}=30$ TeV/n which is about 2 times that at LHC energy according to Table ~\ref{tab:inicond}.
Shown in Fig. \ref{fig:30t-raa} is the nuclear modification factor at mid-rapidity for charged hadron
spectra in $0-5\%$ central Pb+Pb collisions at $\sqrt{s_{NN}}=30.0$
TeV/n with a range of values of initial quark jet transport parameter
$\hat{q}_{0}=2.6-4.0$ GeV$^{2}$/fm at $\tau_{0}=0.6$ fm/$c$ in
the center of the most central collisions(from top to bottom), as
compared to Pb+Pb collisions at $\sqrt{s_{NN}}=2.76$ TeV/n with the
value of $\hat{q}_{0}=1.3 -2.0 $ GeV$^{2}$/fm
and ALICE \cite{Abelev:2012hxa} and CMS data \cite{CMS:2012aa} at
LHC as shown in Fig. \ref{fig:lhc-raa}.  Over the range of $p_T=10-100$ GeV/$c$, the hadron spectra are significantly more suppressed
at $\sqrt{s}=30$ TeV than at LHC due to larger initial values of jet transport parameter. The difference becomes smaller at high
transverse momentum due to different initial jet spectra at two different energies.

In the McGill-AMY approach \cite{Qin:2007rn, Qin:2009bk}, nuclear modification of hadron spectra in nucleus-nucleus collisions  can be calculated by first solving a set of coupled transport rate equations for the hard jet energy/momentum distributions $f(p,t) = {dN(p,t)}/{dp}$ in the hot nuclear medium. The coupled rate equations for quark and gluon jets may generically be written as the following form:
\begin{eqnarray}
\frac{df_j(p,t)}{dt} = \sum_{ab} \! \int \! dk \left[f_a(p+k,t) \frac{d\Gamma_{a\to j}(p+k,k)}{dkdt} - P_j(k,t)\frac{d\Gamma_{j\to b}(p,k)}{dkdt}\right], \ \ \
\end{eqnarray}
In the above equation, $d{\Gamma_{j\to a}(p,k)}/{dk dt}$ represents the transition rate for the process $j\to a$, with $p$ the initial jet energy and $k$ the momentum lost in the process.
The transition rates for radiative processes are taken from Refs. \cite{Arnold:2001ba, Arnold:2002ja, Jeon:2003gi}, and for the collisional processes, the drag and the diffusion contributions are incorporated following Refs. \cite{Qin:2007rn, Qin:2009bk}.
The contributions from energy gain processes are taken into account by the $k<0$ integral.

After solving the above coupled rate equations, one may obtain the medium-modified fragmentation function as follows:
\begin{eqnarray}
\label{mmff} \tilde{D}_{h/j}(z,\vec{r}_\bot, \phi_p) = \sum_{j'} \!\int\! dp_{j'} \frac{z'}{z} D_{h/j'}(z') P(p_{j'}|p_j,\vec{r}_\bot, \phi_p),
\end{eqnarray}
where $z = p_h / p_{j}$ and $z' = p_h / p_{j'}$, with $p_h$ the momentum of the hadron $h$ and $p_{j}$($p_{j'}$) the initial (final) jet momentum.
$D_{h/j}(z)$ is the vacuum fragmentation function, and $P(p_{j'}|p_j,\vec{r}_\bot, \phi_p)$ represents the differential probability for obtaining a parton $j'$ with momentum $p_{j'}$ from a given parton $j$ with momentum $p_j$. This probability distribution depends on the path traveled by the parton and the local medium profiles such as the temperature and flow along that path. Therefore, $P(p_{j'}|p_j,\vec{r}_\bot, \phi_p)$ depends on the the initial jet production location $\vec{r}_\bot$ and the propagation direction $\phi_p$.
Jets are decoupled from the medium when the local temperature of the nuclear medium is below the transition temperature $T_c = 160$~MeV.

By convoluting the medium-modified fragmentation function with the initial jet momentum distribution computed from perturbative QCD, one may obtain the final medium-modified hadron spectra:
\begin{eqnarray}
\frac{d\sigma_{AB\to hX}}{d^2p_T^hdy} \!&=&\! \int d^2\vec{r}_\perp {\cal
P}_{AB}(\vec{r}_\perp)  \sum_{j} \int \frac{dz}{z^2} \tilde{D}_{h/j}(z,\vec{r}_\bot, \phi_p) \frac{d\sigma_{AB\to jX}}{d^2p_T^j dy}.
\end{eqnarray}
In the above equation, ${\cal P}_{AB}(b,\vec{r}_\perp)$ is the probability distribution of the initial jet production position $\vec{r}_\perp$, and is determined from binary collision distribution simulated by the Glauber model. One may fix the propagation direction $\phi_p$ or average over a certain range.

Putting the above ingredients together, one may obtain the hadron yield after medium modification and calculate the nuclear modification factor $R_{AA}$.

\begin{figure}[htb]
\includegraphics[width=8.cm]{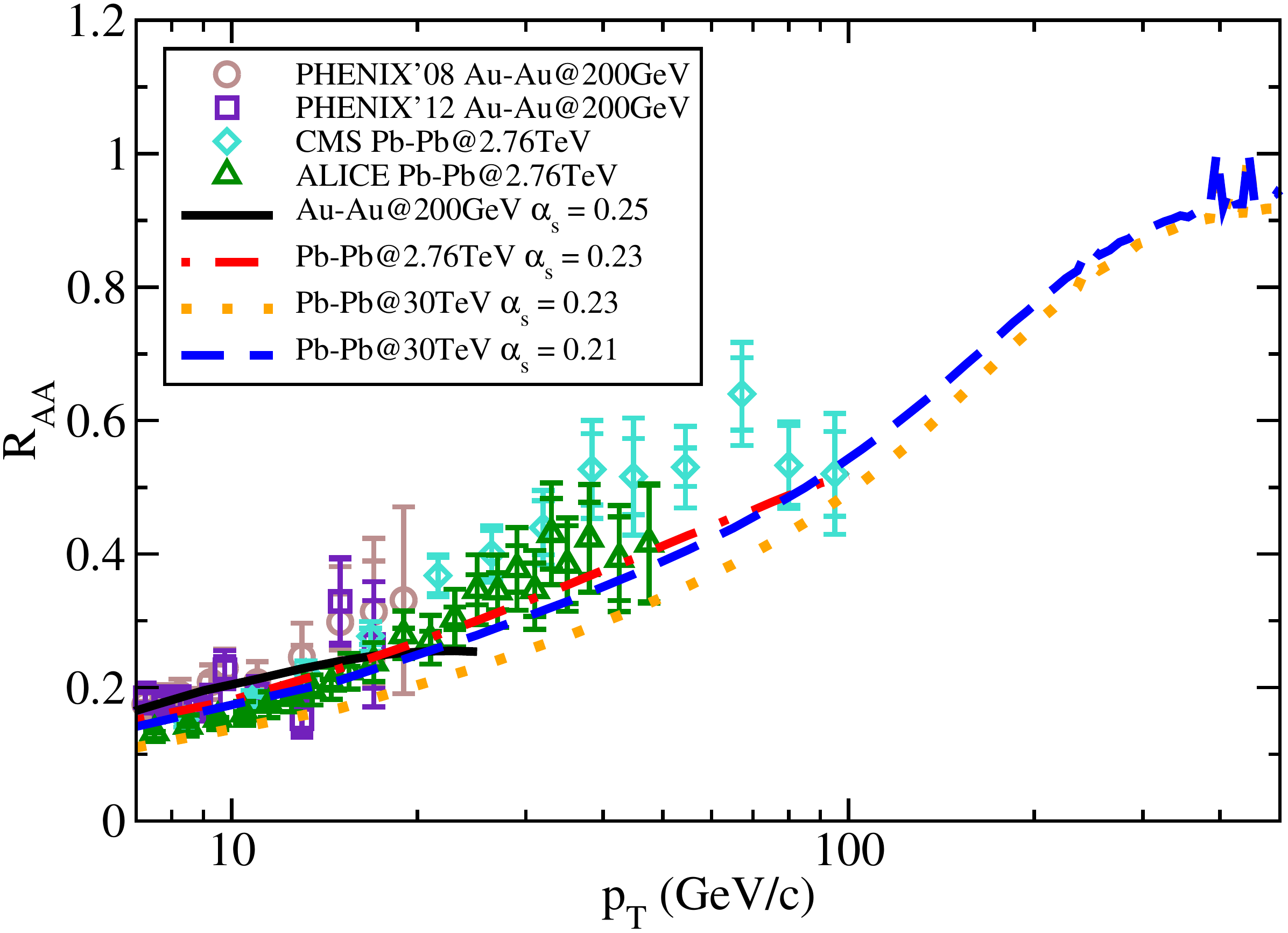}
\caption{(Color online) The nuclear modification factor $R_{AA}$ as a function of $p_T$ for central Au-Au collisions at $200$~GeV/n at RHIC, for central Pb-Pb collisions at $2.76$~TeV/n at the LHC, and for central Pb-Pb collisions at $30$~TeV/n.
} \label{raa_all}
\end{figure}

In Fig. \ref{raa_all}, we show the comparison of the calculated nuclear modification factor $R_{AA}$  from McGill-AMY approach as a function of $p_T$ for: central 0-5\% Au-Au collisions at 200~GeV/n at RHIC, central 0-5\% Pb-Pb collisions at 2.76~TeV/n at the LHC, and central 0-5\% Pb-Pb collisions at 30~TeV/n. Note that in the McGill-AMY model, the model parameter is the strong coupling constant $\alpha_s$ which is usually fitted to the experimental data. For RHIC Au-Au collisions it is obtained as $\alpha_s = 0.25$ by fitting to PHENIX data, and for the LHC $\alpha_s = 0.23$ using CMS and ALICE data. The decreasing of $\alpha_s$ from RHIC to the LHC may be understood as originating from the increasing of the average temperature (or the energy density) of the hot nuclear media produced at RHIC and LHC. To account for such effect when moving from 2.76~ TeV/n Pb-Pb collisions to 30~TeV/n Pb-Pb collisions, we decrease the strong coupling constant from $\alpha_s = 0.23$ to $\alpha_s = 0.21$. One may consider the calculation for 30~TeV/n Pb-Pb collisions using $\alpha_s = 0.23$ as the lower reference bound for the nuclear modification factor $R_{AA}$.

\subsection{Medium modification of reconstructed jets}

The jet quenching or parton energy loss in hot and dense QGP can affect not only hadron suppression
but also in jet modifications in high-energy nuclear collisions \cite{Vitev:2008rz,Vitev:2009rd}.
The study of fully reconstructed jet production in relativistic heavy-ion
collisions plays a very important role in probing the properties of
the QGP formed in Pb+Pb reactions at the LHC \cite{Dai:2012am}.
Full jets in experiments are reconstructed from hadronic energies measured either through tracking or calorimetric
detectors or both with a given jet -inding algorithm \cite{Cacciari:2011ma}. The jet production cross section
with the same jet-finding algorithm can also be calculated within the next-to-leading (NLO ) pQCD, using Monte Carlo packages such
as MEKS \cite{Gao:2012he}. Inclusive differential jet production cross sections in p+p collisions
at NLO accuracy provide the baseline to calculate inclusive jet productions in
heavy-ion collisions \cite{Vitev:2009rd},
\begin{eqnarray}
\frac{1}{\langle N_{\mathrm{bin}}\rangle}\frac{d\sigma^{AA}(R)}{dydE_{T}} & = & \int_{\epsilon=0}^{1}d\epsilon\sum_{q,g}P_{q,g}(\epsilon)\frac{1}{1-(1-f_{q,g})\cdot\epsilon}\frac{d\sigma_{q,g}^{pp\;(CNM)}}{dydE_{T}^{\prime}}.\label{AA cross section}
\end{eqnarray}
In the above expression for jet production cross section in heavy-ion collisions, several cold nuclear matter effects (shadowing, anti-shadowing
and EMC effect) are taken into account through the EPS09 \cite{Eskola:2009uj} parameterization of nuclear
parton distribution functions (nPDF).  The parameter $f_{q,g}$ is the part of the fractional energy loss falling in the jet area, and can be calculated from the angular distribution of medium induced parton energy loss \cite{Vitev:2009rd,Lokhtin:2005px}, $P_{q,g}(\epsilon)$
is the probability that a jet loses energy fraction $\epsilon E_{T}^{\prime}$,
here $E_{T}^{\prime}=E_{T}/[1-(1-f_{q,g})\cdot\epsilon]$ \cite{Vitev:2009rd}.

\begin{figure}[!htb]
\includegraphics[width=0.5\linewidth]{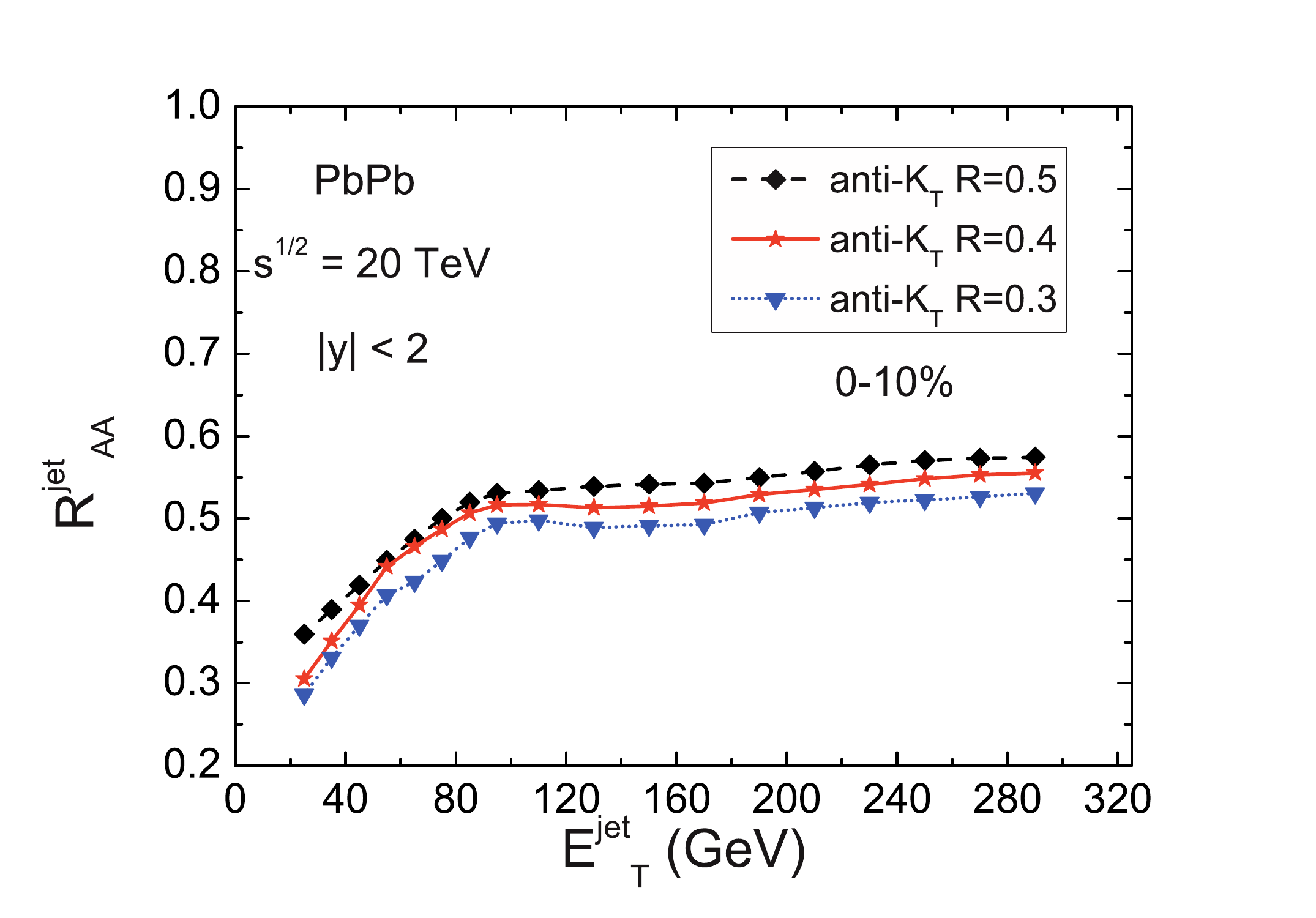}
\caption{Nuclear modification factor $R_{AA}$ for inclusive jet production
as a function of jet transverse energy for different jet radius in
central Pb+Pb collisions at $\sqrt{s_{NN}}=20$ TeV.\label{size}}
\end{figure}

Shown in Fig. \ref{size} are the nuclear modification factors of inclusive jet production,
\begin{equation}
R_{AA}^{\mathrm{jet}}=\frac{d\sigma^{AA}/dydE_{T}}{\langle N_{\mathrm{bin}}\rangle d\sigma^{pp}/dydE_{T}},
\end{equation}
in central Pb+Pb collisions for different jet radius $R=$ 0.3,0.4,0.5 at $\sqrt{s_{NN}}=20$ TeV.
One-dimension longitudinal Bjorken expansion of the QGP fireball
with Glauber transverse distribution and a highest initial temperature $T_{0}=$570 MeV is assumed.
The calculated jet suppression factors increase with
jet transverse energy for all three different jet radii. The suppression factor for a larger jet cone size
is slightly less because more radiated gluon remain inside the jet cone and thus less effective energy loss for
the reconstructed jet.  These calculated suppression factors for reconstructed jets at $\sqrt{s}= 20$ TeV are somewhat similar
to that measured at the current LHC energy \cite{Aad:2014bxa,Aiola2014382} even though the initial parton energy density
or the jet transport parameter is almost a factor of 2 larger. This indicates that the jet suppression
factor is less sensitive to the properties of the medium as compared to the single inclusive hadrons. It is therefore helpful
to explore other observables such as jet shape or profile functions.

\begin{figure}[!htb]
\includegraphics[width=0.5\linewidth]{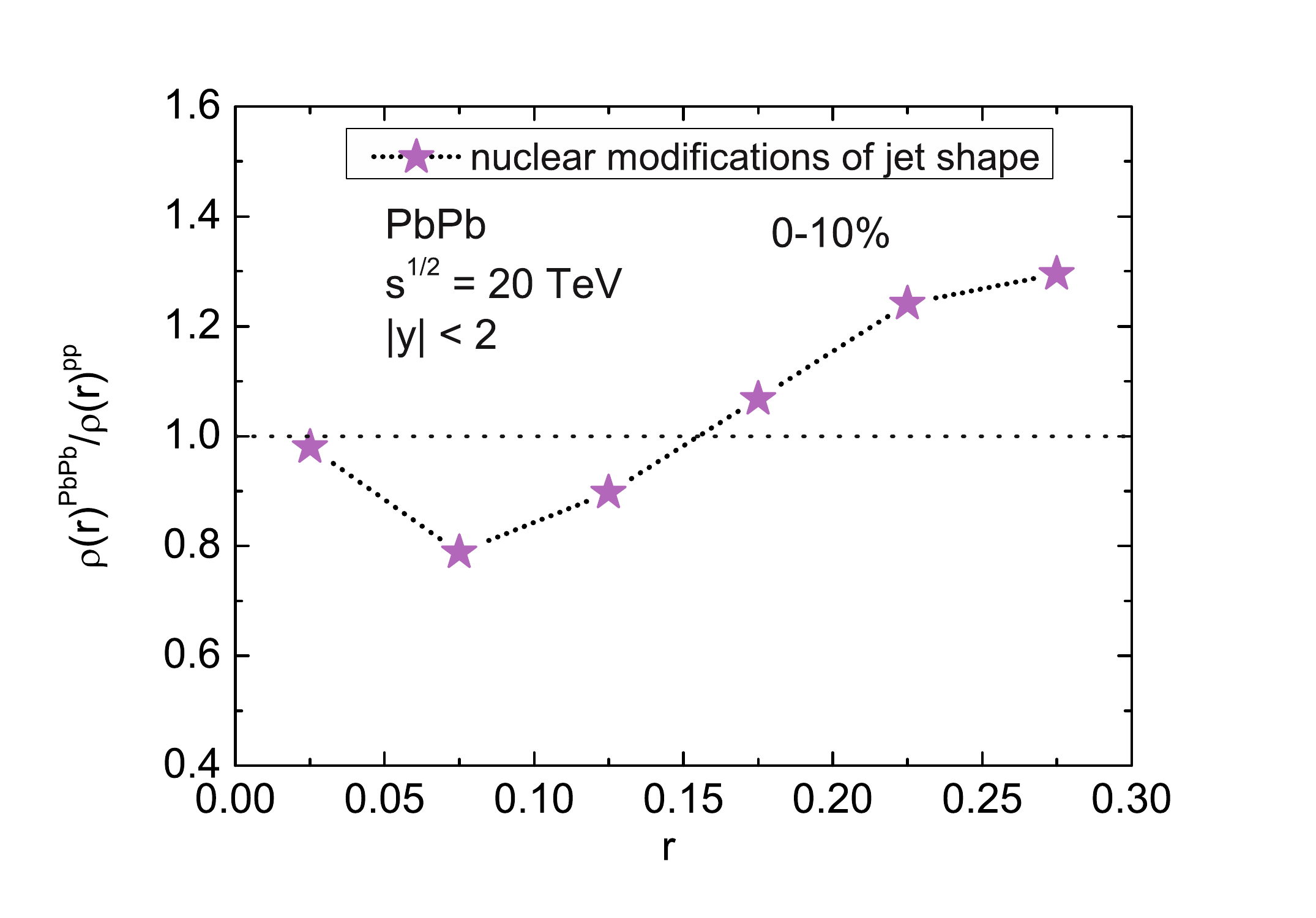}
\caption{Nuclear modification $R_{AA}^{\text{jet-shape}}$ for differential
jet shapes with $R=0.3$ in central Pb+Pb collisions at $\sqrt{s}=20$
TeV. \label{shape}}
\end{figure}

Jet shape, also called jet energy profile, is the internal energy
distribution of a jet. Medium modification of the jet shape in heavy-ion collisions due to multiple scattering
and induced radiation relative to hadron-hadron reactions has shown to be sensitive to jet-medium
interaction \cite{Vitev:2008rz,Vitev:2009rd}. The differential jet shape is defined as

\begin{eqnarray}
\rho(r) & = & \frac{1}{\Delta r}\frac{1}{N^{\mathrm{jet}}}\sum\limits _{\mathrm{jets}}\frac{P_{T}(r-\Delta r/2,r+\Delta r/2)}{P_{T}(0,R)},\qquad\Delta r/2\leq r\leq R-\Delta r/2.
\label{jet shape}
\end{eqnarray}

Jet shapes in hadronic collisions have been studied recently in the
framework of QCD resummation at NLO \cite{Li:2011hy,Li:2012bw}, which
give a decent description of jet profiles in p+p collisions and provide
the baseline for investigating jet shape modification in high-energy nuclear colliisons.
In heavy-ion reactions, the jet energy consists of two parts: the
energy of quenched leading parton ($E_p$) and the redistributed energy of
radiated gluon ($E_g$) inside the jet cone. The total jet energy should be their sum, $E^{\mathrm{jet}}=E^{g}+E^{p}$.
Thus jet shapes in heavy-ion collisions can be calculated as follows
\begin{eqnarray}
\rho^{AA}(r,E_{\mathrm{jet}}) & = & \frac{E_{g}}{E_{\mathrm{jet}}}\rho^{\text{medium}}(r,E_{g})+\frac{E_{p}}{E_{\mathrm{jet}}}\rho^{pp}(r,E_{p}),\label{jet shape AA}
\end{eqnarray}
where $\rho^{\text{medium}}(r,E_{g})$ is calculated from the angular
spectra of medium induced gluon radiation. Furthermore we define the
nuclear modification ratio of jet shapes as
\begin{eqnarray}
R_{AA}^{\text{jet-shape}} & = & \frac{\rho^{AA}(r,E_{\mathrm{jet}})}{\rho^{pp}(r,E_{\mathrm{jet}})}.
\end{eqnarray}
Shown in Fig.~\ref{shape} is the calculated the nuclear modifications of jet shapes in central
Pb+Pb collisions at $\sqrt{s}=20$ TeV. One can see a considerable enhancement of jet shapes in heavy-ion
collisions relative to those in p+p in the region when $r\rightarrow R$ due to induced gluon radiation while
there is some depletion of jet energy distribution at intermediate $r$ due to fixed total jet energy. Such a feature
has been observed in heavy-ion collisions at LHC \cite{Mao201488} and should provide information on jet-medium interaction
at future high-energy heavy-ion colliders.

\section{Medium modification of open heavy mesons}
\label{sec:DRAA}

An alternative candidate of hard probe of the QGP properties is heavy flavor meson. Since the large mass of heavy quarks effectively suppresses their thermal production from the bulk matter, the majority of them are produced at the primordial stage of nuclear collisions through hard scatterings. After that, they propagate through the hot QGP matter with their flavor conserved and therefore serve as a clean probe of the whole evolution history of the QGP fireballs. 

\subsection{Perturbative heavy quark transport}
\label{sec:duke}

In this section, we adopt an improved Langevin approach \cite{Cao:2011et,Cao:2013ita} to simulate the in-medium energy loss of open heavy quarks. The hadronization into heavy mesons is simulated using a hybrid model of fragmentation and coalescence developed in Ref. \cite{Cao:2013ita}.

In the limit of small momentum transfer in each interaction, multiple scatterings of heavy quarks inside a thermalized medium can be described using the Langevin equation. Apart from the collisional energy loss due to these quasi-elastic scatterings, heavy quarks may also lose energy via medium-induced gluon radiation. We modify the classical Langevin equation as follows to simultaneously incorporate these two energy loss mechanisms:
\begin{equation}
\label{eq:modifiedLangevin}
\frac{d\vec{p}}{dt}=-\eta_D(p)\vec{p}+\vec{\xi}+\vec{f}_g.
\end{equation}
The first two terms on the right-hand side of Eq.(\ref{eq:modifiedLangevin}) are inherited from the original Langevin equation, describing the drag force and thermal random force exerted on heavy quarks when they scatter with light partons in the medium background. We assume that the fluctuation-dissipation theorem is still hold between these two terms $\eta_D(p)=\kappa/(2TE)$, in which $\kappa$ is known as the momentum-space diffusion coefficient of heavy quarks: $\langle\xi^i(t)\xi^j(t')\rangle=\kappa\delta^{ij}\delta(t-t')$. The third term $\vec{f}_g$ in Eq.(\ref{eq:modifiedLangevin}) is introduced to describe the recoil force heavy quarks experience when they radiate gluons. The probability of gluon radiation within the time interval $[t,t+\Delta t]$ can be evaluated according to the number of radiated gluons,
\begin{equation}
\label{eq:gluonnumber}
P_\mathrm{rad}(t,\Delta t)=\langle N_\mathrm{g}(t,\Delta t)\rangle = \Delta t \int dxdk_\perp^2 \frac{dN_\mathrm{g}}{dx dk_\perp^2 dt},
\end{equation}
as long as $\Delta t$ is chosen sufficiently small so that $P_\mathrm{rad}(t,\Delta t)<1$ is always satisfied. In our study, the distribution of the medium-induced gluon radiation is taken from the high-twist approach to parton energy loss \cite{Guo:2000nz,Majumder:2009ge,Zhang:2003wk}:
\begin{equation}
\label{eq:gluondistribution}
\frac{dN_\mathrm{g}}{dx dk_\perp^2 dt}=\frac{2 C_A \alpha_s  P(x)\hat{q} }{\pi k_\perp^4} {\sin}^2\left(\frac{t-t_i}{2\tau_f}\right)\left(\frac{k_\perp^2}{k_\perp^2+x^2 M^2}\right)^4,
\end{equation}
where $x$ is the fractional energy taken from the heavy quark by the radiated gluon, $k_\perp$ is the gluon transverse momentum, $P(x)$ is the quark splitting function and $\tau_f={2Ex(1-x)}/{(k_\perp^2+x^2M^2)}$ is the gluon formation time. In Eq.(\ref{eq:gluondistribution}), a quark transport coefficient $\hat{q}$ is utilized, which is related to the diffusion coefficient $\kappa$ by adding the factor of dimension in our work $\hat{q}=2\kappa$. With this assumption, only one free parameter remains in the modified Langevin equation [Eq.(\ref{eq:modifiedLangevin})]. As shown in the earlier work \cite{Cao:2013ita}, $\hat{q}/T^3=5.0$ is chosen to best describe the experimental data of heavy flavor meson at high $p_\mathrm{T}$ at RHIC and LHC.

With this improved Langevin approach, we may simulate the evolution of heavy quarks in relativistic nuclear collisions. The dense QCD medium produced in these collisions is simulated with a (3+1)-dimensional ideal hydrodynamic model \cite{Pang:2012he,Pang:2013pma}. This hydrodynamic calculation provide the space-time evolution of the local temperature and fluid velocity of the QGP fireballs. With these information, for every time step, we boost each heavy quark into the local rest frame of the fluid cell through which it propagates and then update its energy and momentum according to our modified Langevin equation. After that, the heavy quark is boosted back to the global center of mass frame and stream freely until its interaction with the medium for next time step. Before the hydrodynamical evolution commences (at $\tau_0=0.6$~fm/c), heavy quarks are initialized with a MC-Glauber model for their spatial distribution and a leading order pQCD calculation together with the CTEQ5 parton distribution functions \cite{Lai:1999wy} for their momentum distribution. The nuclear shadowing effect in the initial state of nucleus-nucleus collisions is taken into account by using the EPS09 parameterization \cite{Eskola:2009uj}. With these initializations, heavy quarks evolve inside the QGP matter until they reach fluid cells with local temperature below $T_\mathrm{c}$ (165~MeV). Then they hadronize into heavy flavor hadrons within a hybrid model of fragmentation and coalescence. The relative probability of fragmentation and heavy-light quark coalescence is calculated with the Wigner functions in an instantaneous coalescence model \cite{Oh:2009zj}. If a heavy quark is chosen to combine with a thermal parton from the medium, the momentum distribution of the produced hadron is calculated directly with the coalescence model itself. On the other hand, if the heavy quark is chosen to fragment, its fragmentation process is simulated with PYTHIA 6 \cite{Sjostrand:2006za} in which the Peterson fragmentation function is used.

\begin{figure}[tb]
  \includegraphics[width=0.45\textwidth]{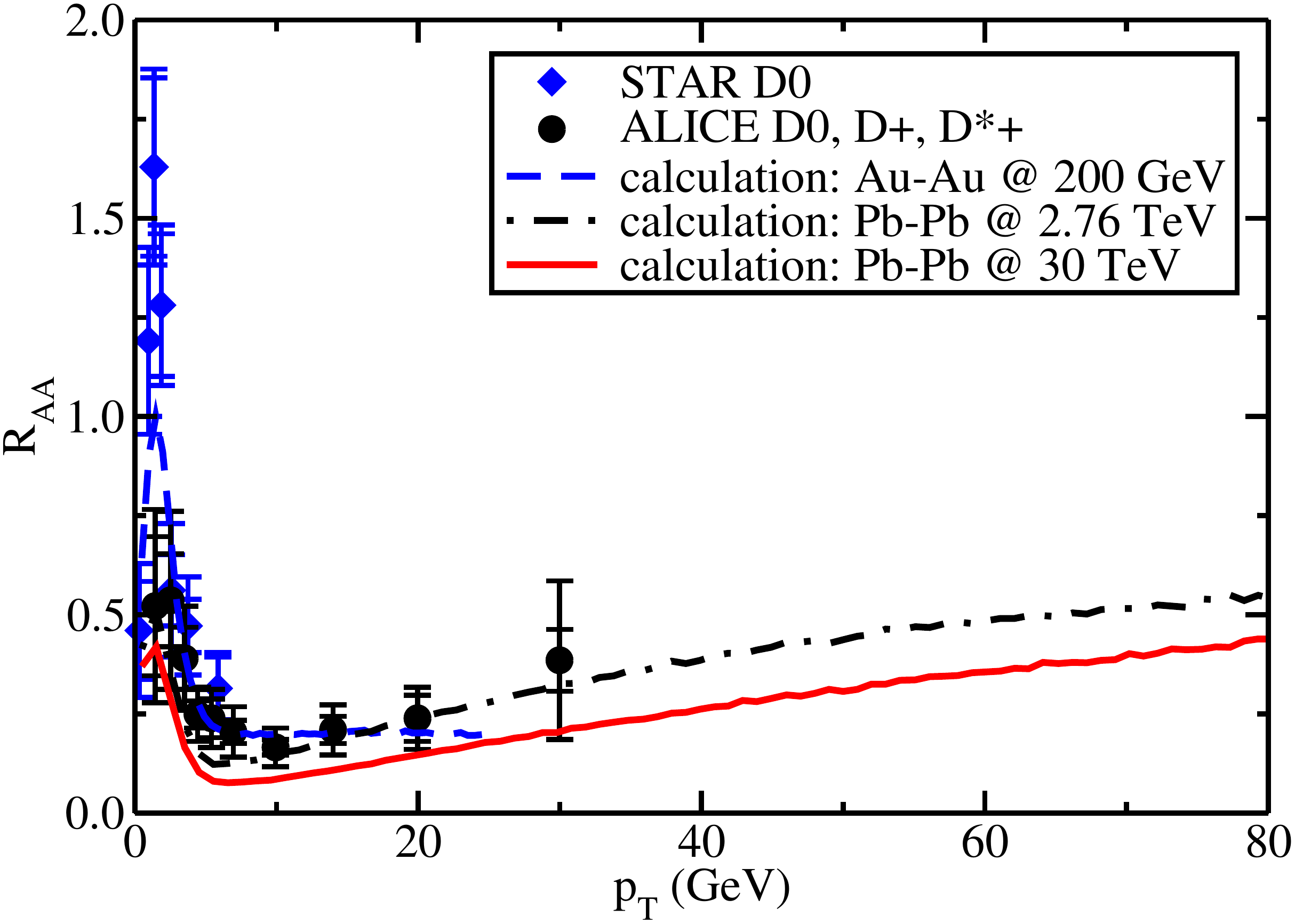}
  \caption{(Color online) The nuclear modification factor $R_\mathrm{AA}$ of $D$ mesons, compared between in central Au-Au collisions at 200 AGeV, in central Pb-Pb collisions at 2.76 ATeV, and in central Pb-Pb collisions at 30 ATeV.}
  \label{fig:plot-DRAA}
\end{figure}

In Fig.\ref{fig:plot-DRAA}, the suppression factors for $D$ meson $R_\mathrm{AA}$ are shown for different collisional energies. For central Au-Au collisions at 200~AGeV, a bump structure in the $D$ meson $R_\mathrm{AA}$ can be observed around 1-2~GeV. This is mainly contributed by the coalescence mechanism in heavy quark hadronization process, which combines low $p_\mathrm{T}$ heavy quarks and light thermal partons and enhances the production of $D$ mesons at medium $p_\mathrm{T}$. Such a bump is significantly suppressed in Pb-Pb collisions at 2.76~ATeV and 30~ATeV due to the strong nuclear shadowing effect for these collisional energies at low $p_\mathrm{T}$. At higher $p_\mathrm{T}$, the $D$ meson $R_\mathrm{AA}$ is relatively flat between 10 and 20~GeV in Au-Au collisions at  200~AGeV, but starts to increase with $p_\mathrm{T}$ in 2.76 and 30~ATeV Pb-Pb collisions. This probably results from the harder initial $p_\mathrm{T}$ spectra of charm quarks produced at the LHC energy than at the RHIC energies. In addition, we observe that $D$ mesons are more suppressed in 30~ATeV than in 2.76~ATeV Pb-Pb collisions, since larger collisional energy leads to higher initial temperature and longer lifetime of the QGP fireballs and therefore increases the total in-medium energy loss of heavy quarks.

\subsection{Non-perturbative heavy quark transport}
\label{sec:tamu}

A non-perturbative transport model for heavy quarks and open heavy-flavor
(HF) mesons in ultrarelativistic heavy-ion collisions was introduced in
Ref.~\cite{He:2011qa}. The HF transport is simulated
with relativistic Langevin simulations with temperature and momentum
dependent transport coefficients computed from $T$-matrix interactions
as described below.  It treats both microscopic HF transport through quark-gluon plasma (QGP)
and hadronization in a strong-coupling scheme. Its applications to HF  phenomenology at
RHIC~\cite{He:2012df,He:2014epa} and LHC~\cite{He:2014cla} result in
fair agreement with existing data for the nuclear modification factor
and elliptic flow of $D$ mesons, $B$ mesons (from non-prompt $J/\psi$)
and HF decay leptons. 

In the deconfined high-temperature phase, heavy-quark (HQ) scattering
with medium partons is calculated using the thermodynamic $T$-matrix
approach~\cite{Riek:2010fk,Huggins:2012dj}, which accounts for all color
channels (e.g., $a=1,8$ for $Q\bar q$), different flavors ($u$, $d$, $s$
and gluons) and partial waves ($l=S,P$) via a Lippmann-Schwinger equation
of the type
\begin{equation}
T_{l,a} = V_{l,a} + \frac{2}{\pi}\int_0^\infty k^2 dk  V_{l,a} G_{2} T_{l,a} \ ;
\end{equation}
here, $G_2$ is the uncorrelated in-medium two-particle propagator which
includes the single-quark self-energies. The potential, $V_{l,a}$, the kernel
of the integral equation, is approximated by the internal energy computed in
thermal lattice-QCD (lQCD)~\cite{Petreczky:2004pz,Kaczmarek:2007pb}, and
incorporates relativistic corrections to recover the correct high-energy
perturbative limit. The use of the internal energy yields better agreement
with lQCD data on, e.g., quarkonium correlators, HQ susceptibilities and
the HQ diffusion coefficient, than the free energy~\cite{Riek:2010py}. As
the color-screening of the potential gradually reduces when approaching
the critical temperature, $T_{\rm pc}\simeq 170$\,MeV, from above, the
remnant confining potential strengthens and the HQ interactions with
light quarks in the QGP start to develop heavy-light ($D$- of $B$-meson like)
resonance correlations.

Once the medium evolution reaches the critical region, the resonant
$Q\bar q$ correlations emerging from the $T$-matrices are utilized as a
hadronization mechanism via the resonance recombination model
(RRM)~\cite{Ravagli:2007xx} on a hydrodynamic hypersurface at $T_{\rm pc}$; 
left-over charm and bottom quarks are hadronized via FONLL
fragmentation~\cite{He:2014cla} as used for the initial spectra in $pp$
(for which EPS09 shadowing has been accounted for~\cite{He:2014cla}).
The $D$ and $B$ mesons thus formed continue to diffuse in the subsequent
hadronic phase via scatterings off bulk hadrons ($\pi$, $K$,
$\eta$, $\rho$, $\omega$, $K^*$, $N$, $\bar N$, $\Delta$ and $\bar\Delta$),
evaluated using effective hadronic lagrangians available from the
literature~\cite{He:2011yi}. Around $T_{\rm pc} $ the resulting diffusion
coefficient for $D$ mesons turns out to be comparable to the $T$-matrix
results for charm quarks on the partonic side.

%\begin{figure} [!t]
%\includegraphics[width=0.6\columnwidth]{MinHe/epsilon_P.pdf}
%\vspace{-0.5cm}
%\caption{(Color online) The energy-momentum anisotropy of the hydrodynamic
%evolution as a function of longitudinal proper time.}
%\label{fig_epsilon_P}
%\end{figure}

\begin{figure} [!t]
\includegraphics[width=0.6\columnwidth]{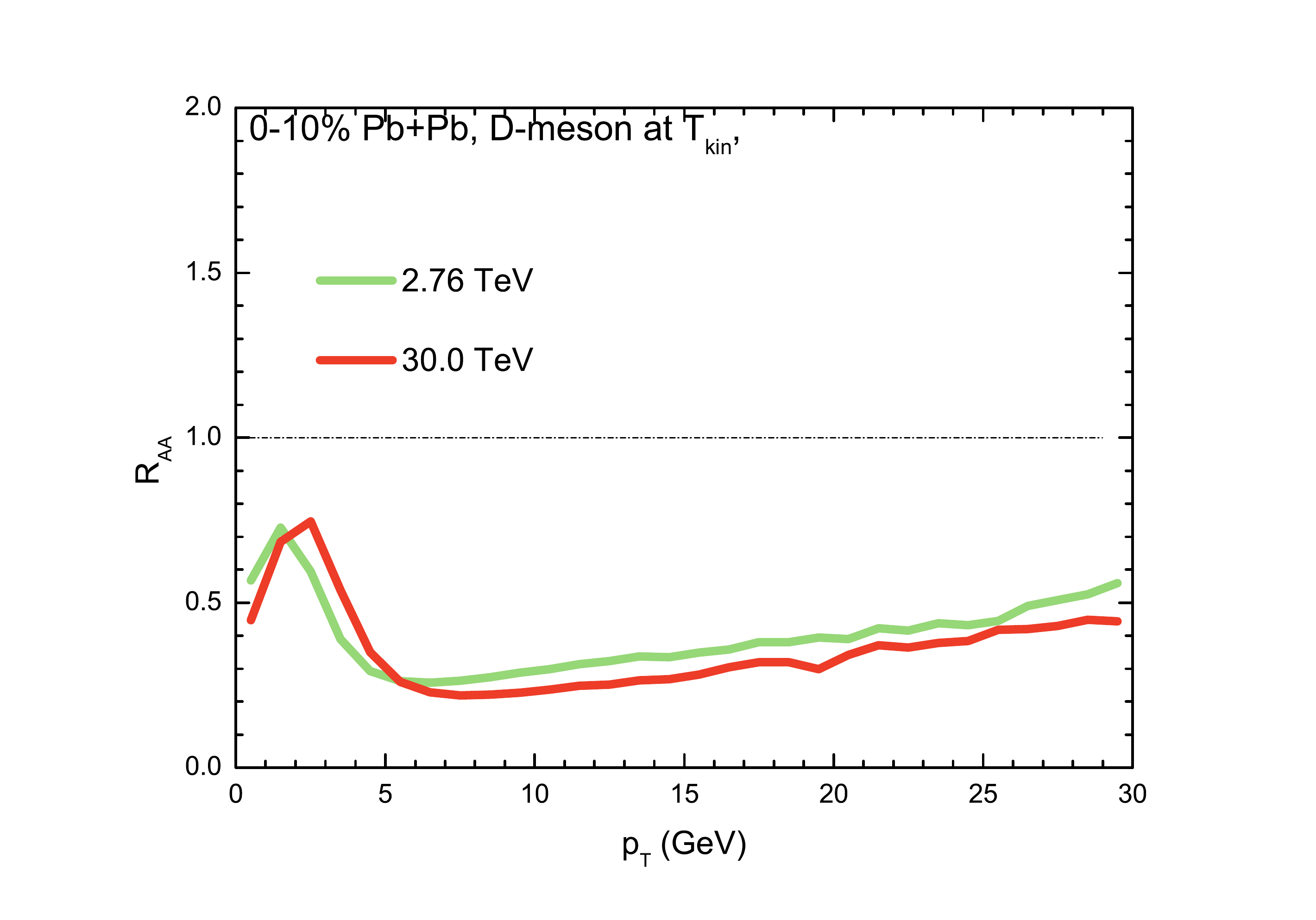}
%\vspace{-0.5cm}
\includegraphics[width=0.6\columnwidth]{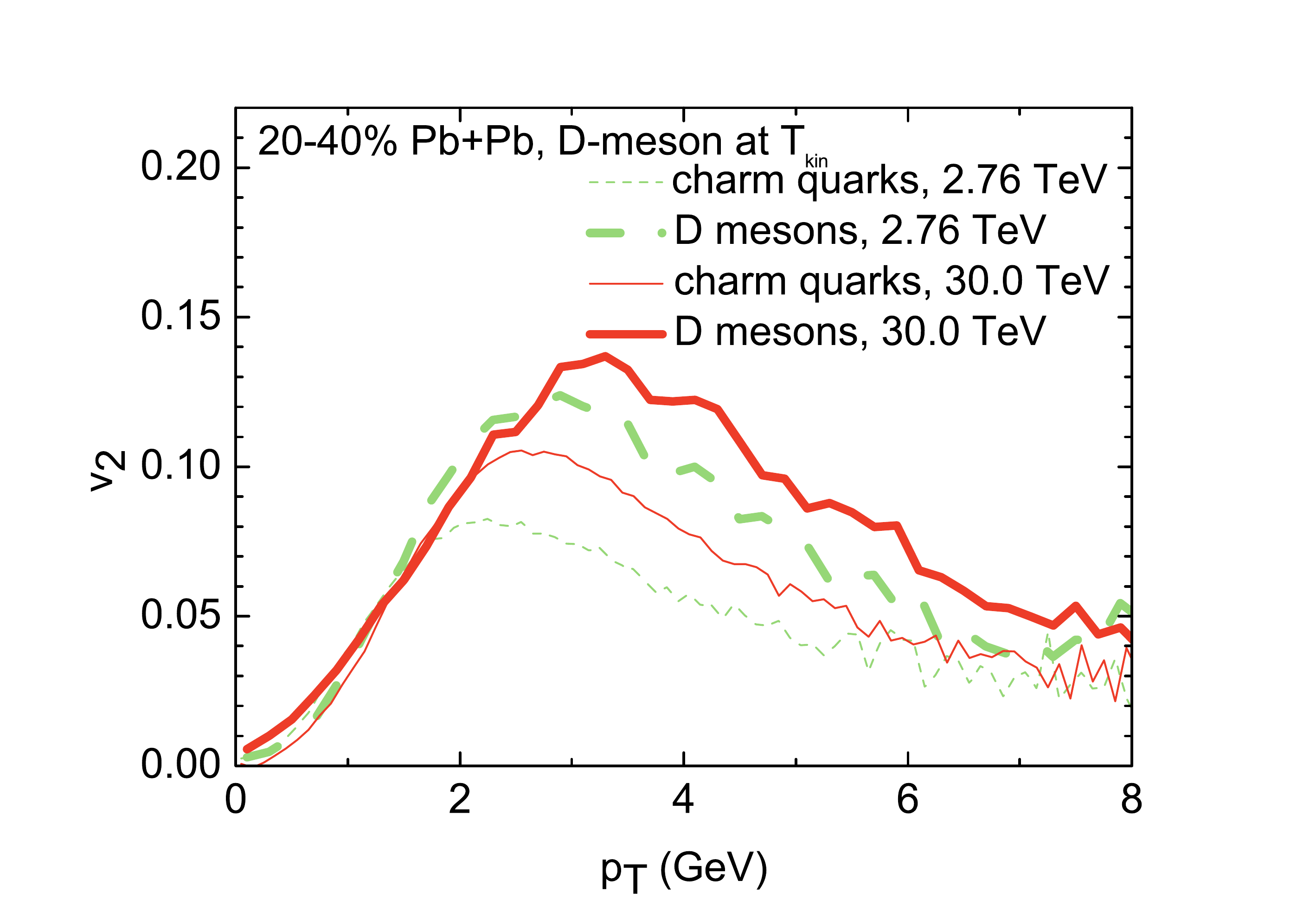}
%\vspace{-0.5cm}
\caption{(Color online) The $R_{\rm AA}$ (upper panel) and $v_2$ (lower panel)
of charm quarks and $D$ mesons for semi-central Pb+Pb collisions at
$\sqrt{s_{NN}}=2.76~{\rm TeV}$ and 30\,TeV.}
\label{fig_cD}
\end{figure}

\begin{figure} [!t]
\includegraphics[width=0.6\columnwidth]{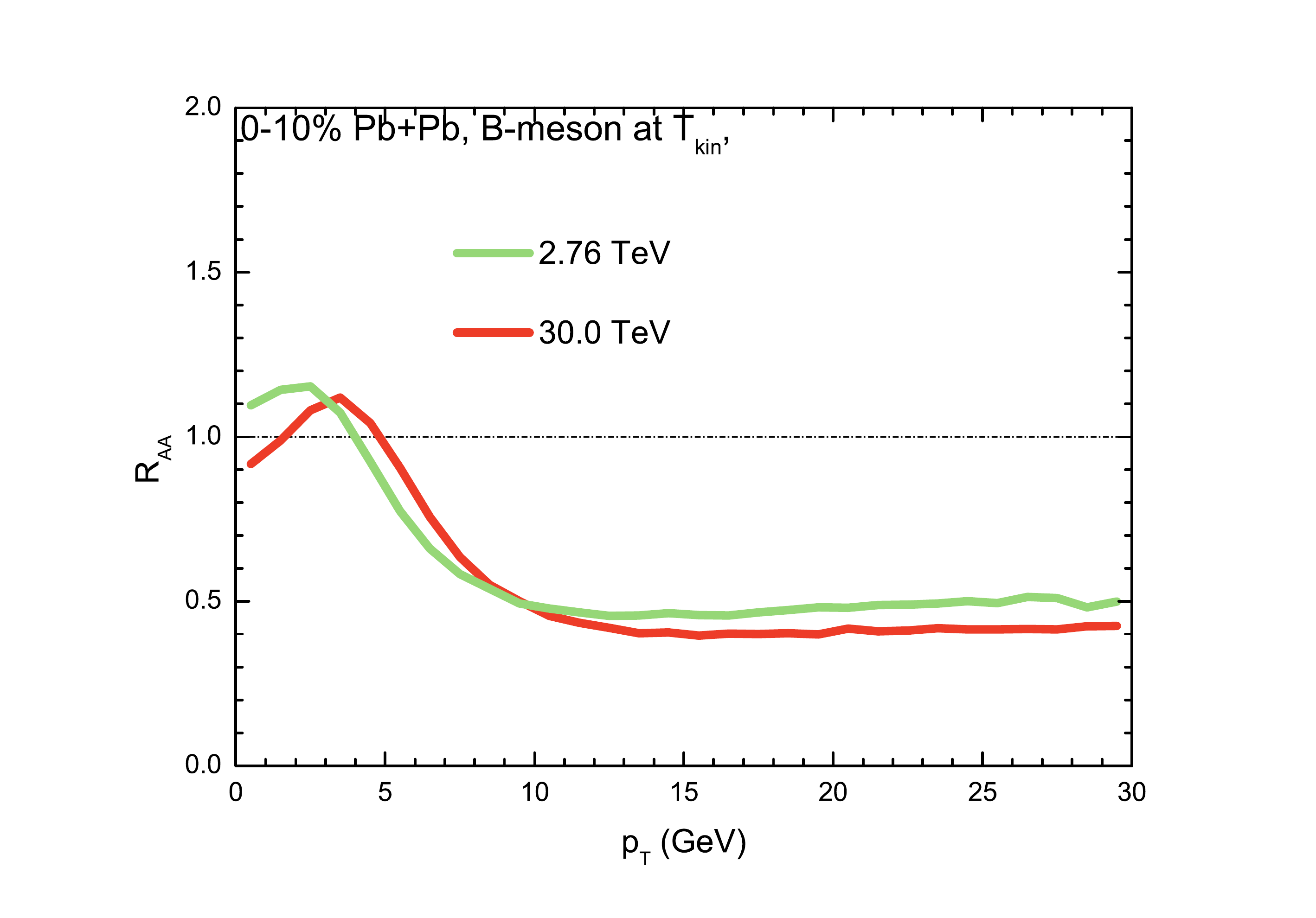}
%\vspace{-0.5cm}
\includegraphics[width=0.6\columnwidth]{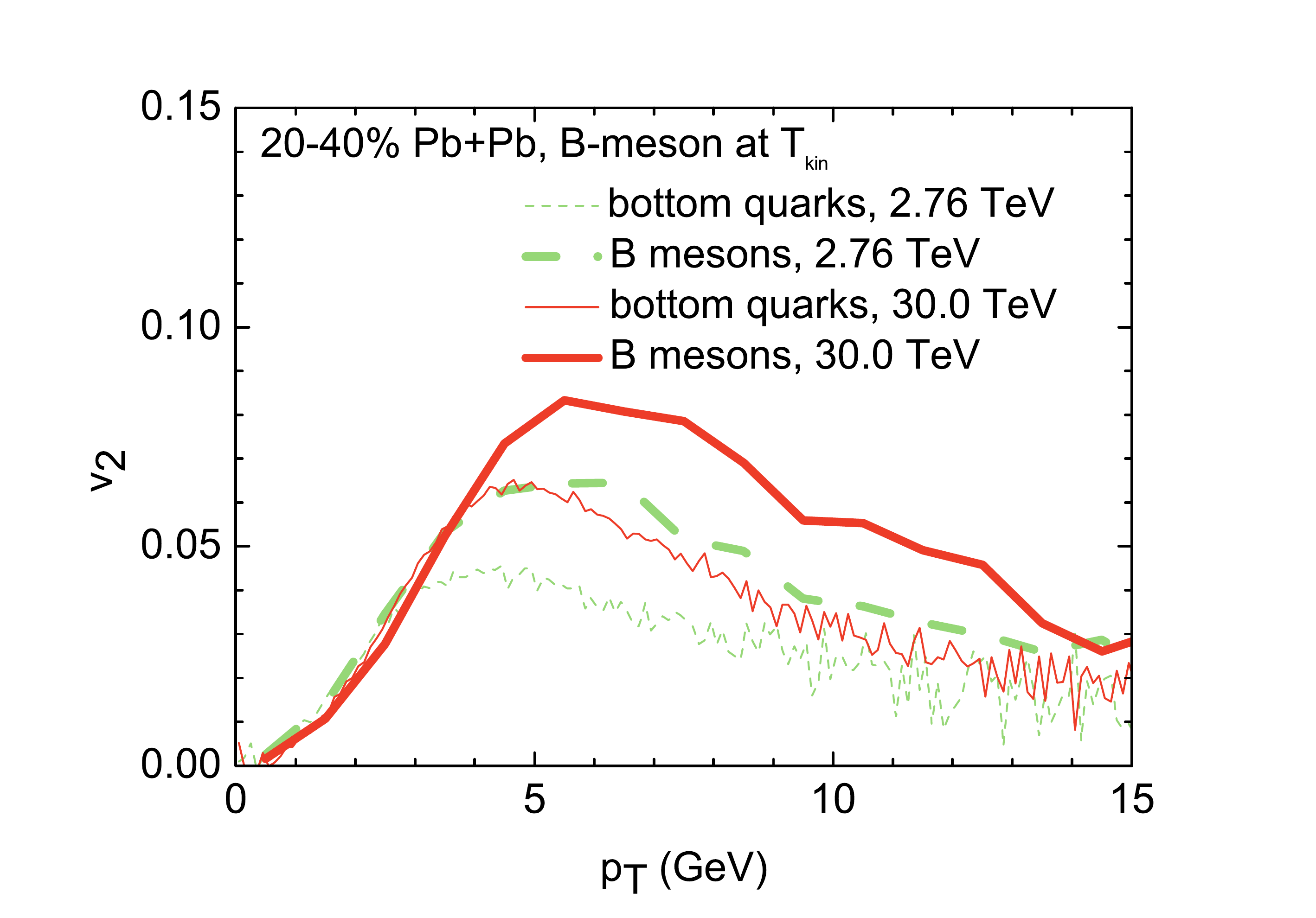}
%\vspace{-0.5cm}
\caption{(Color online) The $R_{\rm AA}$ (upper panel) and $v_2$ (lower panel)
of bottom quarks and $B$ mesons for semi-central Pb+Pb at
$\sqrt{s_{NN}}=2.76~{\rm TeV}$ and 30\,TeV.}
\label{fig_bB}
\end{figure}

The hydrodynamic evolution utilizes the 2+1D ideal hydro code
AZHYDRO~\cite{Kolb:2003dz}, augmented with a modern lQCD equation
of state for the QGP which is matched in a near-smooth transition to a hadron
resonance gas (HRG) at $T_{\rm pc}$=170\,MeV. The chemical
freezeout of hadrons is implemented for temperatures below $T_{\rm ch}$=160\,MeV,
utilizing effective chemical potentials for hadrons stable under strong
interactions. Our hydro tune in Pb+Pb collisions consists of initial conditions
from a Glauber model with an initialization time of 0.4\,fm/$c$ (without
initial flow nor fluctuations), which allows for a reasonable description of
the bulk-hadron spectra and inclusive elliptic flow at kinetic freezeout at
2.76\,TeV~\cite{He:2014cla}. It features a fast build-up of radial flow as well as bulk momentum anisotropy. 
As a result, the bulk $v_2$ gets almost saturated around $T_c$, which helps to develop substantial elliptic 
flow for both heavy quarks and thermal electromagnetic emissions (dileptons and photons).

% The evolution of the energy-momentum anisotropy,
%a key quantity characterizing the build-up of the elliptic flow of the
%hydrodynamic medium, is shown in Fig.~\ref{fig_epsilon_P} both for Pb+Pb
%$\sqrt{s_{NN}}=2.76~{\rm TeV}$ and 30\,TeV semi-central collisions.

In Fig.~\ref{fig_cD} we summarize our predictions for the nuclear modification
factor ($R_{\rm AA}$) and elliptic flow ($v_2$) of charm quarks and $D$ mesons
in 30\,TeV Pb+Pb collisions; the corresponding results for bottom quarks and
$B$ mesons are shown in Fig.~\ref{fig_bB}. The results overall are similar to Pb+Pb collisions
at the current LHC energy. A careful examination of the results shows, however, that the peak of $R_{AA}$ 
at low $p_T$ due to diffusion and parton recombination for hadronization shifts to higher $p_T$ because of the 
higher temperature and radial flow achieved at the higher colliding energy. The predicted suppression factors are
also larger than that given in the perturbative approach. This may be caused by the lack of radiative energy 
loss in this non-perturbative approach which is important at high $p_T$. In this model calculation, heavy quark 
diffusion in QGP contributes to about 60-70\% of the final total $v_2$ of $D/B$ mesons. 
The remaining contribution is due to coalescence of heavy and light quarks during hadronization and interaction of  $D/B$ 
mesons during the hadronic phase. Therefore, interactions of heavy flavor with medium during the entire evolution of the 
medium are all indispensable for $D/B$ mesons to develop large final $v_2$ that could reach as much 
as 12-14\% (for $D$ mesons) and  6-8\% (for $B$ mesons) in semi-central collisions.

\section{$J/\psi$ production}

The suppression of $J/\psi$ in hot medium has been considered as
a probe of the QGP created in the early stage of heavy
ion collisions \cite{Matsui:1986dk}. The nuclear modification factor
$R_{AA}\sim0.3$ in central Au+Au collisions at RHIC \cite{Adare:2006ns}
goes up to about 0.5 in central Pb+Pb collisions at $\sqrt{s_{NN}}=2.76$
TeV at LHC \cite{Palomo:2014zva} due to the increasing contribution
of charmonium regeneration \cite{BraunMunzinger:2000px,Thews:2005vj,Grandchamp:2002wp}.
One can similarly investigate the behavior of nuclear modification factor $R_{AA}$
of $J/\psi$ at tens of TeV, e.g. $\sqrt{s_{NN}}=20$ TeV in the framework of transport
approach \cite{Zhu:2004nw,Yan:2006ve,Liu:2009nb,Zhou:2014kka}.

Considering that charmonium is so heavy and difficult to be thermalized
in heavy-ion collisions, one can use the classical transport equation
to describe the charmonium motion in the medium,
\begin{equation}
\frac{\partial f_{\Psi}}{\partial t}+\frac{{\bf p}}{E_{\Psi}}\cdot{\bf \nabla}f_{\Psi}=-\alpha_{\Psi}f_{\Psi}+\beta_{\Psi},\label{transport}
\end{equation}
where $f_{\Psi}({\bf x},{\bf p},t)$ are the charmonium distribution
functions in phase space for $\Psi=J/\psi,\psi',\chi_{c}$. Considering
the fact that $J/\psi$'s in p+p collisions come from direct production
and decay from excited states $\psi'$ and $\chi_{c}$, one needs
the distributions of $\psi'$ and $\chi_{c}$. The charmonium energy
is denoted by $E_{\Psi}=\sqrt{m_{\Psi}^{2}+{\bf p}^{2}}$, and $\alpha_{\Psi}$
and $\beta_{\Psi}$ are the charmonium dissociation and regeneration
rates. Taking the gluon dissociation $g+\Psi\to c+\bar{c}$ as the
main dissociation process at high temperature, the loss term $\alpha_{\Psi}$
can be expressed as,
\begin{eqnarray}
\alpha_{\Psi}({\bf x},{\bf p},t|{\bf b}) & = & \frac{1}{2E_{\Psi}}\int\frac{d^{3}{\bf k}}{(2\pi)^{3}}\frac{1}{2E_{g}}W_{g\Psi}^{c\bar{c}}({\bf p},{\bf k})f_{g}({\bf x},{\bf k},t)\nonumber \\
 &  & \times\Theta\left(T({\bf x},t|{\bf b})-T_{c}\right),
\end{eqnarray}
with impact parameter ${\bf b}$ describing the centrality of collisions,
gluon energy $E_{g}$, the thermal gluon distribution $f_{g}$ and
the dissociation probability $W_{g\Psi}^{c\bar{c}}$. The dissociation
probability is determined by the dissociation cross section from gluons
whose vacuum value $\sigma_{\Psi}(0)$ is calculated through the operator
production expansion \cite{Peskin:1979va,Bhanot:1979vb}. Temperature
dependent cross section $\sigma_{\Psi}(T)$ can be estimated by taking
into account the geometry relationship between the cross section and the
size of $J/\psi$, $\sigma_{\Psi}(T)=\sigma_{\Psi}(0)\langle r_{\Psi}^{2}\rangle(T)/\langle r_{\Psi}^{2}\rangle(0)$.
The charmonium size $\langle r_{\Psi}^{2}\rangle(T)$ can be calculated
in the potential model \cite{Satz:2005hx}. The step function $\Theta$
means that we considered here only the dissociation (and regeneration)
in the deconfined phase with $T_{c}$ being the critical temperature
for the deconfinement phase transition. Considering the strong interaction
between charm quarks and the high-temperature
medium at colliding energy $\sqrt{s_{NN}}=20$ TeV, one can approximately
take thermal charm quark distribution $f_{c}$ in calculating the
regeneration rate. Since the regeneration process is the inverse process
of the gluon dissociation, the regeneration probability $W_{c\bar{c}}^{g\Psi}$
can be obtained via the detailed balance between the two processes.
In the above transport approach, we have neglected elastic collisions
between charmonium and the medium, since its effect on the momentum
integrated $R_{AA}$ is rather small \cite{Chen:2012gg}.

The local temperature $T({\bf x},t)$ and medium velocity $u_{\mu}({\bf x},t)$
appeared in the thermal gluon and charm quark distributions $f_{g}$
and $f_{c}$ are given by equations of ideal hydrodynamics, $\partial_{\mu}T^{\mu\nu}=0$,
where $T_{\mu\nu}$ is the energy-momentum tensor of the medium. While
the charm quarks are assumed to be thermalized, they do not reach
chemical equilibrium. The space-time evolution of the charm quark
density $n_{c}({\bf x},\tau|{\bf b})$ satisfies the conservation
$\partial_{\mu}(n_{c}u^{\mu})=0$, with the initial density determined
by the nuclear thickness functions $n_{c}({\bf x},\tau_{0}|{\bf b})=[d\sigma_{NN}^{c(\bar{c})}/dy]T_{A}({\bf x}-{\bf b}/2)T_{B}({\bf x}+{\bf b}/2)$.

The shadowing effect becomes extremely important at small $x$ or
high colliding energy $\sqrt{s_{NN}}$. In our calculation we use
the EKS98 package \cite{Eskola:1998df} to take into account of the shadowing
factor $R(x)$. Its value in the dominant kinematic region for charm quark production at $\sqrt{s_{NN}}=20$ TeV
is around $0.8$, which leads to a strong suppression for the regeneration:
the charmonium nuclear modification factor is reduced to $\sim64\%$
due to the shadowing effect! The other cold nuclear matter effects
like Cronin effect \cite{Gavin:1988tw} and nuclear absorption can
also be included in the initial condition of the transport equation
\cite{Zhou:2014kka}.

The initial thermodynamic conditions for the hydrodynamic evolution
is determined by fitting the extrapolated multiplicity of charged hadrons at
$\sqrt{s_{NN}}=20$ TeV. We take the initial thermalization time $\tau_{0}=0.6$
fm/c and the initial temperature at the center of the fireball $T_{0}=540$
MeV for central Pb+Pb collisions \cite{Heinz:2005bw}. The critical
temperature is chosen as $T_{c}=165$ MeV.

\begin{figure}[!hbt]
\includegraphics[width=0.5\textwidth]{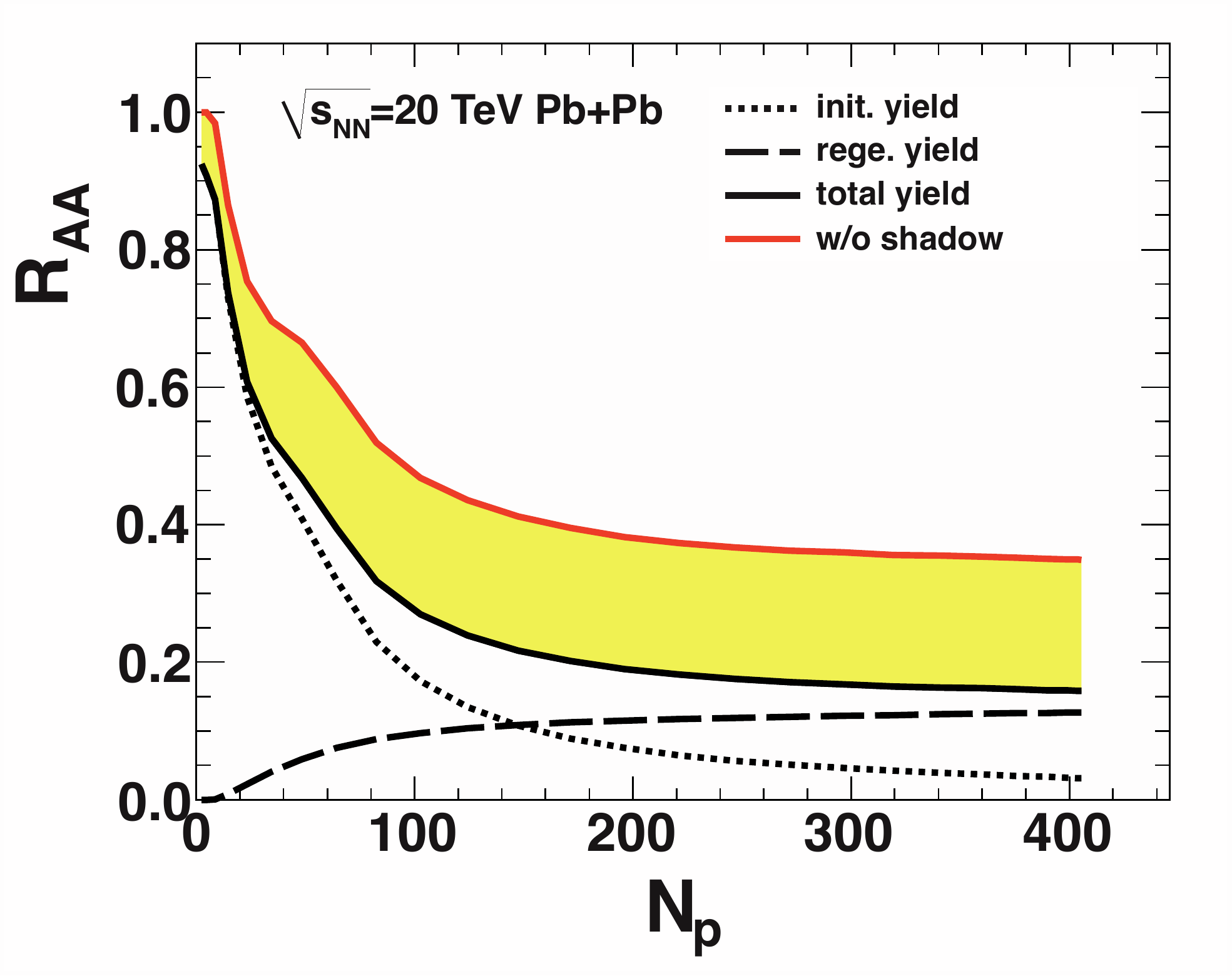} \caption{The prompt $J/\psi$ nuclear modification factor as a function of
centrality at mid-rapidity $|y|<1$ in $\sqrt{s_{NN}}=20$ TeV Pb+Pb
collisions. The dotted, dashed and solid lines are the initial production,
regeneration and full result, respectively. The upper limit of the
band is the full result without shadowing effect. \label{jpsi-fig1}}
\end{figure}

The initial charmonium distribution is in principle fixed by the corresponding
p+p data, modified by the cold nuclear matter effects~\cite{Zhou:2014kka}.
Since there are not yet p+p data at $\sqrt{s_{NN}}=20$ TeV, we use
the simulator PYTHIA \cite{Sjostrand:2000wi} to extract the $J/\psi$
and charm quark production cross sections in central rapidity region
$|y|<1$,

\begin{equation}
\frac{d\sigma_{NN}^{J/\psi}}{p_{t}dp_{t}}=A\frac{n-1}{\langle\bar{p}_{t}^{2}\rangle_{NN}}\left(1+\frac{p_{t}^{2}}{\langle\bar{p}_{t}^{2}\rangle_{NN}}\right)^{-n}
\end{equation}

and $\frac{d\sigma_{NN}^{c\bar{c}}}{dy}=1.4$ mb, where $\langle\bar{p}_{t}^{2}\rangle_{NN}=\langle p_{t}^{2}\rangle_{NN}+a_{gN}l$
is the $J/\psi$ averaged transverse momentum square modified by the
Cronin effect with $\langle p_{t}^{2}\rangle_{NN}=22.69(\text{GeV/c})^{2}$,
$a_{gN}=0.2\:\text{GeV}^{2}/\text{fm}$, $A=2.011\times1.68\times(10)^{-2}$
mb, $n=3.164$, and $l$ being the averaged traveling length of the
two gluons before they fuse into a $J/\psi$.

The prediction of the nuclear modification factor $R_{AA}$ for $J/\psi$
at $\sqrt{s_{NN}}=20$ TeV is shown in Fig.~\ref{jpsi-fig1}. The initially
produced $J/\psi$'s are almost totally dissolved in central collisions
due to the high temperature at mid rapidity. Because of the strong
shadowing effect which reduces the charm quark number by a factor
of $80\%$ and the regenerated $J/\psi$ number by a factor of about
$64\%$, the charmonium regeneration is largely suppressed, and the
full result is only about $15\%$ in central collisions. Considering
the uncertainty of the calculation of the shadowing effect, the maximum
$R_{AA}$ without considering the shadowing effect can reach $35\%$,
see the upper limit of the band in Fig. \ref{jpsi-fig1}.

The small nuclear modification factor for $J/\psi$ at high energies shown here is caused by the complete melting of initially produced charmonia and strong shadowing effect on initial production of charm quarks and regenerated charmonia. However, the case for $\Upsilon$ may be different. While the maximum temperature ($T_0$ = 540 MeV) of the fireball at $\sqrt{s_{NN}}$=20 TeV is several times higher than the $J/\psi$ dissociation temperature $T_d^{J/\psi}\sim 1.5 T_c$, it is around the $\Upsilon$ dissociation temperature  $T_d^\Upsilon\sim 3T_c$. Therefore, most of the initially produced and regenerated $\Upsilon$'s can survive the quark matter. The initial number of produced bottom quarks are also smaller leading to smaller number of regenerated $\Upsilon$'s in the final state. The effect of shadowing on the initial bottom quark production is also expected to be smaller. Therefore,  the nuclear modification factor for $\Upsilon$ is expected to be larger than that for $J/\psi$ and increases with collision centrality.

\section{Electromagnetic emission from heavy-ion collisions}

Electromagnetic observables serve as a clean penetrating probe to the ultra-relativistic heavy-ion collisions. Because of the smallness of the electromagnetic coupling compared to the strong interaction, the produced real and virtual photons suffer negligible final-state interactions. Therefore, the radiated thermal photons and dileptons carry direct dynamical information about the early stage of the fireball evolution, which are complementary to the majority of the hadronic observables.

The thermal dilepton emission rate per unit phase space can be written as
\begin{equation}
\frac{dN_{l^+l^-}}{d^4xd^4q} = -\frac{\alpha_{\rm EM}^2 L(M)}{\pi^3 M^2} \
f^B(q_0;T) \ {\rm Im}\Pi_{\rm EM}(M,q;\mu_B,T) \ ,
\label{rate}
\end{equation}
where the key quantity is the electromagnatic (EM) spectral function
of the QCD medium, ${\rm Im}\Pi_{\rm EM}\equiv \frac{1}{3} g_{\mu\nu}
{\rm Im}\Pi_{\rm EM}^{\mu\nu}$, weighted by the thermal Bose
factor, $f^B$, and the virtual photon propagator, $1/M^2$, with dilepton
invariant mass $M^2=q_0^2-q^2$;
$L(M)$ is a lepton phase-space factor (=1 for vanishing lepton mass).

\begin{figure} [!t]
\includegraphics[width=0.6\columnwidth]{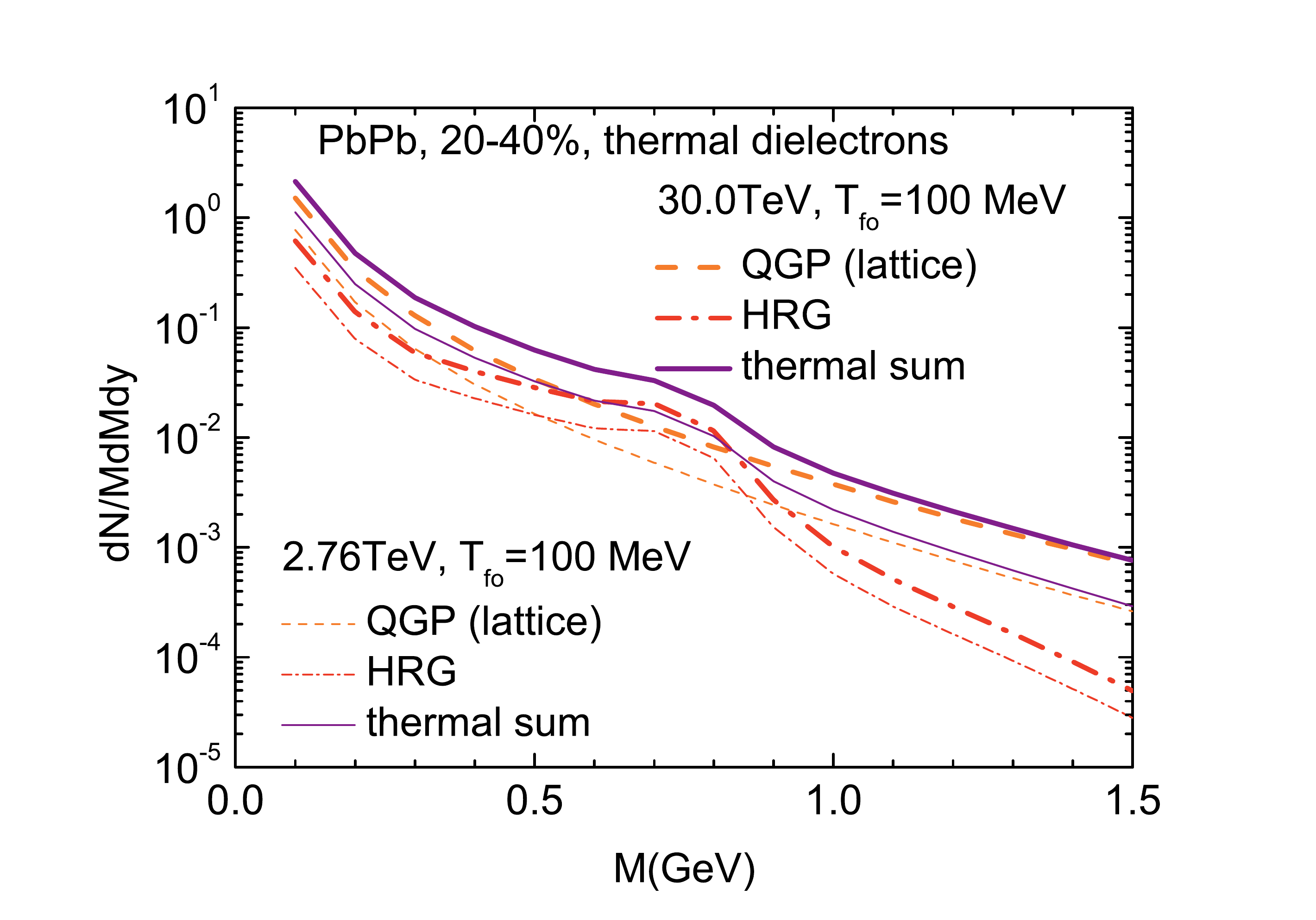}
%\vspace{-0.5cm}
\includegraphics[width=0.6\columnwidth]{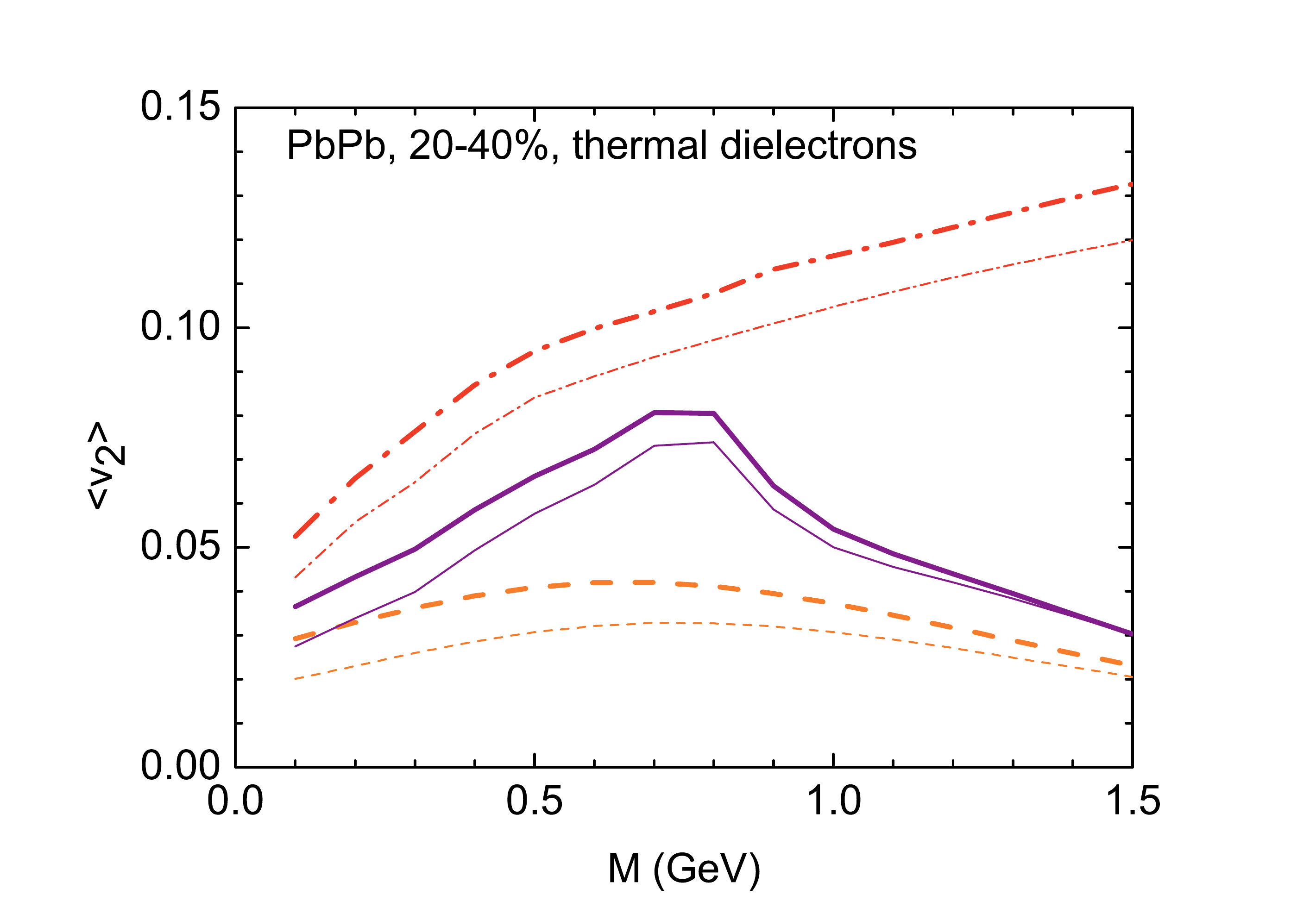}
%\vspace{-0.5cm}
\caption{(Color online) The invariant-mass spectra (upper panel) and integrated
elliptic flow (lower panel) of thermal dielectrons for  semi-central Pb+Pb
collisions at $\sqrt{s_{\rm NN}}=2.76~{\rm TeV}$ and 30\,TeV.}
\label{fig_thermaldielectrons}
\end{figure}

\begin{figure} [!t]
\includegraphics[width=0.6\columnwidth]{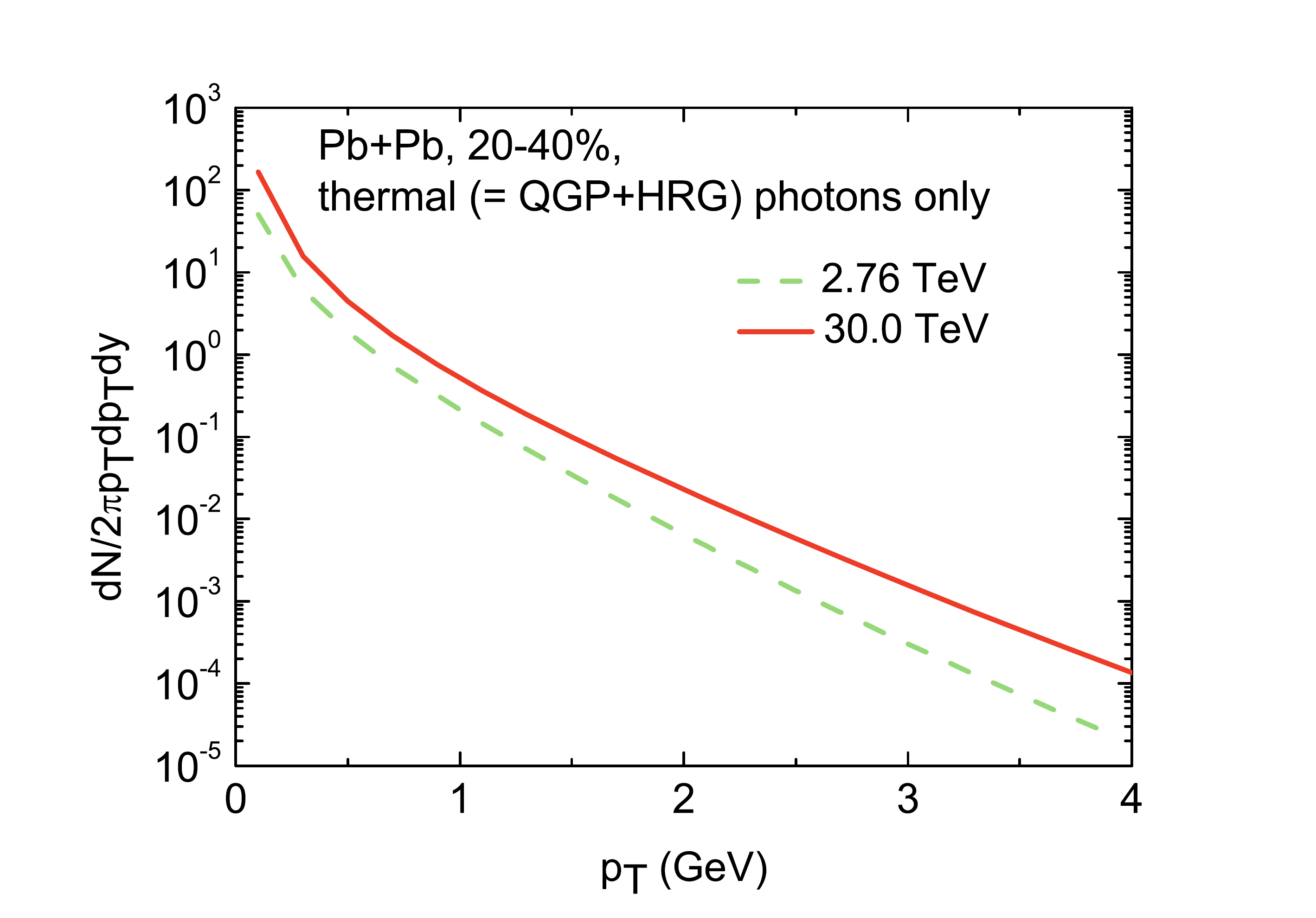}
%\vspace{-0.5cm}
\includegraphics[width=0.6\columnwidth]{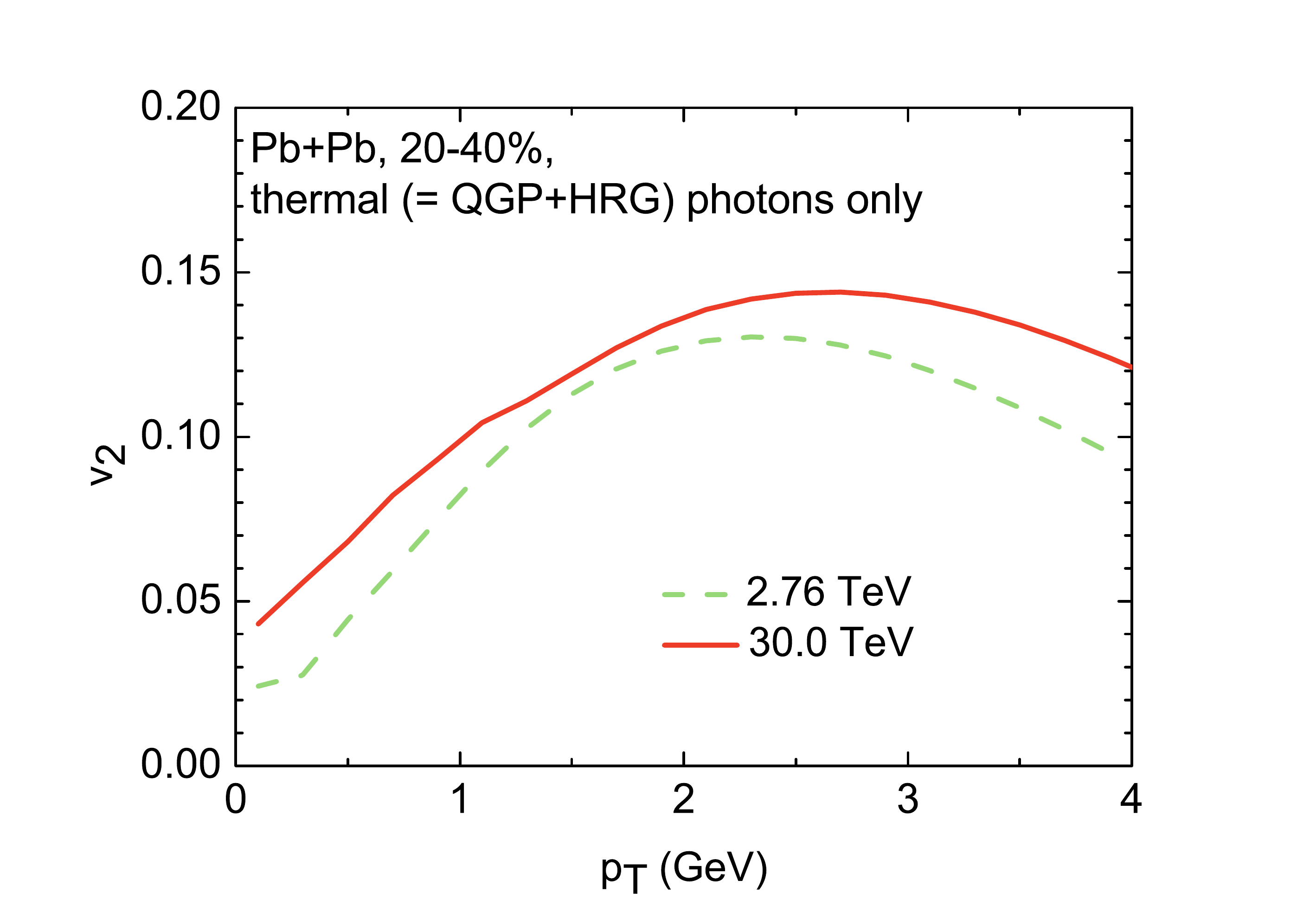}
%\vspace{-0.5cm}
\caption{(Color online) The transverse-momentum spectra (upper panel) and
elliptic flow (lower panel) of thermal photons for semi-central Pb+Pb
collisions at $\sqrt{s_{\rm NN}}=2.76~{\rm TeV}$ and 30\,TeV.}
\label{fig_thermalphotons}
\end{figure}

In the low-mass region, and in hadronic matter, the EM spectral function is
dominated by the $\rho$ meson, i.e., it is essentially proportional
to the imaginary part of the in-medium $\rho$ propagator,
\begin{equation}
D_{\rho}(M,q;\mu_B,T)=\frac{1}{M^2-m_{\rho}^2-\Sigma_{\rho\pi\pi}-\Sigma_{\rho M}-\Sigma_{\rho B}} \ .
\end{equation}
The medium effects are calculated in thermal-field theory through
self-energies~\cite{Rapp:1997fs,Urban:1998eg,Rapp:1999us,Rapp:1999ej} caused by
(a) interactions of the $\rho$'s pion cloud with hadrons from the heat bath
($\Sigma_{\rho\pi\pi}$), e.g., $\pi N\to \Delta$;
(b) resonant $\rho$ scattering off thermal mesons ($\Sigma_{\rho M}$), e.g.,
$\rho\pi\rightarrow a_1$; and
(c) resonant $\rho$ scattering off baryons and anti-baryons ($\Sigma_{\rho B}$),
e.g., $\rho N\rightarrow N^*$. The effective hadronic vertices are constrained by
EM gauge invariance and empirical decay branchings and scattering data in vacuum.
The off-shell dynamics naturally includes subthreshold excitations, such as
$\rho + N \rightarrow N^*(1520)$, which are instrumental in populating the low-mass
strength in the EM spectral function. The generic outcome of these calculations
is a strong broadening of the $\rho$'s spectral shape, with
only small mass shifts (which tend to cancel among the different contributions).
For the dilepton emission in QGP, we use the leading-order pQCD rate augmented
by a lQCD-inspired form factor~\cite{Ding:2010ga} (which yields results
similar to hard-thermal loop calculations~\cite{Braaten:1990wp}),
extended to finite 3-momentum~\cite{Rapp:2013nxa}.
This approach allows for good description of available dielectron emission
spectra at SPS and RHIC energies~\cite{vanHees:2007th,Rapp:2013nxa}.

For thermal photon emission one can find the rates as calculated in
Ref.~\cite{Turbide:2003si}. The hadronic emission was obtained by carrying
the above-described many-body calculations for dileptons to the photon point,
and adding mesonic $t$-channel reactions (which become important at the
photon point) from an effective Yang-Mills lagrangian for the $\pi\rho a_1$
system, plus $\omega$ $t$-channel exchange in $\pi\rho\to\pi\gamma$,
plus $\pi\pi$ and $\pi K$ Bremsstrahlung~\cite{Liu:2007zzw}. In the vicinity
of $T_{\rm pc}$, these rates approximately match the LO QGP
rates~\cite{Arnold:2001ms}, thereby rendering a near continuous photon
emissivity across the transition region~\cite{vanHees:2014ida}, analogous to
the dilepton case.

The predicted invariant-mass and transverse-momentum spectra, as well as elliptic flow
for thermal EM radiation in 30\,TeV Pb+Pb collisions are summarized in
Figs.~\ref{fig_thermaldielectrons} and \ref{fig_thermalphotons} for a medium evolution model 
according to the TAMU-tuned AZHYDRO~\cite{Kolb:2003dz} code as described in Sec.\ref{sec:tamu}. 
We recall that the early saturation of the
energy-momentum anisotropy (cf. discussions in Sec. \ref{sec:tamu}) is instrumental for the final $v_2$ of the thermal
emission, and plays an important role in the understanding of the large
direct-photon $v_2$ as recently observed by PHENIX~\cite{Adare:2011zr} and
ALICE~\cite{Lohner:2012ct}.
We also note that, as discussed in Ref.~\cite{vanHees:2014ida}, the continuous
hadronic freeze-out in the hydrodynamic evolution may underestimate somewhat
the hadronic emission contributions. Nonetheless, compared to the LHC results at 
2.76\,TeV, the thermal low-mass dilepton yield at 30\,TeV increases by about a factor 
of 2 (cf.~Fig.~\ref{fig_thermaldielectrons}), which is in line with the stronger 
than $N_{\rm ch}$ scaling found in previous calculations~\cite{Rapp:2013nxa}, 
and with existing dilepton data at SPS and RHIC. In fact, this behavior allows to
utilize the low-mass thermal radiation yield as a unique measure to infer the 
lifetime of the fireball to within $\sim$10\%~\cite{Rapp:2014hha}. 
At higher $p_T$, {\it e.g.}, in the thermal photon spectra around $p_T\simeq2$\,GeV
(cf.~Fig.~\ref{fig_thermalphotons}),
the increase in yield becomes even larger due to the increase in radial flow
at the higher collision energy.

To take into account of the fluctuation in the initial conditions of the hydrodynamic evolution of the medium on direct photon spectra in Pb+Pb collisions at $\sqrt{s_\mathrm{NN}} = 30$ TeV we employ event-by-event {\tt iEBE-VISHNU} framework \cite{Shen:2014vra}. The fluctuating initial entropy density profiles are generated using Monte-Carlo Glauber (MCGlb) and MCKLN models. The spatial configuration of the nucleon positions inside the lead nucleus are sampled with realistic two-body nucleon-nucleon correlations \cite{Alvioli:2009ab}. In MCGlb model, the collision-by-collision multiplicity fluctuation is implemented based on the  phenomenological KNO scaling observed in p-p collisions \cite{Khachatryan:2010nk}. For both initial conditions models, the event centrality is determined by sorting 1 million minimum bias collision events according to their initial total entropy.  The generated entropy density are then evolved using (2+1)-d viscous hydrodynamic code, {\tt VISH2+1} \cite{Song:2007ux}, starting at $\tau_0 = 0.6$ fm. The hydrodynamic equations are numerically solved with a lattice QCD based equation of state (EoS), {\tt s95p-v0-PCE} \cite{Huovinen:2009yb}, which implemented partial chemical equilibrium (PCE) below $T_\mathrm{chem} = 165$ MeV.  MCGlb initial conditions are evolved with specific shear viscosity, $\eta/s = 0.08$ and initial density profiles from MCKLN model are propagated with $\eta/s = 0.20$. These two sets of runs gave reasonable description of hadronic flow measurements in Pb+Pb collisions at $\sqrt{s_\mathrm{NN}} = 2.76$ TeV \cite{Qiu:2011hf,Shen:2011eg}. Here, we use them to extrapolate to higher collision energy. The final kinetic freeze-out is chosen to be $T_\mathrm{dec} = 120$ MeV. The overall normalization factor is fixed to fit the estimated final charged hadron multiplicity, $dN^\mathrm{ch}/d\eta \vert_{\vert \eta \vert < 0.5} = 2700$ at 0-5\% most central centrality. 

Thermal photons radiation is then calculated from these calibrated hydrodynamic medium above $T = 120$ MeV. In the QGP phase, the full leading order $O(\alpha_s \alpha_\mathrm{EM})$ photon emission rate is used \cite{Arnold:2001ms}, which includes Compton scattering, quark-anti quark annihilation, and the effective ``1 $\rightarrow$ 2'' collinear emission. In the hadron gas phase, photon produced through meson-meson reactions in a hadronic $(\pi, K, \rho, \omega, K^*, a_1)$ gas \cite{Turbide:2003si}, through the medium broadened $\rho$-spectral function, and through $\pi+\pi$ bremsstrahlung \cite{Liu:2007zzw,Heffernan:2014mla} are taken into account. Because the hydrodynamic medium is assumed to be slightly out-of-equilibrium, shear viscous corrections to the photon production rates are included in the 2 to 2 scattering processes in the QGP phase \cite{Shen:2014nfa} and in all the mesonic reaction channels in the hadronic phase \cite{Dion:2011pp}. The viscous corrections to the other channels have not been derived in theory yet. We use the QGP photon emission rate for the temperature region above $180$ MeV and switch to hadron gas rate below. In each collision event, the thermal photon spectrum is calculated by convoluting the photon emission rates with the hydrodynamic medium,
\begin{equation}
E \frac{d N^{\mathrm{th},\gamma}}{d^3 p} = \int  d^4 x \left( q \frac{d R}{d^3 q}(q, T(x)) \right) \bigg\vert_{q = p \cdot u(x)}.
\end{equation}
The anisotropy flow coefficients of the thermal photon momentum distribution are computed using the scalar-product method, $v_n\{\mathrm{SP}\}$. We correlate every produced thermal photon with the reference flow vector constructed using all charged hadrons \cite{Shen:2014lpa},
\begin{equation}
v_n\{\mathrm{SP}\}(p_T) = \frac{\langle \frac{dN^\gamma}{dy p_T dp_T} v^\gamma_n (p_T) v^\mathrm{ch}_n \cos (n(\Psi_n^\gamma(p_T) - \Psi_n^\mathrm{ch}))\rangle}{\langle \frac{dN^\gamma}{dy p_T dp_T} \rangle v_n^\mathrm{ch}\{2\}}.
\end{equation}

The prompt photons in Pb+Pb collisions at $\sqrt{s} = 30$ A TeV are estimated using the $N_\mathrm{coll}$-scaled the photons in p+p collisions at the same collision energy. The direct photon production in p+p collisions are calculated using the Next-to-leading-order (NLO) pQCD. The factorization scales in the parton distribution function, $\mu_f$, and fragmentation function, $\mu_D$, are chosen at 2 GeV, which also sets the lower limit for the calculable $p_T$, via the employed scale variations, $\mu = 2 p_T$. The nuclear effects, such as shadowing and isospin effects, in the parton distribution function are not included in the current estimation because their effects are genuinely small and the uncertainty becomes large at such a high collision energy. 

\begin{figure}[ht!]
\centering
\begin{tabular}{cc}
   \includegraphics[width=0.45\linewidth]{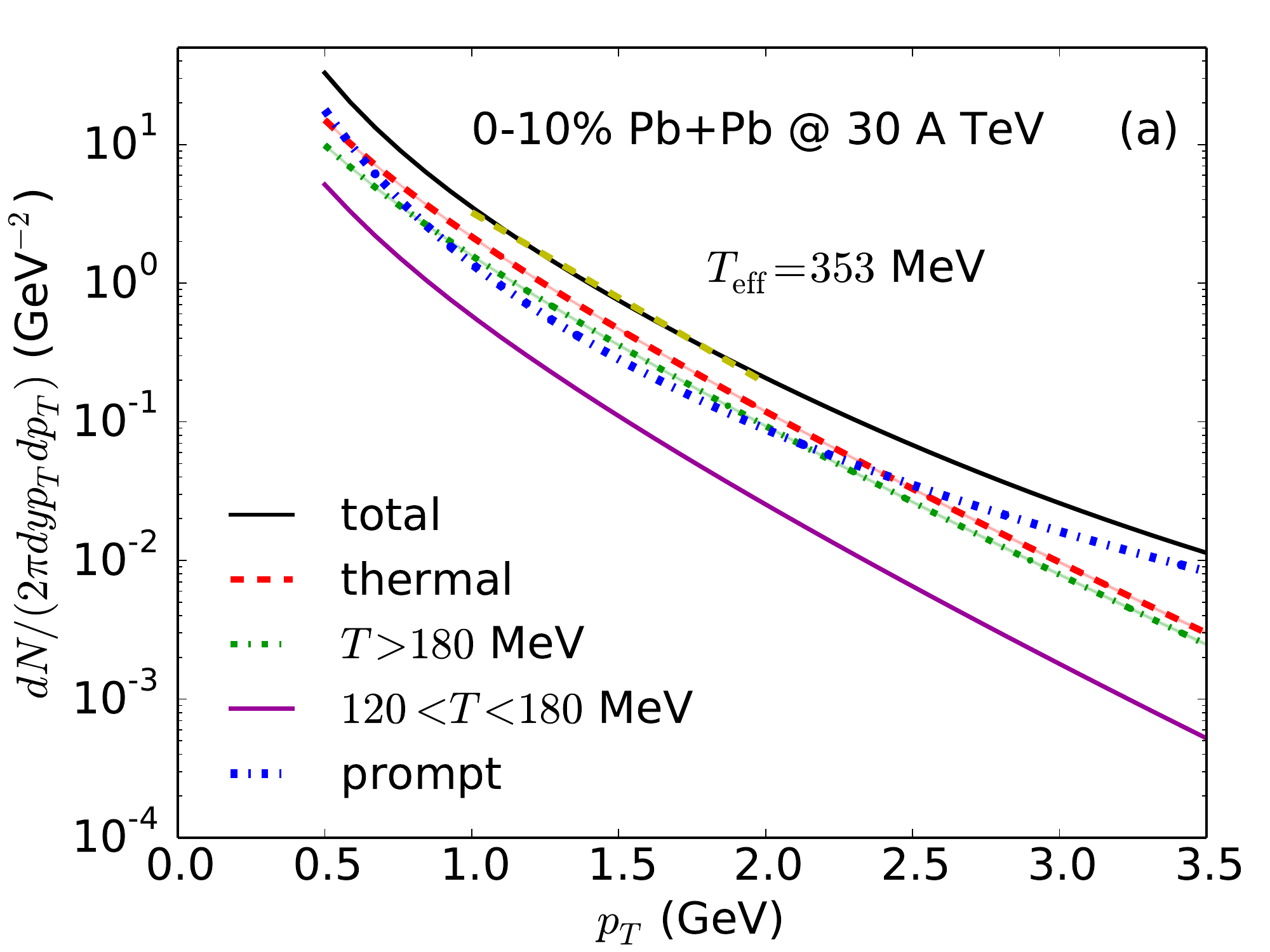} & 
   \includegraphics[width=0.45\linewidth]{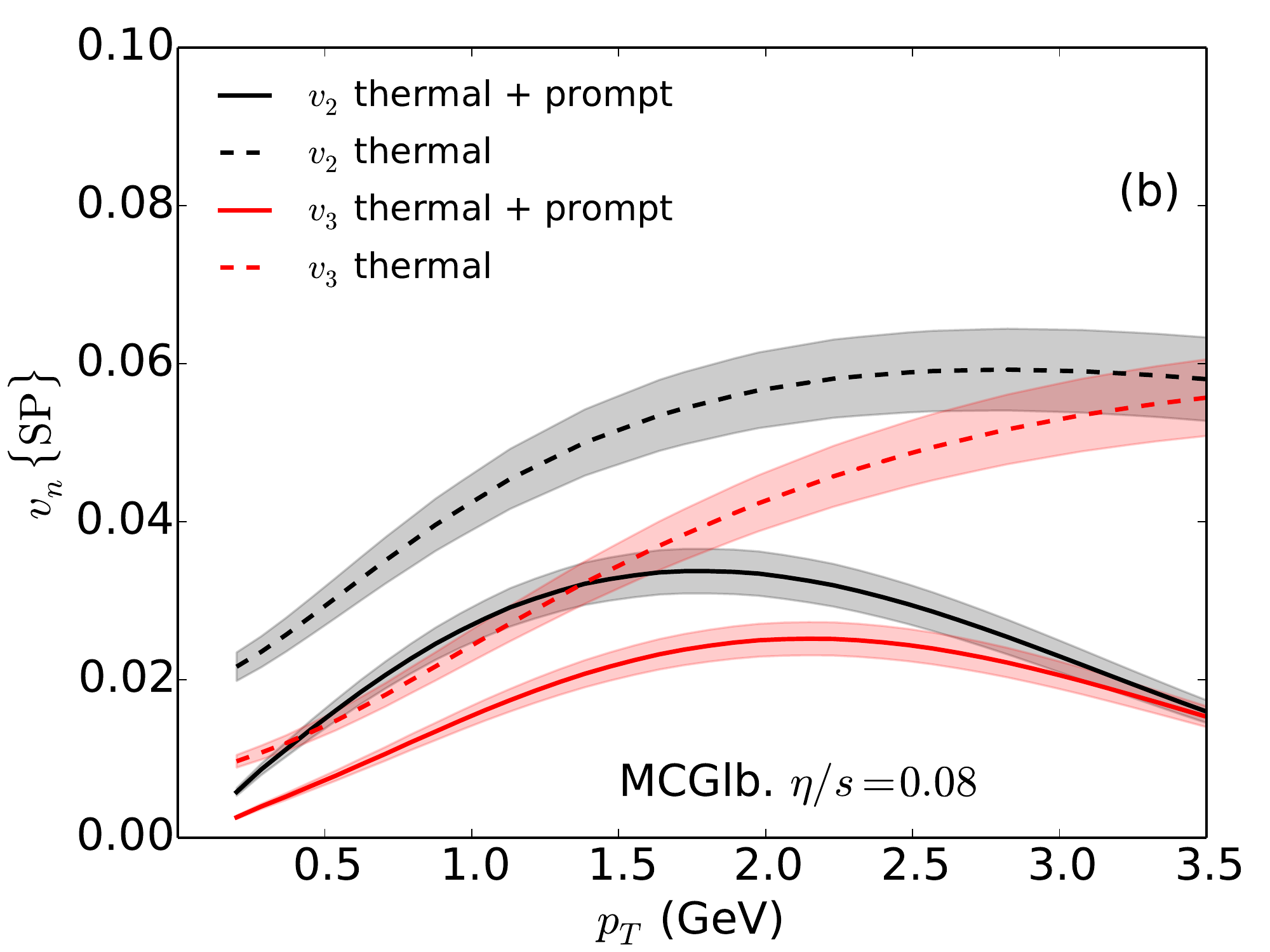}
\end{tabular}
\caption{Direct photon spectra and anisotropic flow coefficients $v_{2,3}\{\mathrm{SP}\}$ at 0-10\% centrality Pb+Pb collisions at $\sqrt{s_\mathrm{NN}} = 30$ TeV using MCGlb model with $\eta/s = 0.08$. Individual contributions of thermal photons are shown. The prompt photons are estimated using $N_\mathrm{coll}$-scaled photon spectrum in p-p collisions at the same collision energy. For 0-10\% centrality, $N_\mathrm{coll} = 2018 \pm 1$. }
\label{photon-fig1}
\end{figure}
%=======================================
%
%=======================================
\begin{figure}[b!]
\centering
\begin{tabular}{cc}
   \includegraphics[width=0.45\linewidth]{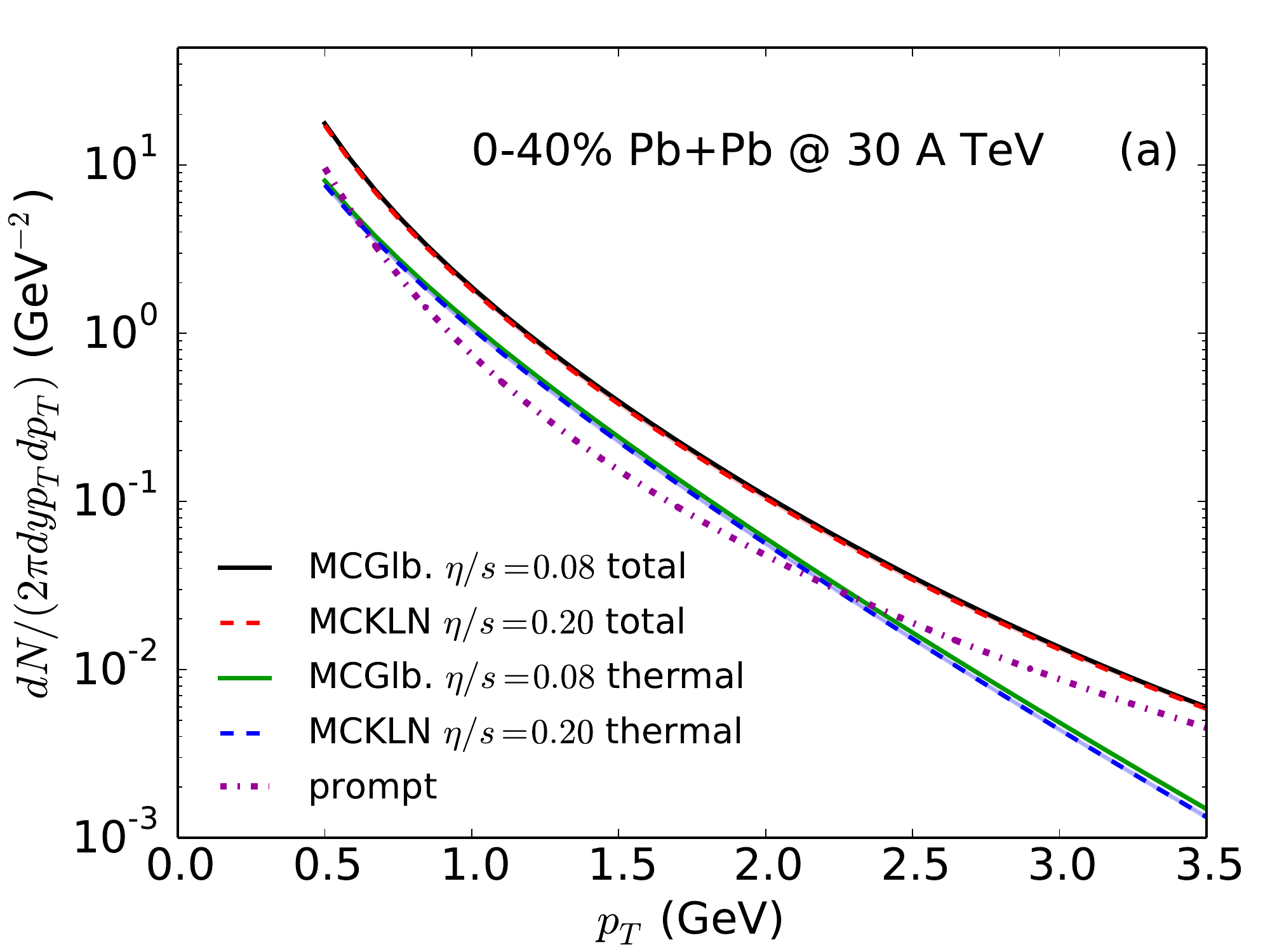} & 
   \includegraphics[width=0.45\linewidth]{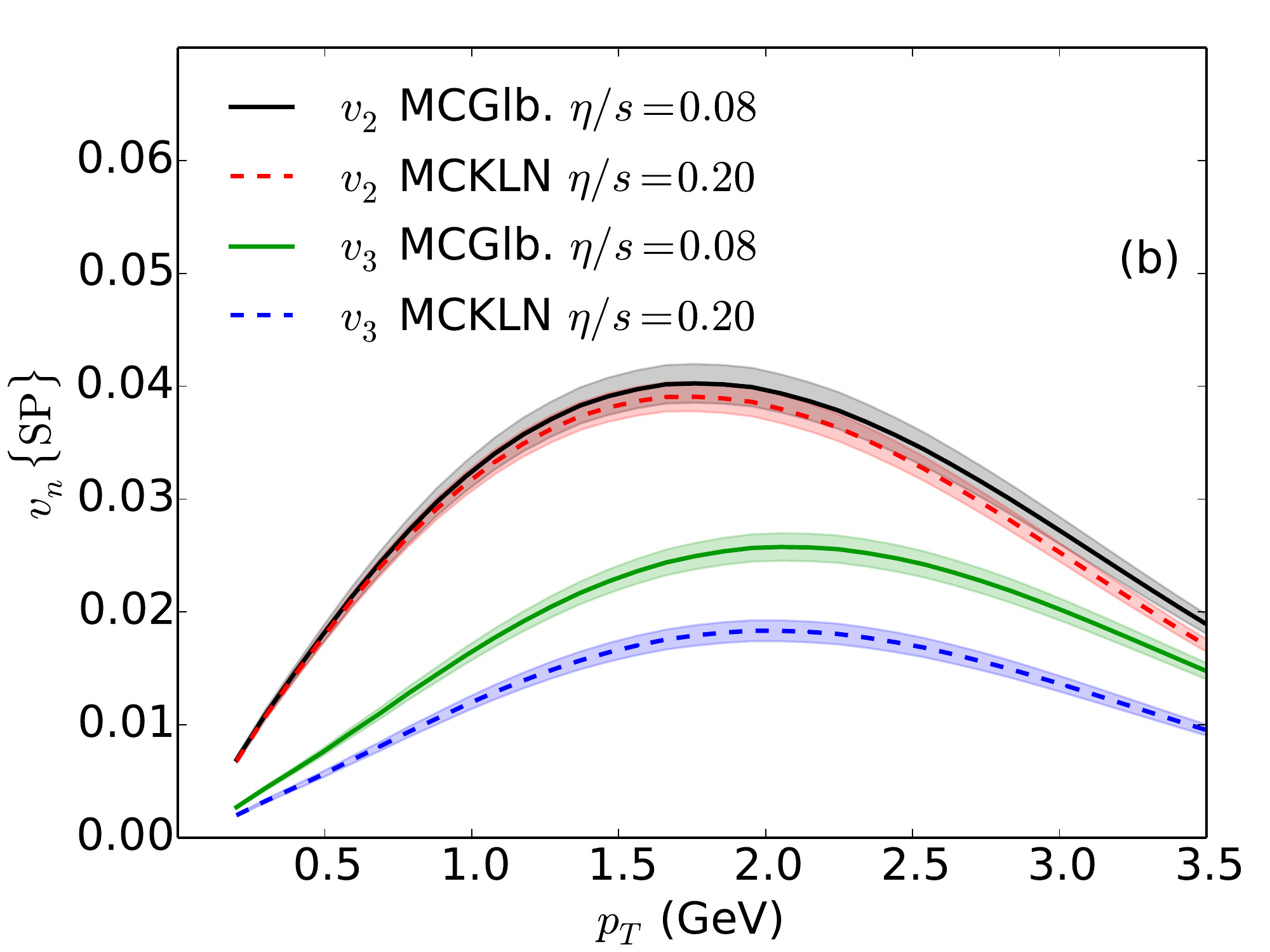}
\end{tabular}
\caption{Prediction of direct photon spectra and anisotropic flow coefficients $v_{2,3}\{\mathrm{SP}\}(p_T)$ at 0-40\% centrality Pb+Pb collisions at $\sqrt{s_\mathrm{NN}} = 30$ TeV from MCGlb model with $\eta/s = 0.08$ and MCKLN model with $\eta/s = 0.20$. The prompt photons are estimated using $N_\mathrm{coll}$-scaled photon spectrum in p-p collisions at the same collision energy. For 0-40\% centrality, $N_\mathrm{coll} = 1092 \pm 1$.}
\label{fig2}
\end{figure}
%=======================================

Fig.~\ref{photon-fig1}a shows the the direct photon spectrum in 0-10\% centrality Pb+Pb collisions at $\sqrt{s_\mathrm{NN}} = 30$ TeV. We find that the thermal signal exceed the prompt photon contribution for $p_T<2.5$ GeV. Most the of the thermal photons come from the high temperature region $T > 180$ MeV. The hadronic phase, $120 < T < 180$ MeV contributes about 10\% to the total thermal photons. Because of the strong hydrodynamic radial flow and high peak temperature of the fireball, the inverse slope of the direct photon spectrum reaches 353 MeV, which is $\sim$ 50 MeV higher than the inverse slope of direct photon spectrum at 2.76 A TeV \cite{Wilde:2012wc,Shen:2013vja} . In Fig.~\ref{photon-fig1}b, we show the direct photon anisotropic flow coefficients, $v_{2,3}\{\mathrm{SP}\}(p_T)$. Thermal components are shown for comparison. The thermal photon anisotropic flows are smaller than hadronic ones as shown in ~\ref{subsec-flow}. This is because the most of thermal photons are emitted from early $T > 180$ MeV region, where the hydrodynamic flow has not fully developed yet. Thus they carry less flow anisotropy compared with the hadrons. The triangular flow of direct photon are driven by the event-by-event fluctuation. Its signal is comparable with elliptic flow in the 0-10\% central collisions. Comparing the $v_n$ of thermal and direct photons, we find the prompt photons dilute $\sim 50\%$ of the flow anisotropy in the final direct photon signals. 

Fig.~\ref{fig2} shows our calculations of the direct photons emitted in 0-40\% centrality bin for two different sets of initial conditions. The two initial conditions with their corresponding specific shear viscosity give very close predictions for the direct photon spectrum and elliptic flow coefficient, $v_2\{\mathrm{SP}\}(p_T)$. The MCKLN initial conditions with a larger $\eta/s$ produce a smaller direct photon triangular flow.

%=======================================
\begin{figure}[h!]
\centering
\begin{tabular}{ccc}
  \includegraphics[width=0.33\linewidth]{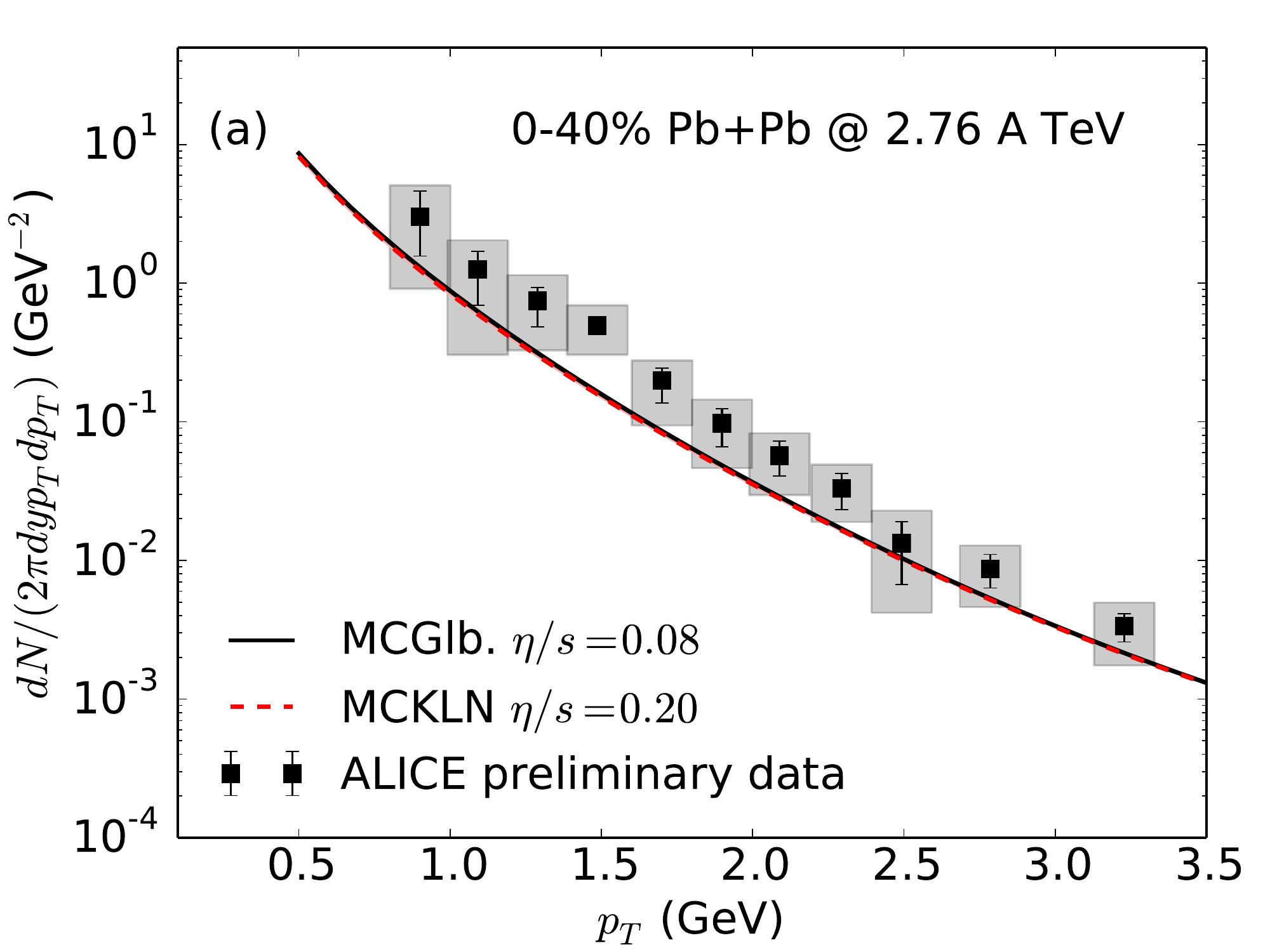} &
   \includegraphics[width=0.33\linewidth]{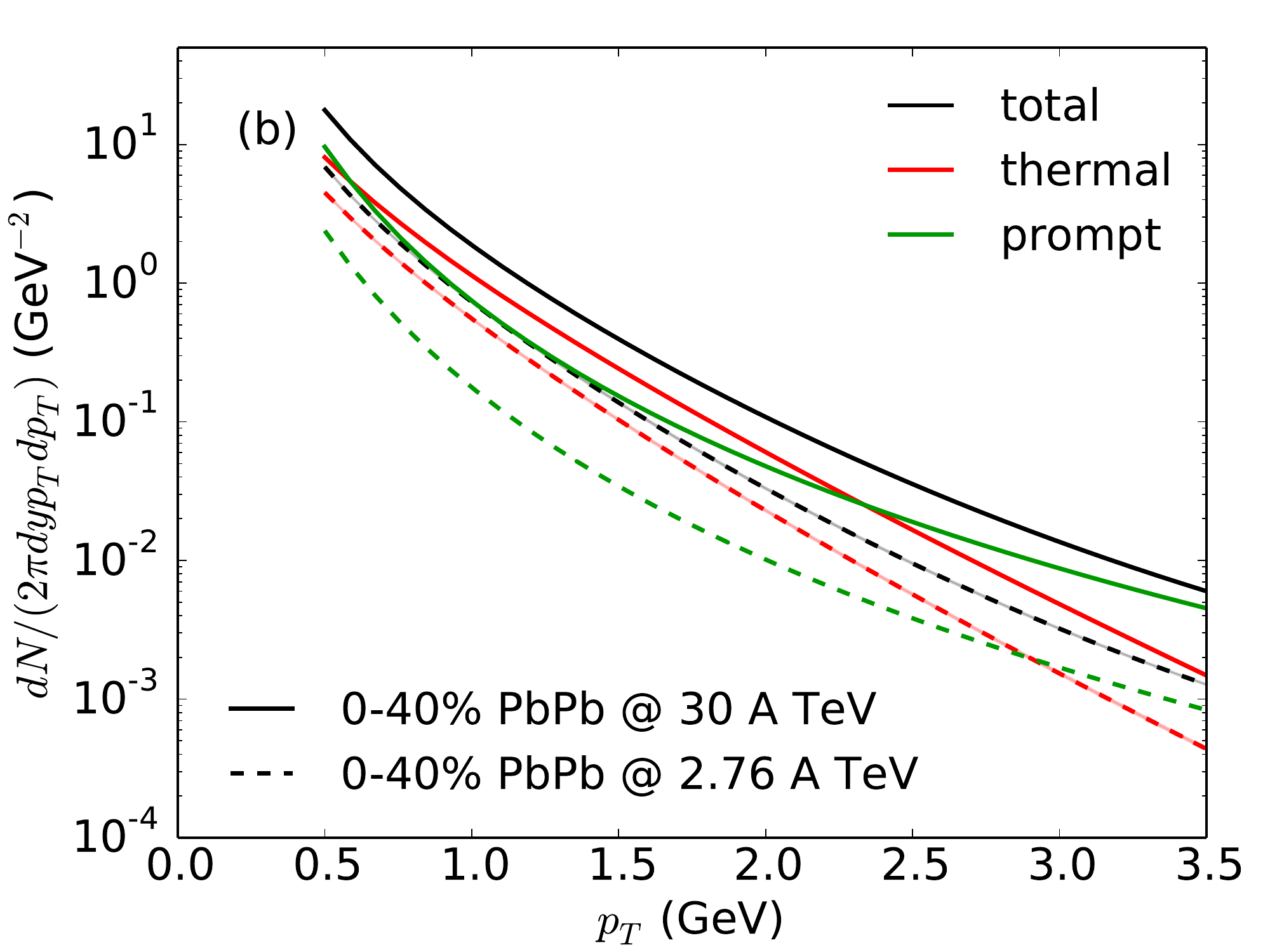}  &
    \includegraphics[width=0.33\linewidth]{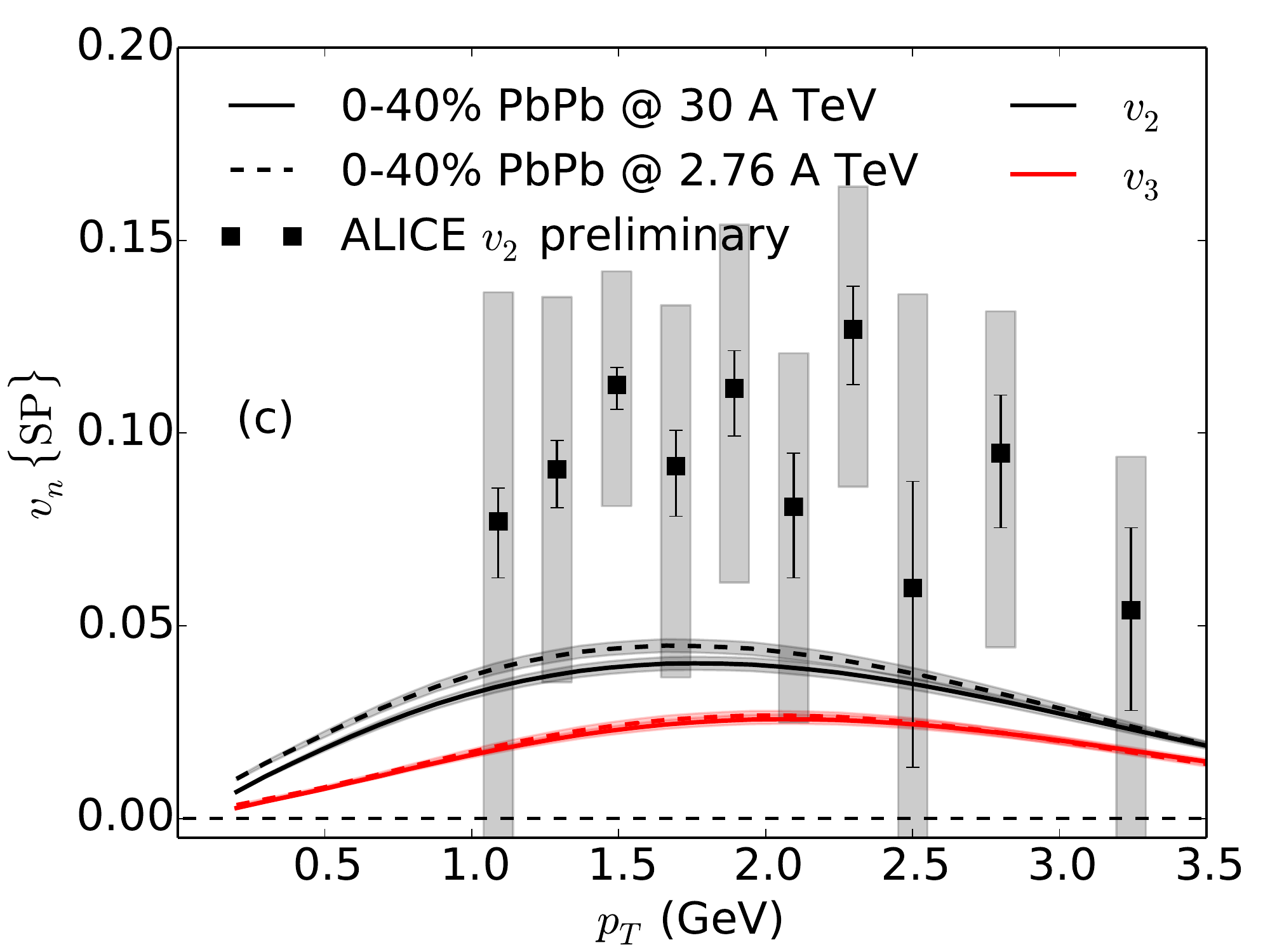} 
\end{tabular}
\caption{Panel (a): Theory calculations of direct photon spectra compared with ALICE preliminary measurement in 0-40\% Pb+Pb collisions at 2.76 A TeV \cite{Wilde:2012wc}. Panel (b): Comparisons of the individual component in direct photon spectra in 0-40\% Pb+Pb collisions at 30 A TeV and at 2.76 A TeV. Panel (c): Direct photon $v_{2,3}\{\mathrm{SP}\}(p_T)$ in 0-40\% Pb+Pb collisions at 30 A TeV and at 2.76 A TeV. Direct photon $v_2\{\mathrm{SP}\}(p_T)$ at 2.76 A TeV is compared with ALICE preliminary measurement \cite{Lohner:2012ct}.}
\label{fig3}
\end{figure}
%=======================================

In Fig.~\ref{fig3}, we compare the direct photon spectra and their anisotropic flow coefficients in 0-40\% Pb+Pb collisions at $\sqrt{s} = 30$\,$A$\,TeV with those at $\sqrt{s} = 2.76$\,$A$\,TeV available at current LHC experiments. Because of $\sim 70\%$ more entropy in the system, the space-time volume of the hydrodynamic medium is considerably larger compared with the fireball at $\sqrt{s} = 2.76$\,$A$\,TeV. The lifetime of the fireball is $\sim 30\%$ longer. Therefore, there are about 2.5 times thermal photons produced compared to current LHC energy. 
However, we find a even large increase of the prompt photons, about a factor of 4, compared to $\sqrt{s} = 2.76$\,$A$\,TeV. The ratio of thermal/prompt reduces as the collision energy increases. Because of this large prompt component, the direct photon anisotropic flow coefficients are slightly smaller compared to current LHC energy.  
The final produced direct photon spectrum at $\sqrt{s} = 30$\,$A$\,TeV is roughly 3.5 times of the photon produced at $\sqrt{s} = 2.76$\,$A$\,TeV.

%=======================================
\begin{figure}[h!]
\centering
\begin{tabular}{ccc}
   \includegraphics[width=0.33\linewidth]{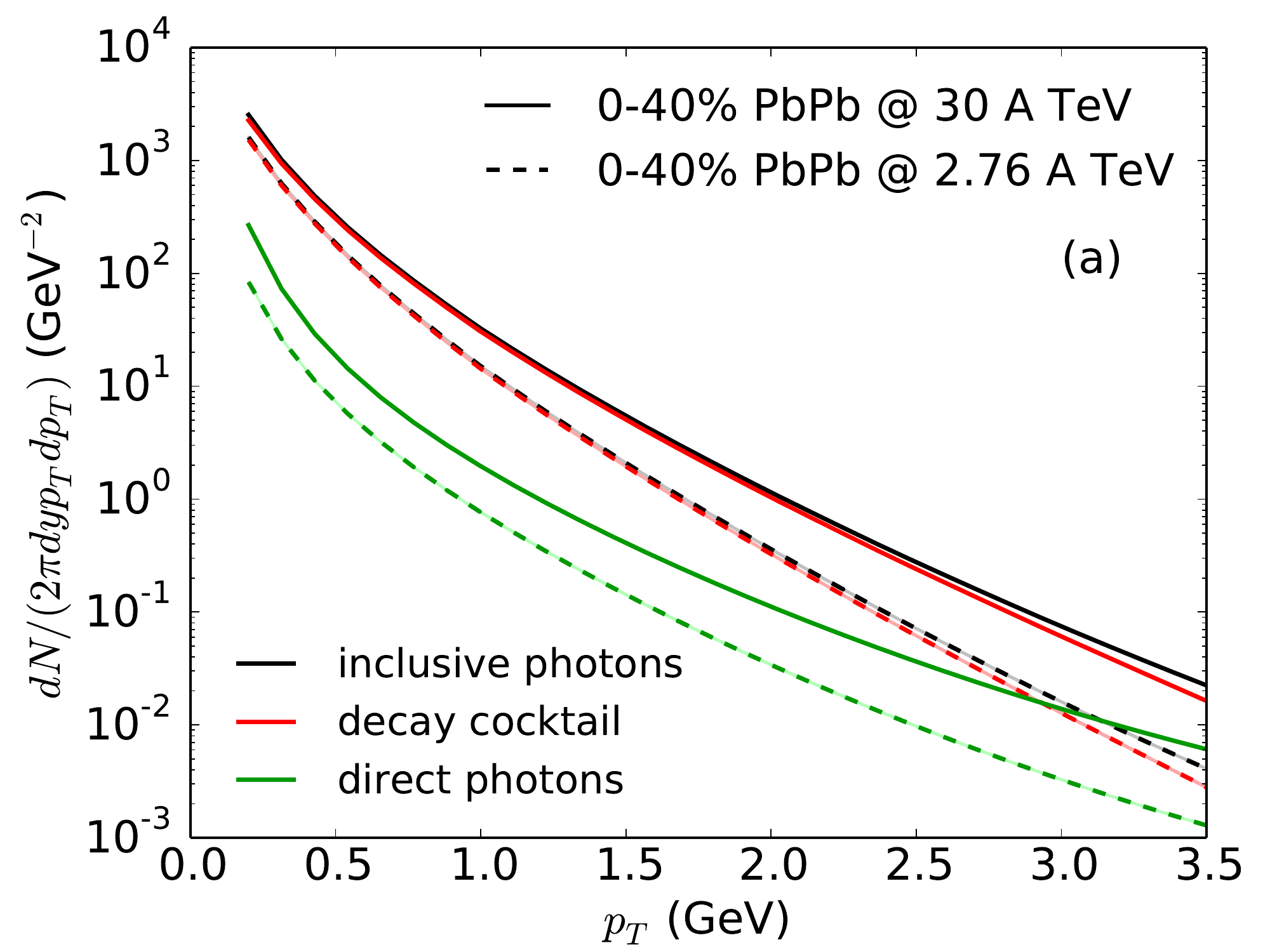} & 
   \includegraphics[width=0.33\linewidth]{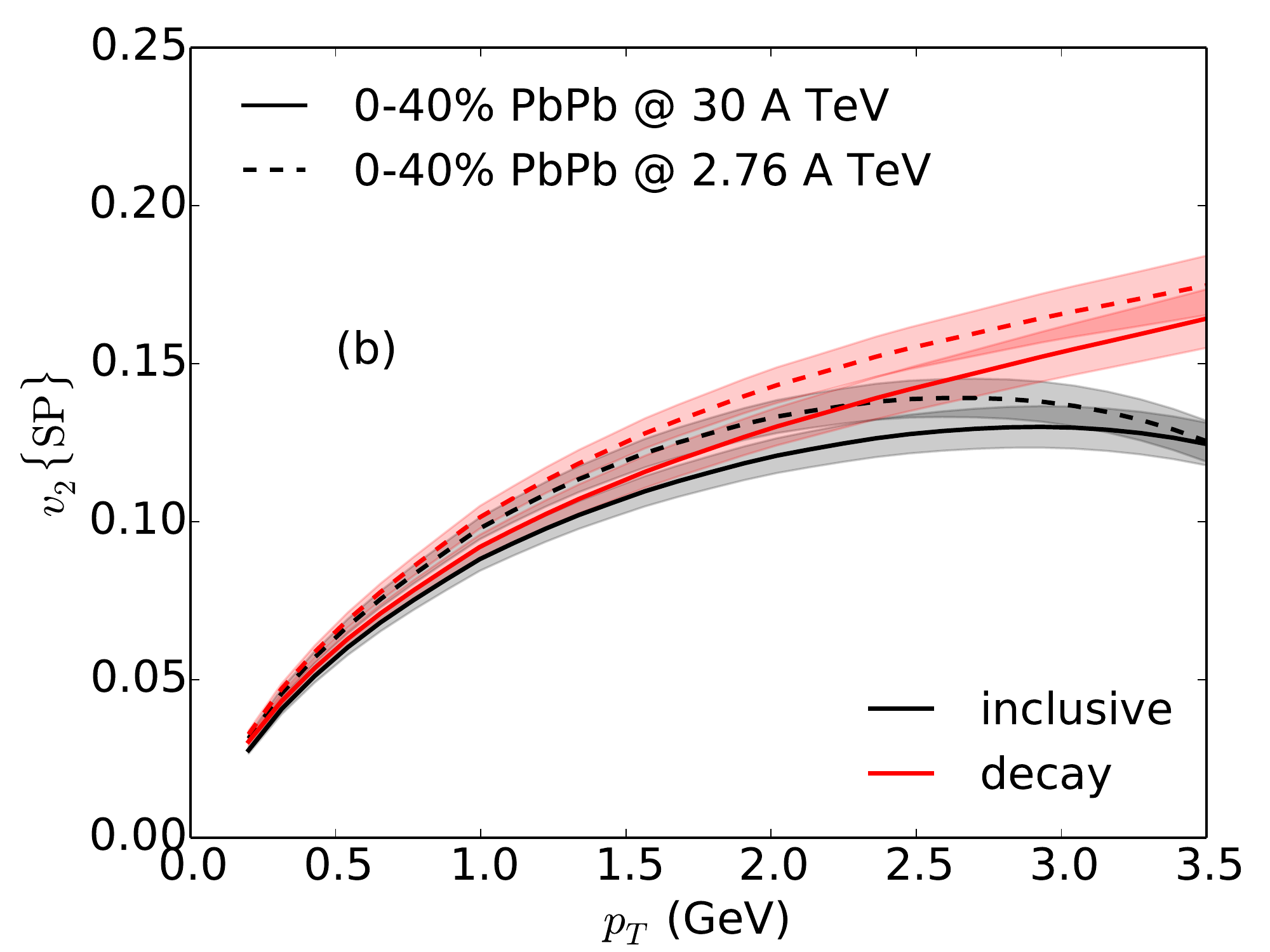} &
   \includegraphics[width=0.33\linewidth]{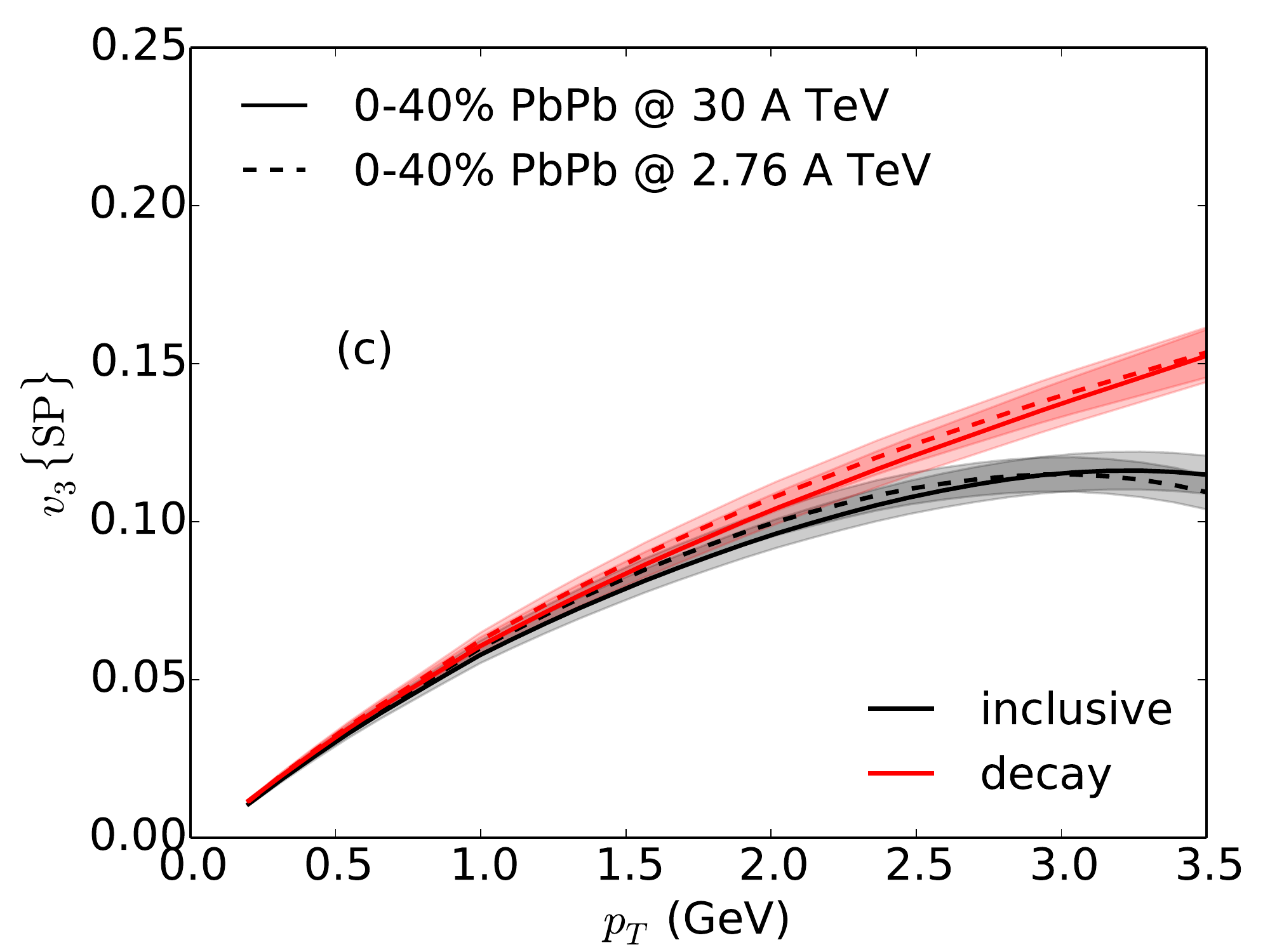}
\end{tabular}
\caption{Comparisons of the inclusive, decay, and direct photon spectra and $v_n\{\mathrm{SP}\}(p_T)$ between 0-40\% Pb+Pb collisions at 30 A TeV and at 2.76 A TeV.}
\label{fig4}
\end{figure}
%=======================================
In Figs.~\ref{fig4}, we compare the inclusive photon spectra as well as the decay cocktail between Pb+Pb collisions at 30 A TeV and at 2.76 A TeV. In Fig.~\ref{fig4}a, we find that the increase of direct photon production is larger compared to the increase in the inclusive photons. The signal to background ratio for direct photons increases as collision energy increases. This makes the direct photon measurement easier at 30 A TeV. In Figs.~\ref{fig4}b,c, we make predictions for the inclusive and decay photon $v_{2,3}\{\mathrm{SP}\} (p_T)$. Similar to hadrons, the $p_T$-differential anisotropic flows of inclusive and decay photons are very close the ones in Pb+Pb collisions at 2.76 A TeV.

%%%%%%%%%%%%%%%%%%%%%%%%%%%%%
\section{Summary}
%%%%%%%%%%%%%%%%%%%%%%%%%%%%
Since the discovery of the strongly coupled QGP at RHIC about a decode ago, experimental
and theoretical efforts in high-energy nuclear physics have been focused on the quantitative study
of the properties of the sQGP at extremely high temperatures. These include the extraction of the
shear viscosity to entropy density ratio of the bulk QGP medium, the jet transport parameter
for energetic jets propagating inside the QGP medium, and the diffusion coefficient for
heavy-flavor particles in QCD matter as formed in high-energy heavy-ion collisions at both RHIC
and LHC. The future frontier of heavy-ion collisions will be at both lower and very-high colliding 
energy regimes. For the latter, such as that in the beam energy scan (BES) program at RHIC and 
at FAIR, one expects to reach the highest baryon density in heavy-ion collisions, to explore
the phase structure of QCD matter, in particular search for signals of a critical end-point in 
the QCD phase transition. At the high-energy frontier, one expects to increase the initial 
temperatures that are currently possible at RHIC and LHC in the central region of the two 
colliding nuclei. Under these conditions, the properties of QGP medium might approach that 
of weakly interacting quarks and gluons. According to predictions by pQCD, the ratio of shear 
viscosity to entropy density and the heavy-flavor diffusion coefficient (scaled by $1/T$) should 
increase, while the jet transport parameter (scaled by $T^3$) should decrease.

Using HIJING and CGC models, we have estimated the final charged-hadron multiplicity in central 
Pb+Pb collisions at $\sqrt{s}=30$ TeV to be about 70\% larger than at the current LHC energy 
($\sqrt{s}=2.76$ TeV). Assuming the initial thermalization time to be the same as at the
LHC, $\tau_0=0.6$ fm/$c$, the initial temperature of a thermalized QGP at $\sqrt{s}=30$ TeV will 
be about $T_0\approx 560$ MeV. Based on our calculations of the anisotropic flow of charged hadrons 
using an event-by-event 3+1D ideal hydrodynamic model with fluctuating initial conditions, we 
expect to see strong signals of higher harmonic flow which should provide stringent constraints 
on the shear viscosity. We have also calculated the suppression factors for charged hadrons with 
large transverse momentum in central Pb+Pb collisions at $\sqrt{s}=30$ TeV within two different 
approaches to energy loss. The suppression factor is found to continue to decrease over a large 
range of transverse momenta, and thus provides sufficient sensitivity to determine the jet transport
parameter at such high initial temperatures. Though the suppression of full jet production is 
not as sensitive to the increase in initial temperature, the jet profile function is found to 
be significantly modified and should provide additional constraints on properties of the QGP 
medium. For open heavy flavor, both the high-$p_T$ suppression and the elliptic flow are expected 
to increase by about 20\%. The final $J/\psi$  yield in Pb+Pb collisions at $\sqrt{s}=30$\,TeV
is predicted to be more strongly dominated by regeneration from the recombination of initially 
produced charm quarks, due to a near-complete suppression of the initially produced $J/\psi$. 
The final $J/\psi$ nuclear modification factor, however, might turn out to be smaller than that 
at LHC due to an expected suppression of initial charm quark production by the gluon nuclear 
shadowing. It will therefore be essential to determine gluon shadowing from p+A and e+A collisions 
in order to reliably quantify the mechansims for $J/\psi$'s regeneration in the QGP medium.
The calculations of electromagnetic radiation from the medium show a more pronounced increase in 
yields as the final spectra receive significant contributions throughout the entire fireball 
evolution. For example, low-mass dilepton yields increase by a factor of $\sim$2 and allow for 
a ``measurement" of the increased fireball lifetime at higher colliding energies. 

To conclude, based on the calculations presented here, a systematic study of the above 
experimental observables at a future very high energy heavy-ion collider will 
provide us with an opportunity to significantly improve our understanding of the properties 
of the QGP and, in particular, open a window on the weakly interacting limit 
of QGP at very high temperature.

\section*{Acknowledgements}
Work in this paper is partially supported by the NSFC under Grant No. 11175071, No. 11221504,  No. 11305089,  No. 11322546, No. 11375072, No. 11435001, No. 11435004,  China MOST under Grant No. 2014DFG02050, No. 2015CB856900,
the Major State Basic Research Development Program in China (No. 2014CB845404 and No. 2014CB845403), 
the Natural Sciences and Engineering Research Council of Canada, the US National Science Foundation under grant number PHY-1306359, the Director, Office of Energy Research, Office of High Energy and Nuclear Physics, Division of Nuclear Physics, of the U.S. Department of Energy under Contract Nos. DE-AC02-05CH11231, DE-SC0012704 and within the framework of the JET Collaboration.  BJS is also supported by a DOE Office of Science Early Career Award.

%\bibliographystyle{apsrev}
%\bibliographystyle{h-physrev5}
%\bibliography{ref}

\end{document}